\documentclass[pdflatex,sn-nature]{sn-jnl}

\usepackage{graphicx}%
\usepackage{multirow}%
\usepackage{amsmath,amssymb,amsfonts}%
\usepackage{amsthm}%
\usepackage{mathrsfs}%
\usepackage[title]{appendix}%
\usepackage{textcomp}%
\usepackage{manyfoot}%
\usepackage{tabularx}
\usepackage{booktabs}%
\usepackage{algorithmicx}%
\usepackage{algpseudocode}%
\usepackage{listings}%
\usepackage{makecell}
\usepackage{enumerate}

\usepackage{ifthen}
\usepackage[table]{xcolor}
\usepackage{graphics,graphicx,epsfig}
\usepackage{epsf,epstopdf,wrapfig}
\usepackage{amssymb,amsfonts,amsmath,mathtools}
\usepackage[export]{adjustbox}
\usepackage[english]{babel}
\include{epsf}
\usepackage{ifthen}
\usepackage{algorithm2e}
\usepackage{rotating} % For rotating the table
\usepackage{soul} % for highlighting
\usepackage{subfigure}
\usepackage{comment}
\usepackage{multirow}

\definecolor{ocre}{RGB}{10,100,185}
\usepackage[normalem]{ulem}

\newcommand{\RNum}[1]{\uppercase\expandafter{\romannumeral #1\relax}}

\newcommand{\model}{\text{model}}
\newcommand{\prop}{\text{prop}}

\newcommand{\mt}{{\text{mt}}}
\newcommand{\wt}{{\text{wt}}}
\newcommand{\from}{{\text{from}}}
\newcommand{\tto}{{\text{to}}}

\newcommand{\proteinmpnn}{ProteinMPNN 0.02}
\newcommand{\proteinmpnnNoise}{ProteinMPNN 0.30}

\newcommand{\HERMESPy}{HERMES-{\em fixed} 0.00}

\newcommand{\HERMESPyFtRelaxed}{HERMES-{\em amortized} 0.00}

\newcommand{\HERMESPyNoise}{HERMES-{\em fixed} 0.50}

\newcommand{\HERMESPyNoiseFtRelaxed}{HERMES-{\em amortized} 0.50}
\newcommand{\HERMESPyNoiseRelaxed}{HERMES-{\em relaxed} 0.50}

\newcommand{\HERMESPyFtRos}{HERMES-{\em fixed} 0.00 + Ros}

\newcommand{\HERMESPyNoiseFtRos}{HERMES-{\em fixed} 0.50 + Ros}

\newcommand{\HERMESPyFtCdna}{HERMES-{\em fixed} 0.00 + cDNA117k}

\newcommand{\HERMESPyFtRelaxedFtCdna}{HERMES-{\em amortized} 0.00 + cDNA117k}

\newcommand{\HERMESPyNoiseFtCdna}{HERMES-{\em fixed} 0.50 + cDNA117k}

\newcommand{\HERMESPyNoiseFtRelaxedFtCdna}{HERMES-{\em amortized} 0.50 + cDNA117k}

\newcommand{\HERMESPyFtCdnaESM}{HERMES-{\em fixed} 0.00 + cDNA117k train ESMFold}

\newcommand{\HERMESPyNoiseFtCdnaESM}{HERMES-{\em fixed} 0.50 + cDNA117k train ESMFold}
\newcommand{\HERMESPyUntrainedFtCdna}{HERMES-{\em fixed} Untr. 0.00 + cDNA117k}
\newcommand{\HERMESPyNoiseUntrainedFtCdna}{HERMES-{\em fixed} Untr. 0.50 + cDNA117k}

\newcommand{\HERMESPyFtSkempiEasy}{HERMES-{\em fixed} 0.00 + SKEMPI Easy}
\newcommand{\HERMESPyNoiseFtSkempiEasy}{HERMES-{\em fixed} 0.50 + SKEMPI Easy}
\newcommand{\HERMESPyFtSkempiMedium}{HERMES-{\em fixed} 0.00 + SKEMPI Medium}
\newcommand{\HERMESPyNoiseFtSkempiMedium}{HERMES-{\em fixed} 0.50 + SKEMPI Medium}
\newcommand{\HERMESPyFtSkempiHard}{HERMES-{\em fixed} 0.00 + SKEMPI Difficult}
\newcommand{\HERMESPyNoiseFtSkempiHard}{HERMES-{\em fixed} 0.50 + SKEMPI Difficult}

\newcommand{\EQ}{\begin{equation}}
\newcommand{\EE}{\end{equation}}
\newcommand{\EQA}{\begin{eqnarray}}
\newcommand{\EEA}{\end{eqnarray}}

\newcommand{\yellowl}{\cellcolor{yellow!25}}

\newcommand{\greenl}{\cellcolor{green!20}}
\newcommand{\green}{\cellcolor{green!50}}

\theoremstyle{thmstyleone}%
\theoremstyle{thmstyletwo}%

\theoremstyle{thmstylethree}%

\begin{document}

\title[HERMES: Holographic Equivariant neuRal network model for Mutational Effect and Stability prediction]{HERMES: Holographic Equivariant neuRal network model for Mutational Effect and Stability prediction }

%%=============================================================%%
%% GivenName	-> \fnm{Joergen W.}
%% Particle	-> \spfx{van der} -> surname prefix
%% FamilyName	-> \sur{Ploeg}
%% Suffix	-> \sfx{IV}
%% \author*[1,2]{\fnm{Joergen W.} \spfx{van der} \sur{Ploeg} 
%%  \sfx{IV}}\email{iauthor@gmail.com}
%%=============================================================%%

\author*[1]{\fnm{Gian Marco} \sur{Visani}}\email{gvisan01@cs.washington.edu}

\author[1]{\fnm{William} \sur{Galvin}}

\author[2, 3]{\fnm{Zac} \sur{Jones}}

\author[4]{\fnm{Michael N.} \sur{Pun}}

\author[1]{\fnm{Eric} \sur{Daniel}}

\author[4]{\fnm{Kevin} \sur{Borisiak}}

\author[5]{\fnm{Utheri} \sur{Wagura}}

\author*[1, 4, 6, 7, 8, 9]{\fnm{Armita} \sur{Nourmohammad}}\email{armita.nourmohammad@yale.edu}

\affil*[1]{\orgdiv{Paul G. Allen School of Computer Science and Engineering}, \orgname{University of Washington}, \orgaddress{\city{Seattle}, \state{WA}, \country{USA}}}

\affil[2]{\orgdiv{Department of Biochemistry}, \orgname{University of Washington}, \orgaddress{\city{Seattle}, \state{WA}, \country{USA}}}

\affil[3]{\orgname{Institute for Protein Design}, \orgaddress{\city{Seattle}, \state{WA}, \country{USA}}}

\affil[4]{\orgdiv{Department of Physics}, \orgname{University of Washington}, \orgaddress{\city{Seattle}, \state{WA}, \country{USA}}}

\affil[5]{\orgdiv{Department of Physics}, \orgname{Massachusetts Institute of Technology}, \orgaddress{\city{Cambridge}, \state{MA}, \country{USA}}}

\affil[6]{\orgdiv{Department of Applied Mathematics}, \orgname{University of Washington}, \orgaddress{\city{Seattle}, \state{WA}, \country{USA}}}

\affil[7]{\orgname{Fred Hutchinson Cancer  Center}, \orgaddress{\city{Seattle}, \state{WA}, \country{USA}}}

\affil[8]{\orgname{Yale Center for Systems and Engineering Immunology}, 
\orgname{Yale University}\orgaddress{\city{New Haven}, \state{CT}, \country{USA}}}

\affil[9]{\orgdiv{Departments of Immunobiology and Biomedical Engineering}, \orgname{Yale University}, \orgaddress{\city{New Haven}, \state{CT}, \country{USA}}}

\abstract{
Accurately predicting how amino acid substitutions alter protein function is a central challenge in biology, with applications from interpreting disease variants to engineering vaccines and therapeutic proteins. We introduce HERMES, a family of fast, structure-based models that predict mutational effects from the local three-dimensional atomic environment around each residue. Pre-trained on the masked amino-acid prediction task, HERMES shows strong zero-shot performance for predicting changes in thermodynamic stability and protein–protein binding affinity. We find that this pre-training induces a bias toward substitutions with similar size to the wild-type. To address this, we develop an amortized fine-tuning strategy that incorporates packing flexibility, substantially reducing size-based bias while preserving sensitivity to mutational effects. We demonstrate that HERMES  can then be fine-tuned on experimental measurements without adding parameters or relying on costly data augmentation, achieving performance competitive with state-of-the-art stability predictors.  Finally, we show that HERMES identifies antigen-stabilizing mutations across multiple viral envelope proteins, enabling computationally efficient, structure-guided vaccine design. Together, these results establish HERMES as a practical and accurate framework for structure-based mutational effect prediction.
}


\maketitle

\section{Introduction}

\noindent Understanding the effects of amino acid substitutions on protein function is fundamental to biological discovery and engineering, with applications spanning disease variant identification~\citep{gerasimavicius_identification_2020,blaabjerg_rapid_2023}, enzyme optimization and engineering~\citep{ishida_effects_2010,wang_d3distalmutation_2021}, viral escape prediction~\citep{thadani_learning_2023,Luksza2014-pg,Neher2014-pf}, antibody engineering~\citep{hie_efficient_2024}, and vaccine antigen stabilization~\citep{mclellan_structure-based_2013, gonzalez_general_2024, bakkers_efficacious_2024, milder_universal_2022, phan_conserved_2022}.

Mutational effects on thermodynamic stability and binding affinity are particularly well-studied, as stability is typically prerequisite for function~\citep{rocklin_global_2017} and most biological processes are mediated by binding events. While these effects can be measured experimentally via denaturation assays~\citep{lindorff-larsen_linking_2021}, surface plasmon resonance~\citep{karlsson_spr_2004}, or deep mutational scanning~\citep{fowler_deep_2014,Kinney2019-wc,starr_deep_2020}, such experiments remain laborious despite recent throughput improvements~\citep{tsuboyama_mega-scale_2023}.

Computational approaches offer an alternative.  Molecular dynamics simulations can accurately capture short-time (nano seconds) responses, but are limited in predicting substantial conformational changes often induced by mutations~\citep{gapsys_accurate_2016}. Physics-based energy functions, including FoldX~\citep{schymkowitz_foldx_2005} and Rosetta~\citep{kellogg_role_2011}, remain widely used to predict mutation-induced stability changes~\citep{gonzalez_general_2024}, yet are often slow and  inaccurate~\citep{blaabjerg_rapid_2023}. In recent years, machine learning models have shown considerable progress in this area: models trained to predict amino acid propensities at a given residue from surrounding sequence~\citep{riesselman_deep_2018, meier_language_2021} or structural context~\citep{pun_learning_2024, diaz_stability_2024, blaabjerg_rapid_2023, li_predicting_2020,benevenuta_antisymmetric_2021, dieckhaus_transfer_2024} can approximate mutational effects across phenotypes. 

Recent work improves these pre-trained baselines by modifying the pre-training objective~\citep{gordon_protein_2024,gong_evolution-inspired_2024}, and by fine-tuning to predict phenotype-specific mutational effects~\citep{dieckhaus_transfer_2024, diaz_stability_2024, blaabjerg_rapid_2023, luo_rotamer_2022}, with thermodynamic stability as a frequent target for structure-based models~\citep{dieckhaus_transfer_2024, diaz_stability_2024, blaabjerg_rapid_2023}. Notable examples include RaSP~\citep{blaabjerg_rapid_2023}, which fine-tunes a 3D convolutional neural network (CNN) on Rosetta-computed $\Delta\Delta G$ values~\citep{chaudhury_pyrosetta_2010}; Stability-Oracle~\citep{diaz_stability_2024}, a graph transformer fine-tuned on experimental stability measurement from the Megascale dataset~\citep{tsuboyama_mega-scale_2023}; and ThermoMPNN~\citep{dieckhaus_transfer_2024}, which fine-tunes the inverse folding model ProteinMPNN~\citep{dauparas_robust_2022} on a different subset of Megascale.

Here, we present HERMES, a family of structure-based models built upon our previous H-CNN architecture~\citep{pun_learning_2024}. Like H-CNN, HERMES employs a 3D rotationally equivariant, all-atom CNN architecture, but incorporates implementation improvements yielding a $\sim 2.75\times$ speedup and an adaptable architecture enabling fine-tuning for arbitrary downstream tasks. We first pre-train HERMES on an inverse folding objective (i.e., predicting a residue's amino acid identity from its surrounding atomic neighborhood within a 10 \AA~radius), then fine-tune for predicting mutational effects. The HERMES family comprises three model variants that differ in their treatment of structural context: HERMES-{\em fixed}, which holds the local environment static during prediction; HERMES-{\em relaxed}, which enables local structure relaxation through explicit side-chain repacking; and HERMES-{\em amortized}, which implicitly encodes the structural flexibility associated with different substitutions through amortization. We systematically analyze the biases of these models with respect to the physicochemical properties of mutating residues and characterize their utility across different tasks.

On thermodynamic stability benchmarks, we show that fine-tuned HERMES models match or exceed state-of-the-art performance. However, we identify a ``wild-type preference bias" introduced by the pre-training objective that is only partially eliminated by fine-tuning. We also demonstrate HERMES' utility for structure-based vaccine design, where stabilizing viral envelope glycoproteins in their metastable pre-fusion conformation is essential for presenting neutralizing antibody epitopes~\citep{mclellan_structure-based_2013}. Identifying stabilizing mutations traditionally requires domain expertise and costly experimental iteration. While computational approaches using Rosetta~\citep{gonzalez_general_2024, phan_conserved_2022} or machine learning (e.g., ReCAP~\citep{bakkers_efficacious_2024}) have emerged, HERMES offers significant computational efficiency over Rosetta and, unlike ReCAP, is publicly available. When evaluated on 33 known stabilizing mutations across 5 viral envelopes, HERMES ranks 19 within the top 3 predicted substitutions at each position, with particularly strong performance on mutations that stabilize independently without synergistic interactions. Lastly, we demonstrate that HERMES can be fine-tuned to predict mutational effects on protein-protein binding affinity, benchmarking competitively against existing models.

Our code is open source at \url{https://github.com/StatPhysBio/hermes/tree/main}, and allows users to both run the models presented in this paper, and easily fine-tune HERMES models on their data.

\begin{figure*}[ht!]
    \centering
    \includegraphics[width=1.0\linewidth]{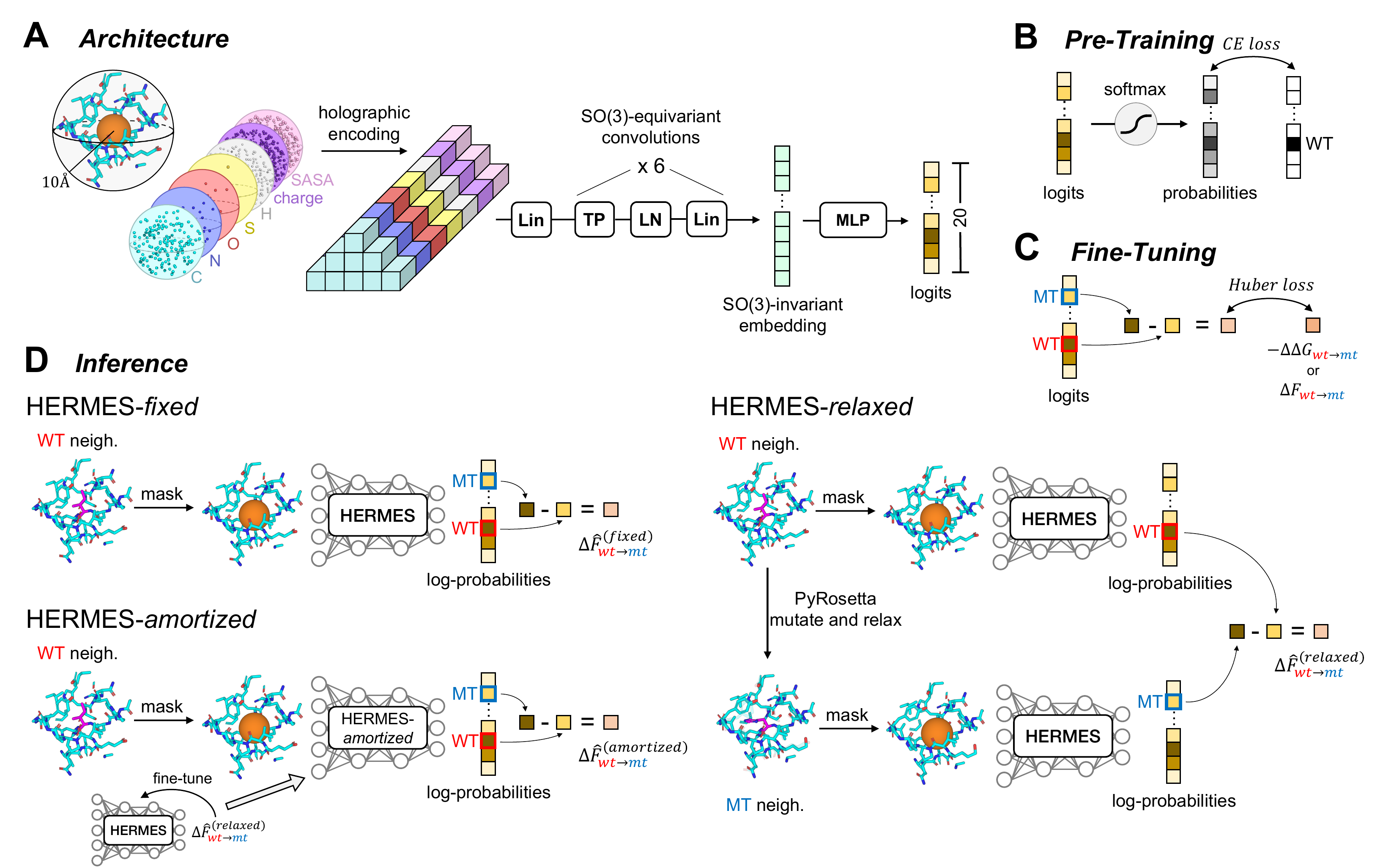}    \caption{\textbf{Overview of HERMES.} 
    \textbf{(A)}   Model architecture: HERMES takes as input an all-atom structural neighborhood (10~\AA{} radius) around a masked focal residue. Each atom is represented by its 3D coordinates, element type, partial charge, and solvent-accessible surface area (SASA). The neighborhood is projected onto a Zernike Fourier basis (spherical hologram) and processed by {rotation (SO(3))} equivariant layers to produce a rotation-invariant embedding, which is mapped to 20 amino-acid-specific logits (Methods).\textbf{(B)} 
    Pre-training: models are trained to predict the identity of the masked focal residue from its surrounding {atomic} neighborhood; logits in (A) are converted to amino-acid propensities (probabilities) via a softmax. \textbf{(C)} 
    Fine-tuning for mutational effects: model weights are optimized to regress the logit difference between mutant and wild-type amino acids to the {corresponding} experimental mutational effect. \textbf{(D)} Inference protocols. HERMES-{\em fixed} scores a substitution as the difference between the the mutant and wild-type amino-acids logits from a single forward pass, conditioned on the masked wild-type neighborhood $X_\wt$. HERMES-{\em relaxed} conditions the mutant term on an approximate mutant neighborhood $\hat X_{\wt\to\mt}$ generated in-silico by introducing the mutation on the wild-type structure and locally relaxing the structure {with} Rosetta~\citep{chaudhury_pyrosetta_2010}. {HERMES-{\em amortized}} distills the relaxed protocol by fine-tuning on HERMES-{\em relaxed} predictions, enabling fast fixed-style inference while retaining relaxation-aware behavior.} 
    \label{fig:HERMES}
\end{figure*}

\section{Model}

\noindent{\bf HERMES architecture and pre-training on masked amino acid classification.}
HERMES is a 3D, rotationally equivariant convolutional neural network that predicts the propensity of the 20 different canonical amino acids at a masked (removed) {focal} residue, given its local atomic neighborhood within the protein structure (Fig.~\ref{fig:HERMES}A). Building on our prior atomistic model H-CNN~\citep{pun_learning_2024}, HERMES achieves faster inference and supports task-specific fine-tuning (Methods). Atomic neighborhoods are featurized by atom type (including added hydrogens), partial charge, and solvent-accessible surface area, and then projected onto an orthonormal Zernike Fourier basis centered at the masked C-$\alpha$ to form a \textit{holographic encoding}. SO(3)-equivariant layers of the HERMES neural network map this encoding to a rotation-invariant embedding, which a final MLP converts into amino-acid propensities; additional details on Fourier-space SO(3) models are provided in the Methods and in~\citep{visani_holographic-vae_2024,visani_h-packer_2024}.

For pre-training, we train HERMES to recover the identity of the masked residues on ProteinNet's CASP12 set, filtered at 30\% sequence identity~\citep{alquraishi_proteinnet_2019} (Fig.~\ref{fig:HERMES}B). Each model is an ensemble of 10 independently trained networks (3.5M parameters each), with predictions averaged at inference. To improve robustness for zero-shot applications, we additionally train models with 0.50~\AA~coordinate noise. Pre-training performance is summarized in Table~\ref{table:aa_wt_cls}.\\

\noindent{\bf Zero-shot prediction of mutational effects with pre-trained HERMES.}
The pre-trained HERMES model outputs $P(aa| X_{i,aa})$, the probability of assigning amino acid  $aa$ at residue $i$, given  the surrounding atomic environment (neighborhood) $X_{i,aa}$ in the structure. Crucially,  $X_{i,aa}$ depends on the $aa$: the neighborhood geometry has the fingerprint of the masked amino acid during pre-training.

Conditional models are widely used for zero-shot prediction of mutational effects on protein function (e.g., \citep{riesselman_deep_2018, meier_language_2021, pun_learning_2024}). We approximate the effect of wild-type to mutant ($\wt \to \mt$) substitution at residue $i$ by  the log-likelihood ratio between the wild-type  and the mutant  amino acids, conditioned on the respective local atomic neighborhoods $X_{i,\wt}$ and $X_{i,\mt}$ (omitting the residue index $i$ for clarity):
\begin{equation}
\begin{aligned}
  \Delta \hat F_{\wt \to \mt}  \propto  \log P(\mt | X_{\mt}) - \log P(\wt | X_{\wt})
\end{aligned}
\label{eq:log_ratio_wt_and_mt}
\end{equation}
Here, the neighborhood $X$ depends on the identity of the original amino acid ($\wt$ or $\mt$), suggesting the potential need for having access to  mutant structures for such predictions. The $\hat\cdot$ {(hat)} indicates the model-predicted mutational effects, as opposed to experimental measurements (no hat).

When a mutant structure is unavailable, neighborhoods can be relaxed {\em in silico} (e.g., by Rosetta~ \citep{chaudhury_pyrosetta_2010}), though this procedure is computationally expensive. A common alternative is to score all mutations using a single (typically wild-type) structure~\citep{dauparas_robust_2022,diaz_stability_2024,pun_learning_2024}. Here, we {consider} both of these protocols: {\bf HERMES-{\em fixed}} evaluates mutational effects using the wild-type structure only (for both $\mt$ and $\wt$ propensities), while {\bf HERMES-{\em relaxed}} evaluates the propensity of the mutant amino acid in the Rosetta-relaxed $\mt$ neighborhood $\hat X_{\wt \to \mt}$  starting from the available $\wt$ structure (Fig.~\ref{fig:HERMES}D); see Methods for details. The resulting estimates for mutational effects from these two approaches follow, 
\begin{equation}
\begin{aligned}
\Delta \hat F_{\wt \to \mt}^{\text{(HERMES-{\em fixed})}} &\propto \log P\left(\mt \mid X_{\wt}\right) - \log P\left(\wt \mid X_{\wt}\right)\\
\Delta \hat F_{\wt \to \mt}^{\text{(HERMES-{\em relaxed})}} &\propto \log P\left(\mt \mid \hat X_{\wt \to \mt}\right) - \log P\left(\wt \mid X_{\wt}\right).
\end{aligned}
\label{eq:deltaF_variants}
\end{equation}

\noindent{\bf Fine-tuning HERMES to predict protein function.}
Context-conditioned amino-acid likelihoods provide useful zero-shot proxies for mutational effects across diverse functions~\citep{riesselman_deep_2018, meier_language_2021, pun_learning_2024}, but supervised  models for specific functions can perform better in practice. Building on prior work~\citep{blaabjerg_rapid_2023, diaz_stability_2024}, we fine-tune HERMES models directly on mutational effect data. Unlike approaches that train a separate regression head~\citep{blaabjerg_rapid_2023, diaz_stability_2024, dieckhaus_transfer_2024}), we update the model end-to-end so that its \emph{predicted scores} themselves align with measurements. Specifically, as shown in Fig.~\ref{fig:HERMES}C,
we make the predicted HERMES-{\em fixed} $\Delta \hat F_{\wt \to \mt}^{\text{(HERMES-{\em fixed})}}$ (eq.~\ref{eq:deltaF_variants}) regress over the experimentally measured mutational effects $\Delta F_{\wt\to \mt}$ by minimizing a robust Huber loss,
\begin{equation}
   \mathcal{L} =  \text{HuberLoss}(\Delta \hat F_{\wt \to \mt}^{\text{(HERMES-{\em fixed})}}, \Delta F_{\wt \to \mt})
\label{eq:loss_function}
\end{equation}
which stabilizes training under outliers while calibrating predictions to the function of interest; see Methods for details.\\

\noindent{\bf Hermes-{\em amortized} to encode structural flexibility in HERMES-{\em fixed} by amortization.}
Because the structural-relaxation step in HERMES-\emph{relaxed} is computationally costly ($\sim 66$ times slower than HERMES-{\em fixed} on a single CPU / A40 GPU, Table~\ref{table:inference_speed}), we distill HERMES-\emph{relaxed} predictions into the model via a fine-tuning procedure. Concretely, we fine-tune the model so that the mutational-effect predictions produced with the fast HERMES-\emph{fixed} protocol regress to the corresponding HERMES-\emph{relaxed} predictions {via Eq.~\ref{eq:loss_function}} (Fig.~\ref{fig:HERMES}D). We perform this fine-tuning on a small subset of neighborhoods extracted from the pre-training proteins ($\sim$15k neighborhoods, 0.5\% of the total). The resulting amortized model HERMES-{\em amortized} runs at HERMES-\emph{fixed} speed (effectively the same protocol, just different model weights) yet closely matches HERMES-\emph{relaxed} performance (Fig.~\ref{fig:relaxed_vs_ft_relaxed}).\\

\section{Results}

\subsection{Predicting mutational effects on thermodynamic fold-stability}
\label{sec:results_stability}

The chief task we evaluated the HERMES models on was to predict mutational effects on thermodynamic folding stability.
Thermodynamic folding stability is defined as the change in Gibbs Free $\Delta G$ energy upon folding. Thus, the effect of a mutation $\wt \rightarrow \mt$ on folding stability is denoted by $\Delta \Delta G_{\wt \rightarrow \mt} = \Delta G_\mt - \Delta G_\wt$.\\
\begin{figure*}[ht!]
    \centering
    \includegraphics[width=0.80\textwidth]{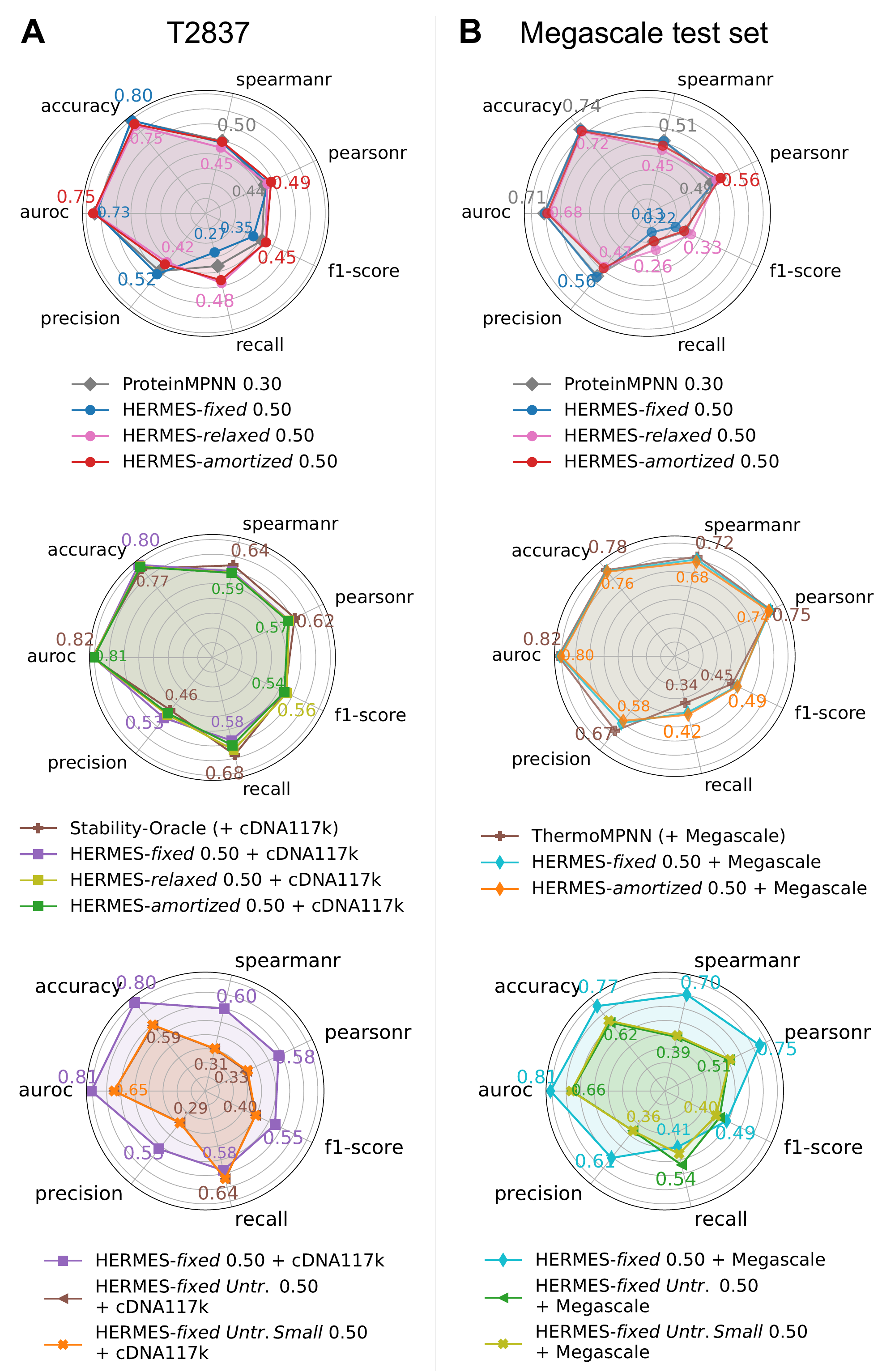}
    \caption{\textbf{Predicting mutational effects  on thermodynamic folding stability.} Stabilizing-versus-destabilizing classification metrics are computed using $\Delta\Delta G < 0$ (experimental) and $\Delta \log p > 0$ (predicted) as  cutoffs for  stabilizing mutations.
    \textbf{(A)} T2837 results: zero-shot models (top) and models fine-tuned or only trained on cDNA117k (middle and bottom). \textbf{(B)} Megascale test set results: zero-shot models (top) and models fine-tuned or only trained on the Megascale training set (middle and bottom). Model names indicate the architecture, the coordinate-noise amplitude used, and when applicable, the fine-tuning dataset (listed after ``$+$"); \textit{Untr.} is short for \textit{Untrained}, indicating models that had no pre-training and were instead only trained on stability effects. Only models trained with coordinate noise are shown; the noise amplitude is indicated within each model name {as standard deviation in~\AA~units}. Results for models trained without noise are provided in Fig.~\ref{fig:radial_plots__no_noise}.
    }
   \label{fig:radial_plots__noise}
\end{figure*}

\noindent {\bf Training and test Data.} To enable a direct comparison with recent structure-based stability predictors, we fine-tuned and evaluated HERMES on the same benchmark splits used by RaSP~\citep{blaabjerg_rapid_2023}, Stability-Oracle~\citep{diaz_stability_2024}, and ThermoMPNN~\citep{dieckhaus_transfer_2024}. Specifically, we (i) trained on the RaSP Rosetta-derived stability dataset and tested on the experimental benchmark used in RaSP, (ii) used Stability-Oracle's curated cDNA117k training set and T2837 test set; training on 117k $\Delta\Delta G$ values from ref.~\citep{tsuboyama_mega-scale_2023} filtered by enforcing $\leq 30\%$ sequence identity to the test set, and (iii) trained on ThermoMPNN's Megascale \textit{train} split and evaluated on its Megascale \textit{test} split, both from ref.~\citep{tsuboyama_mega-scale_2023}; training on 216k  and testing on 28k  mutation effects with 25\% sequence-identity cutoff. Because these splits are defined by the original studies, our results are directly comparable across methods. We additionally note that the Megascale \textit{train} split is not de-duplicated against T2837, and six of the T2837 proteins ($\sim 5\%$) have $>90\%$ sequence-similar homologs in the Megascale training set; see Methods for more details on these training and test sets.\\

\noindent\textbf{Noise and side-chain relaxation improve zero-shot model predictions.} We first evaluated the \emph{zero-shot} HERMES models (no fine-tuning on experimental data) on the T2837 and Megascale test sets (Fig.~\ref{fig:radial_plots__noise}, \ref{fig:radial_plots__no_noise}). Pre-training on structures with Gaussian coordinate noise (0.5~\AA{} s.d.) improves performance, consistent with prior work~\citep{dauparas_robust_2022, pun_learning_2024}. Relative to HERMES-\emph{fixed}, the packing-aware HERMES-\emph{relaxed} achieves significantly higher recall (0.48 vs. 0.27; p-value $<$ 0.01), with only a slight loss in precision, leading to a higher overall F1-score. HERMES-\emph{amortized} performs similarly to HERMES-\emph{relaxed}: on T2837, and on Megascale, it shows slightly lower recall and F1 while still outperforming HERMES-\emph{fixed}; the p-values for the significance of these performance differences are reported in Fig.~\ref{fig:t2837_and_megascale_pairwise_pvalues_permutation}.

\begin{figure*}
    \centering
    \includegraphics[width=0.9\textwidth]{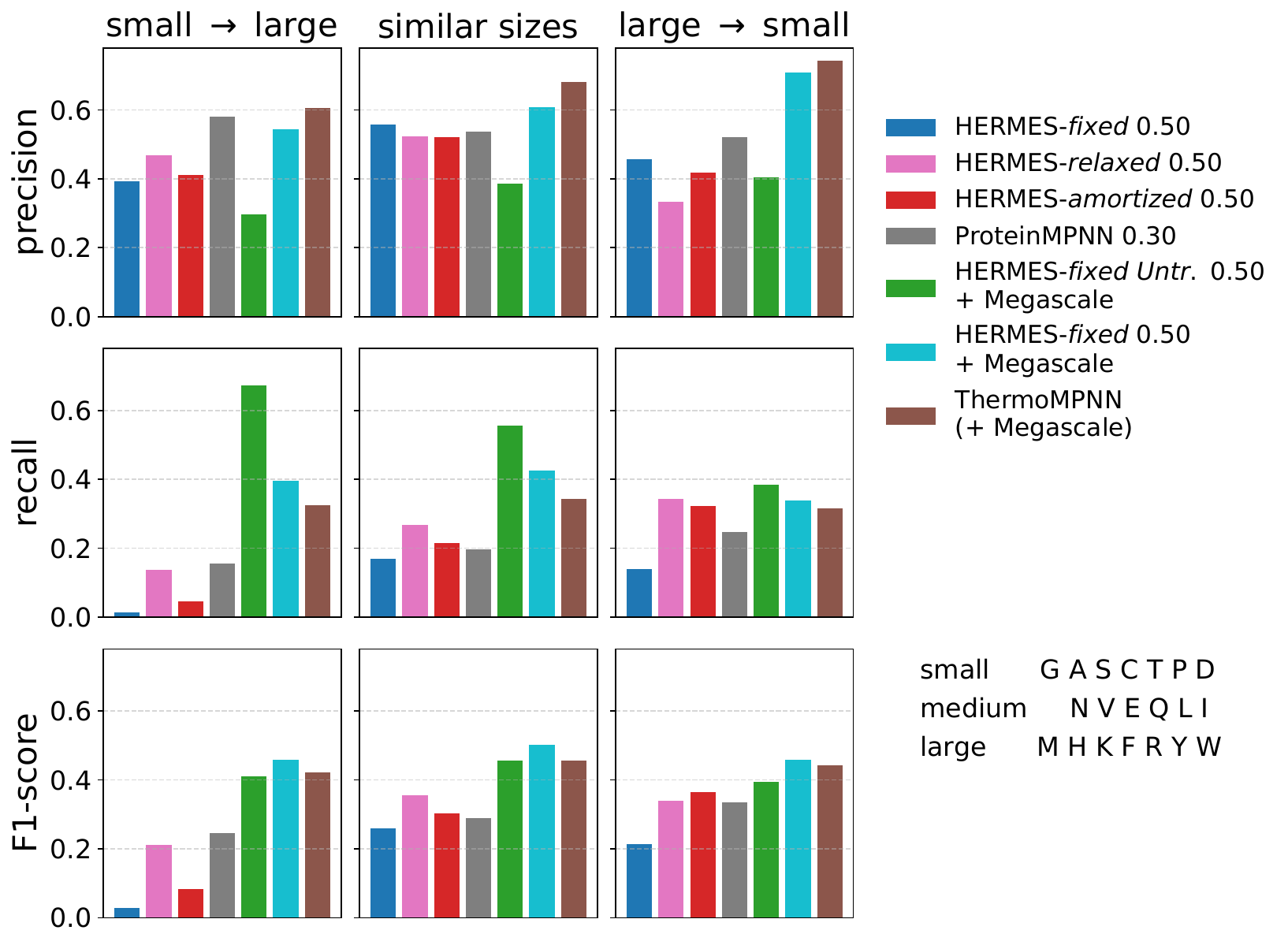}
    \caption{\textbf{Impact of amino acid size changes on  predicting mutational effects on protein stability.} 
    Amino acids are grouped into three size classes based on van der Waals volume (as listed), and predictions are stratified by wild-type and mutant size classes; ``similar sizes" denote substitutions within the same class.
    Stabilizing-versus-destabilizing classification metrics (rows) are then computed and shown for each stratum (columns), using $\Delta\Delta G < 0$ (experimental) and $\Delta \log p > 0$ (predicted) as cutoffs for stabilizing mutations.
    P-values for pairwise model comparisons are shown in Fig.~\ref{fig:pairwise_pvalues_permutation_bucketed_by_sizereduced_precision_recall_f1}, and p-values comparing each model's performance on small$\to$large vs. large$\to$small  substitutions are shown in Fig.~\ref{fig:pvalues__bucketed_by_size__between_small_large_buckets__vertical}. 
    Model names indicate the architecture, the coordinate-noise amplitude used during pre-training {as standard deviation in~\AA~units}, and when applicable, the fine-tuning dataset (listed after ``$+$").
}    \label{fig:megascale_precision_recall_f1_by_size_cutoff_0p0}
\end{figure*}

To test whether packing awareness preferentially improves predictions for substitutions that perturb local packing, we stratified mutations by residue size. Wild-type and mutant residues were assigned to small, medium, or large classes using normalized van der Waals volumes~\citep{mei_new_2005}, and we evaluated stabilizing-versus-destabilizing classification performance within each $\wt \to \mt$ size-transition group (Fig.~\ref{fig:megascale_precision_recall_f1_by_size_cutoff_0p0}).

Switching from HERMES-{\em fixed} to HERMES-{\em relaxed} yielded the largest recall gains in identifying stabilizing substitutions between residues with substantially different sizes. This improvement was most  pronounced for large$\to$small substitutions: HERMES-{\em fixed} is constrained by the rigid wild-type cavity, preventing it from recognizing that mutations that create voids could be stabilizing, whereas HERMES-{\em relaxed} can repack neighbors to fill the space and stabilize the smaller side chain. While precision changes varied across size categories, F1-score improved in all categories, with the largest gains occurring for  small$\to$large substitutions, where relaxation can reorganize the local environment to accommodate bulkier side chains that would otherwise clash. Interestingly, HERMES-\emph{amortized}, which learned packing implicitly through fine-tuning, showed comparable improvements for large$\to$small substitutions but much weaker gains for small$\to$large substitutions. This could suggest that stabilizing small$\to$large substitutions may be underrepresented in the fine-tuning training set; the p-values associated with the significance of performance differences between models within each size-transition category, and within models across different size-transition categories are reported in Figs.~\ref{fig:pvalues__bucketed_by_size__between_small_large_buckets__vertical},~\ref{fig:t2837_and_megascale_pairwise_pvalues_permutation}.

For comparison, we evaluated ProteinMPNN~\citep{dauparas_robust_2022}
on the same tasks.  ProteinMPNN outperformed HERMES-{\em fixed}, and performed on par with HERMES-\emph{relaxed} on both small$\to$large and large$\to$small mutations (Fig.~\ref{fig:megascale_precision_recall_f1_by_size_cutoff_0p0},~\ref{fig:pairwise_pvalues_permutation_bucketed_by_sizereduced_precision_recall_f1}). We attribute this  performance to ProteinMPNN's input representation: by conditioning only on backbone coordinates and residue identities, ProteinMPNN is less constrained by the explicit wild-type side-chain geometry compared to the all-atom HERMES-{\em fixed} model. This architectural choice enables implicit reasoning about side-chain flexibility, achieving results comparable to  using explicit relaxations  in HERMES-{\em relaxed}. Notably, ProteinMPNN performed better than HERMES-{\em amortized} on small$\to$large substitutions.\\

\noindent {\bf Fine-tuned HERMES models achieve state-of-the-art performance for stability effect prediction.}
When fine-tuning on the respective stability effect datasets, HERMES outperformed RaSP (Fig.~\ref{fig:rasp_exp}), and matched the performance of Stability-Oracle (Fig.~\ref{fig:radial_plots__noise}A,~\ref{fig:t2837_broken_down_and_comparison}) and ThermoMPNN (Fig.~\ref{fig:radial_plots__noise}B).
These results underscore both the effectiveness of the HERMES architecture and the importance of fine-tuning data for test-time accuracy. HERMES was also robust to using ESMFold-resolved structures~\citep{lin_evolutionary-scale_2023} for either fine-tuning or inference (Fig.~\ref{fig:radial_plots__esmfold}), supporting practical use cases where only computationally predicted structures are available.

Removing pre-training on wild-type amino-acid classification significantly degraded performance: HERMES models trained {\em only} for stability prediction performed poorly when trained on either cDNA117k or Megascale, even after reducing capacity from 3.5M parameters to 50k to mitigate overfitting (Fig.~\ref{fig:radial_plots__no_noise}). A similar failure mode was reported for ThermoMPNN on the Megascale training data~\citep{dieckhaus_transfer_2024}.

Finally, fine-tuning on stability effects largely eliminated size-dependent bias, yielding comparable recall and F1 scores for small$\to$large and large$\to$small substitutions (Figs.~\ref{fig:megascale_precision_recall_f1_by_size_cutoff_0p0},~\ref{fig:pvalues__bucketed_by_size__between_small_large_buckets__vertical}).\\

\begin{figure*}
    \centering
    \includegraphics[width=1.0\textwidth]{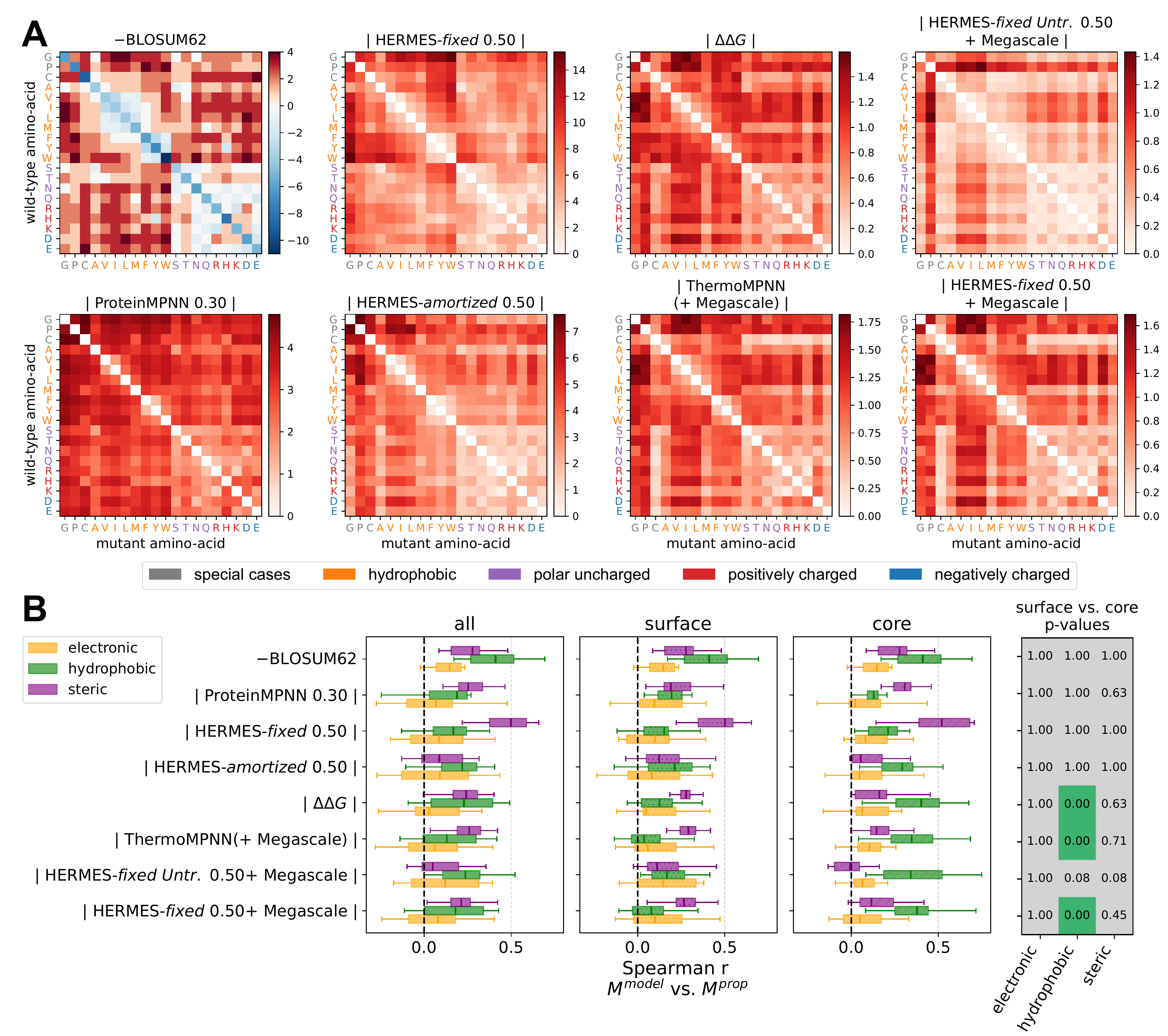}
    \caption{\textbf{Uncovering mutation preferences via model-averaged substitution matrices.}
    \textbf{(A)} 
    Heatmaps of model-averaged substitution matrices $M^{\model}$ computed by averaging over the Megascale test set, shown alongside BLOSUM62 and and the experimental matrix of mean $|\Delta\Delta G|$ values across mutations. 
    Spearman correlations between matrices are reported in Fig.~\ref{fig:spearmanr_between_model_average_substitution_matrices}. Core- and surface-restricted matrices (core: SASA $< 1 \AA^2$;  surface: SASA $> 3 \AA^2$) are shown in Fig.~\ref{fig:average_prediction_matrices__abs_symm__part_1}~and~\ref{fig:average_prediction_matrices__abs_symm__part_2}.
    \textbf{(B)} For each model and site subset (all, surface, core), boxplots summarize Spearman correlations between $M^{\model}$ and property-difference matrices $M^{\prop}$, grouped by property class (color). P-values from two-tailed t-tests comparing the surface vs. core correlations within each property class are shown on the right. P-values for between-model comparisons of property-class correlations are shown in Fig.~\ref{fig:correlation_to_aa_properties__significance}.
      }
    \label{fig:correlations_to_aa_properties}
\end{figure*}

\noindent \textbf{Biophysical interpretation of HERMES predictions for mutational  stability effects.}
We sought to quantify how much models' mutation preferences could be explained by amino-acid physicochemical properties. Specifically, we asked which properties are shared by amino-acid pairs that the model treats as interchangeable, i.e., substitutions with $\Delta \hat{F} \approx 0$ on average.

To do this, we constructed  model-averaged substitution matrices $M^{\model}$ whose $(\alpha,\beta)$ entry represents how neutrally the model treats exchanging amino acids $\alpha$ and $\beta$. For each pair $(\alpha,\beta)$, we computed the model-predicted mean {\em absolute} effect, $M^{\model}_{\alpha, \beta} = \langle |\Delta \hat{F}^{(\model)}_{\alpha,\beta} |\rangle$, where $\langle \cdot \rangle$
denotes averages over all $\alpha\to \beta$ and $\beta\to \alpha$ substitutions in the Megascale test set. This procedure yields a symmetric, non-negative $20\times 20$ matrix for each model,  in which values approaching zero indicate greater predicted interchangeability. For comparison, we constructed an analogous matrix using experimentally measured $|\Delta\Delta G|$ values from the Megascale test set ($M^{|\Delta\Delta G|}$), and include the BLOSUM62 substitution matrix as an additional reference.

Following ref.~\cite{pyo_data-driven_2025}, we also constructed symmetric, nonnegative matrices of \emph{absolute amino-acid property differences}. Specifically, we considered 50 quantitative properties spanning (i) hydrophobic, (ii) electronic, and (iii) steric categories (Table~\ref{table:aa_properties}), yielding 50 property-specific matrices   $M^{\prop}$ with the $(\alpha,\beta)$ entry: $M^{\prop}_{\alpha,\beta} = |\prop_\alpha - \prop_\beta|$. Smaller values indicate greater similarity between the two amino acids with respect to the specified property.

To quantify which biophysical properties each model tends to preserve, we computed the Spearman correlation between the model-averaged substitution  matrix
 $M^{\model}$ and each of property-specific amino acid distance  matrix $M^{\prop}$ (Fig.~\ref{fig:correlations_to_aa_properties}B). 
Consistent with the size-stratified analysis (Fig.~\ref{fig:megascale_precision_recall_f1_by_size_cutoff_0p0}), HERMES-{\em fixed} predicted amino-acid interchangeability align most strongly with  steric properties, particularly average buried-residue volume (Spearman $r =0.66$) and normalized van der Waals volume (Spearman $r = 0.64$). This correlations was significantly stronger than for the other models (p-values in Fig.~\ref{fig:correlation_to_aa_properties__significance}).
In contrast, BLOSUM62, derived from evolutionary substitution frequencies, preferentially preserves hydrophobic properties. When restricting $M^{\model}$ to core residues (SASA $< 1$\AA$^2$), the  interchangeability matrix associated with stability-fine tuned models {(e.g. $M^{\text{HERMES-\textit{fixed} 0.50 + Megascale}}$)} showed stronger alignment with hydrophobic properties, to levels comparable to BLOSUM62 and consistent with the experimental matrix $M^{|\Delta\Delta G|}$.
Zero-shot models instead did not show a comparable increase in hydrophobic preservation when restricting to the core (Fig.~\ref{fig:correlations_to_aa_properties}B).\\

\begin{figure*}[ht!]
    \centering
    \includegraphics[width=0.9\textwidth]{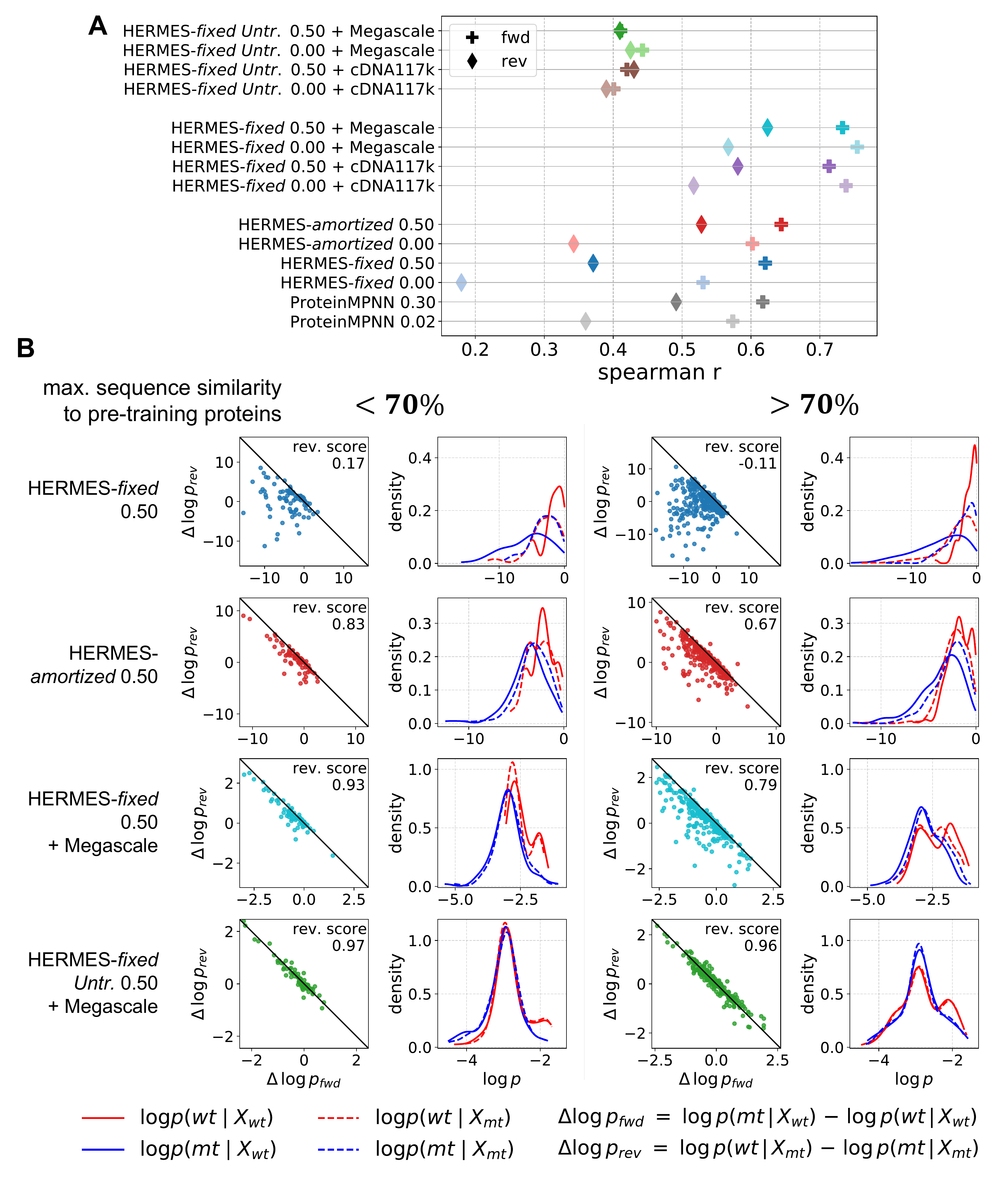}
    \caption{
    \textbf{Identifying model biases with structure-conditioned reversibility}
    \textbf{(A)} 
   Spearman correlations between experimental stability changes ($\Delta\Delta G$ and model predictions on the Ssym dataset are shown. For each model, we report correlations for forward substitutions ($\Delta \log p_{fwd}$ vs. $\Delta\Delta G$) and the reverse substitutions ($\Delta \log p_{rev}$ vs. $-\Delta\Delta G$). {\bf(B)} Ssym proteins are stratified by their maximum sequence identity to pre-training proteins ($\geq 70\%$ vs. $< 70\%$). For each split (columns) and four representative HERMES variants (rows), the left panel shows the scatter plots for $\Delta \log p_{fwd}$ vs. $\Delta \log p_{rev}$; an unbiased model should exhibit strong anti-correlation, summarized by the reversibility score (rev.; higher is more reversible; Eq.~\ref{eq:rev_score}). Reversibility is highest for the model not pre-trained on wild-type amino-acid classification (green; bottom row) and is consistently higher for the low-similarity subset.  Fig.~\ref{fig:ssym_antisymmetry_reversibility_score_results} shows the  reversibility scores across all models in (A). In each column, the right panel shows the distributions of the log-probabilities that make up $\Delta \log\,p_{fwd} = \log p (\mt | X_\wt) - \log p (\wt|X_\wt)$ (solid lines) and $\Delta \log\,p_{rev} = \log p (\wt | X_\mt) - \log p (\mt|X_\wt)$ (dashed lines). All models except for  the one that was \textit{not} pre-trained on wild-type amino-acid classification (last row) exhibit elevated $\log p(\wt\,|X_{\wt})$.
    }
    \label{fig:ssym_figure}
\end{figure*}
   
\noindent{\bf Reversibility and path-independence of HERMES predictions with respect to mutations.}
Mutational effects are, in principle, reversible: the effect of substituting a  residue from amino acid $\alpha$ to $\beta$ should be equal in magnitude and opposite in sign to the reverse change, i.e., $\Delta F_{\alpha \to \beta} = -\Delta F_{\beta\to \alpha}$. Moreover, equilibrium quantities such as protein stability free energy are state functions. As such, the net effect of a multi-step substitution depends only on the initial and final amino acids, not on the mutational path. For a path $\alpha \to \beta \to \gamma$, this implies: $\Delta F_{\alpha\to \gamma} = \Delta F_{\alpha\to \beta} + \Delta F_{\beta\to \gamma}$.

HERMES predicts mutation effects as differences in amino-acid–specific log-probabilities (logits) $\log p(aa | X_{aa})$ (both zero-shot  and after  fine-tuning). As log-probability differences under a fixed structural context, these predictions satisfy reversibility by construction:
\EQ 
\Delta \hat F^{(\text{model})}_{\alpha\to \beta} = \log P(\beta | X_{\beta}) - \log P(\alpha |X_{\alpha}) =  -\Delta \hat F^{(\text{model})}_{\beta\to \alpha}
\label{eq:reversibility}
\EE
where we set $X_{\alpha}= X_{\beta}$ for HERMES-{\em fixed} and HERMES-{\em amortized}, and  $X_{\beta}= \hat X_{\alpha\to \beta}$ for HERMES-{\em relaxed}. Path-independence follows similarly. 
In contrast, other structural models such as Stability-Oracle~\citep{diaz_stability_2024} require explicit $19\times$ data augmentation (termed ``thermodynamic permutation augmentation" in ref.~\citep{diaz_stability_2024}) to enforce these properties.

Alternatively, reversibility can be assessed in a structure-conditioned manner~\citep{blaabjerg_rapid_2023, dieckhaus_transfer_2024}, where the effects of forward $\alpha\to \beta$ and reverse $\beta \to \alpha$ substitutions are computed using the outgoing  structural contexts, i.e., $\alpha\to \beta$ is conditioned on $X_{\alpha}$ and $\beta\to\alpha$ on $X_{\beta}$ (Methods). Under this transformation, we do not automatically expect a ``structure-conditioned reversibility", as forward and reverse transitions are conditioned on different structures. However, a well-balanced model should have near-reversibility in this setting, making this a stringent test of model bias. We measured structure-conditioned reversibility on the Ssym dataset, which contains wild-type and single-mutant structures for 352 mutations across 19 proteins~\citep{pancotti_predicting_2022}.
For each mutation, we computed a ``forward" effect ($\wt \to\mt$, conditioned on the wild-type structure) and a ``reverse" effect ($\mt \to\wt$, conditioned on the mutant structure).

We found that zero-shot HERMES models and ProteinMPNN predict stability effects more accurately for  forward substitutions ($\wt \to\mt$) than for the reverse direction, consistent with previous observations~\citep{blaabjerg_rapid_2023, dieckhaus_transfer_2024} (Fig.~\ref{fig:ssym_figure}A). Adding coordinate noise during pre-training and fine-tuning on stability effects both reduced this forward–reverse disparity, but did not eliminate it. Models trained {\em only} on stability effects showed little or no disparity, albeit with substantially lower overall accuracy.
Across all pre-trained models, the predicted stability effects of forward mutations from the wild-type tend to have larger magnitudes than those of the corresponding reverse mutations (Fig.~\ref{fig:ssym_figure}B, scatterplots). This bias stems from the elevated log-probabilities assigned to wild-type amino acids in wild-type structure neighborhoods $\log p(\wt | X_{\wt})$ (Fig.~\ref{fig:ssym_figure}B, density plots), suggesting a wild-type preference that may reflect pre-training memorization. Consistently, stratifying Ssym proteins by sequence similarity to the pre-training set (Methods) yielded a modest reduction in wild-type preference for lower-similarity proteins (Fig.~\ref{fig:ssym_figure}B). A more detailed characterization of this bias is left to future work.

\begin{figure*}[t!]
    \centering
    \includegraphics[width=1.0\linewidth]{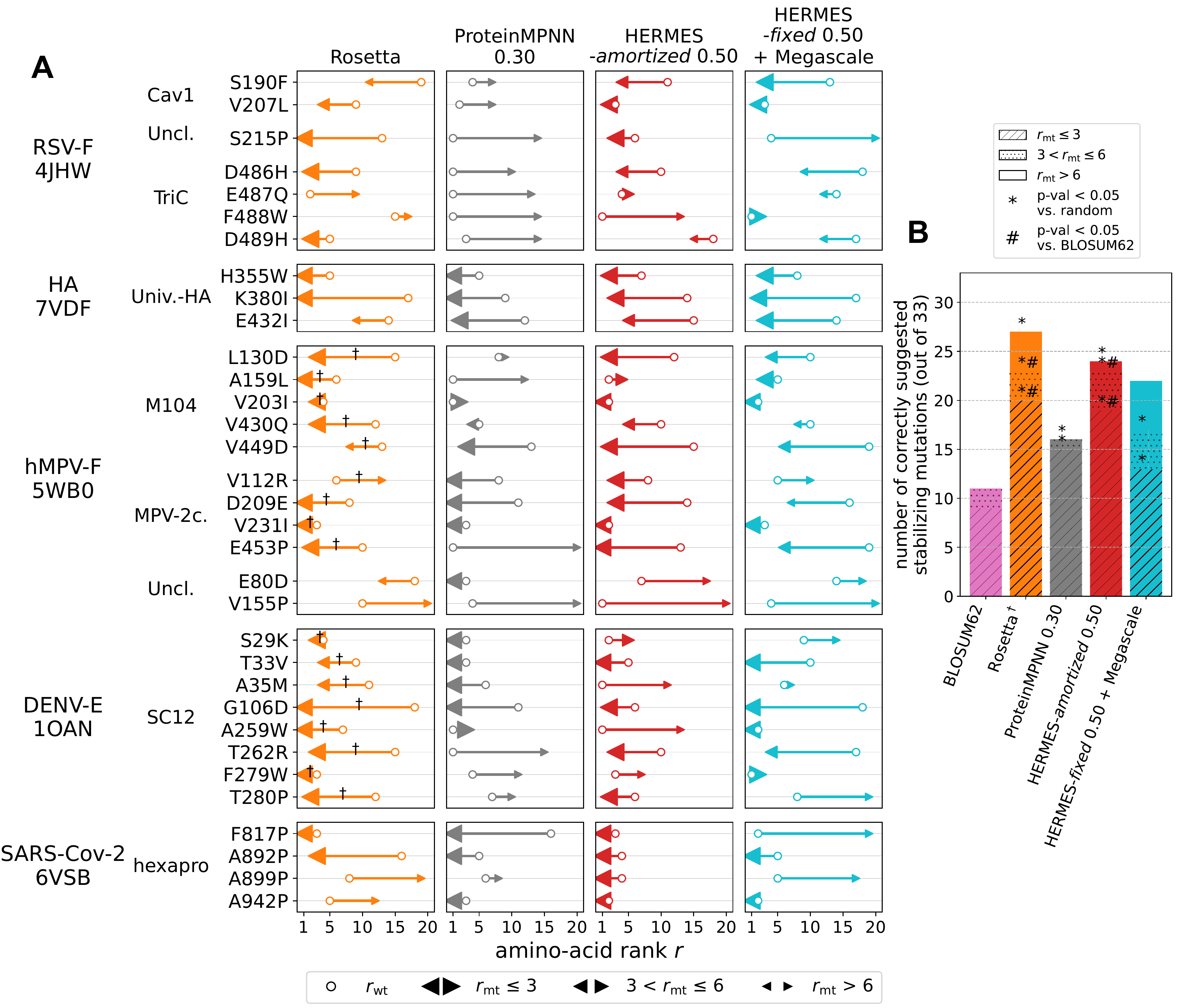}
    \caption{\textbf{Predicting antigen-stabilizing mutations with HERMES.}
    \textbf{(A)} Model recall is evaluated on 33 previously reported antigen-stabilizing mutations (rows) across five viral antigens. For each antigen, the PDB structure used for scoring is indicated. Mutations are specified as wild-type$\to$mutant substitutions at the annotated site. Four models (columns) are compared, as labeled above each column; see Table~\ref{table:antigen_results} for a more extensive comparison of models on this task. Arrows depict the change in predicted rank from the wild type (open circle) to the stabilizing mutant (arrow tip); arrow size indicates whether the mutant ranks in the top 3 (large), ranks 4--6 (medium), or ranks $>6$ (small). The $\dagger$ symbols mark mutations originally proposed as stabilizing by Rosetta. ``Univ.-HA" stands for ``Universal-HA"; ``MPV-2c." stands for MPV-2cREKR; ``Uncl." stands for ``Uncleaved Prefusion-Closed".  {\bf (B)} Counts of correctly prioritized antigen-stabilizing mutations ($r_\mt < r_\wt$), stratified by predicted rank group: strongly suggested ($r_\mt \leq 3$), moderately suggested ($3 < r_\mt \leq 6$), or weakly suggested ($6\leq r_\mt$) are shown for  each model. For comparison, BLOSUM62 is used to rank substitutions into the  ($r_\mt \leq 3$) or ($3 < r_\mt \leq 6$) groups; under this scheme the wild-type residue is always ranked first. Statistical enrichment is assessed by comparing the number of strong/moderate recalls to a random-ranking baseline and to BLOSUM62 (denoted by ($*$) and ($\#$), respectively, when binomial tests corrected p-value$<0.05$; see Methods).
    We note that all structures were considered in their native multimeric state, generating symmetric partners with PyMOL when necessary.
    }
    \label{fig:antigen_figure_main_results}
\end{figure*}

\subsection{Antigen stabilization  with HERMES  for vaccine design}
Structure-based vaccine design seeks to increase vaccine efficacy by stabilizing viral envelope glycoproteins (hereafter, ``antigens") in their metastable pre-fusion conformation. Stabilization enriches presentation of neutralization-relevant epitopes and can bias elicited immune responses toward protective specificities~\citep{mclellan_structure-based_2013, gonzalez_general_2024, bakkers_efficacious_2024, milder_universal_2022, phan_conserved_2022}.

To test whether HERMES can identify  antigen-stabilizing mutations, 
we benchmarked model performances on 33 previously reported antigen-stabilizing mutations drawn from five viral antigens: Influenza HA~\citep{milder_universal_2022} (3 mutations), RSV-F~\citep{mclellan_structure-based_2013, lee_rational_2024} (7 mutations), hMPV-F~\citep{gonzalez_general_2024, bakkers_efficacious_2024, phan_conserved_2022} (11 mutations), DENV-E~\citep{phan_conserved_2022} (8 mutations), and SARS-CoV-2 spike protein~\citep{hsieh_structure-based_2020} (4 mutations). We scored mutations using zero-shot and stability-fine-tuned variants of ProteinMPNN and HERMES. Importantly, none of the stabilized variants in this benchmark appeared in the training data of any HERMES model, either during pre-training or during stability fine-tuning.
 
To quantify performance, we emulated a simple model-guided selection workflow in which a practitioner considers substitutions at a site in descending order of model scores. For each known stabilizing mutation, we record its rank $r_\mt$ among all 20 amino acids at that position and compare it to the wild-type rank $r_\wt$. We posit that a practitioner would select a particular mutant for experimental validation if (1) its rank $r_\mt$ is better (lower) than that of the wild-type $r_\wt$, and (2) its rank is among the top-scoring candidates (lower end). We categorize each predicted-stabilizing mutation (i.e., with $r_\mt < r_\wt$) as strongly suggested ($r_\mt \leq 3$), moderately suggested ($3 < r_\mt \leq 6$), or weakly suggested ($r_\mt > 6$) (Fig.~\ref{fig:antigen_figure_main_results} {and Table~\ref{table:antigen_results}}). This scheme would group mutations with similar physico-chemical properties into the same rank class. Consequently, models are generally not penalized for ranking a biophysically similar alternative above a known stabilizing mutant, making our performance metrics robust to fine-grained rank differences in different models (see Fig.~\ref{fig:antigen_heatmap} and SI for a more detail discussion on antigen-stabilizing mutations).

\begin{figure*}[t!]
    \centering
    \includegraphics[width=0.9\linewidth]{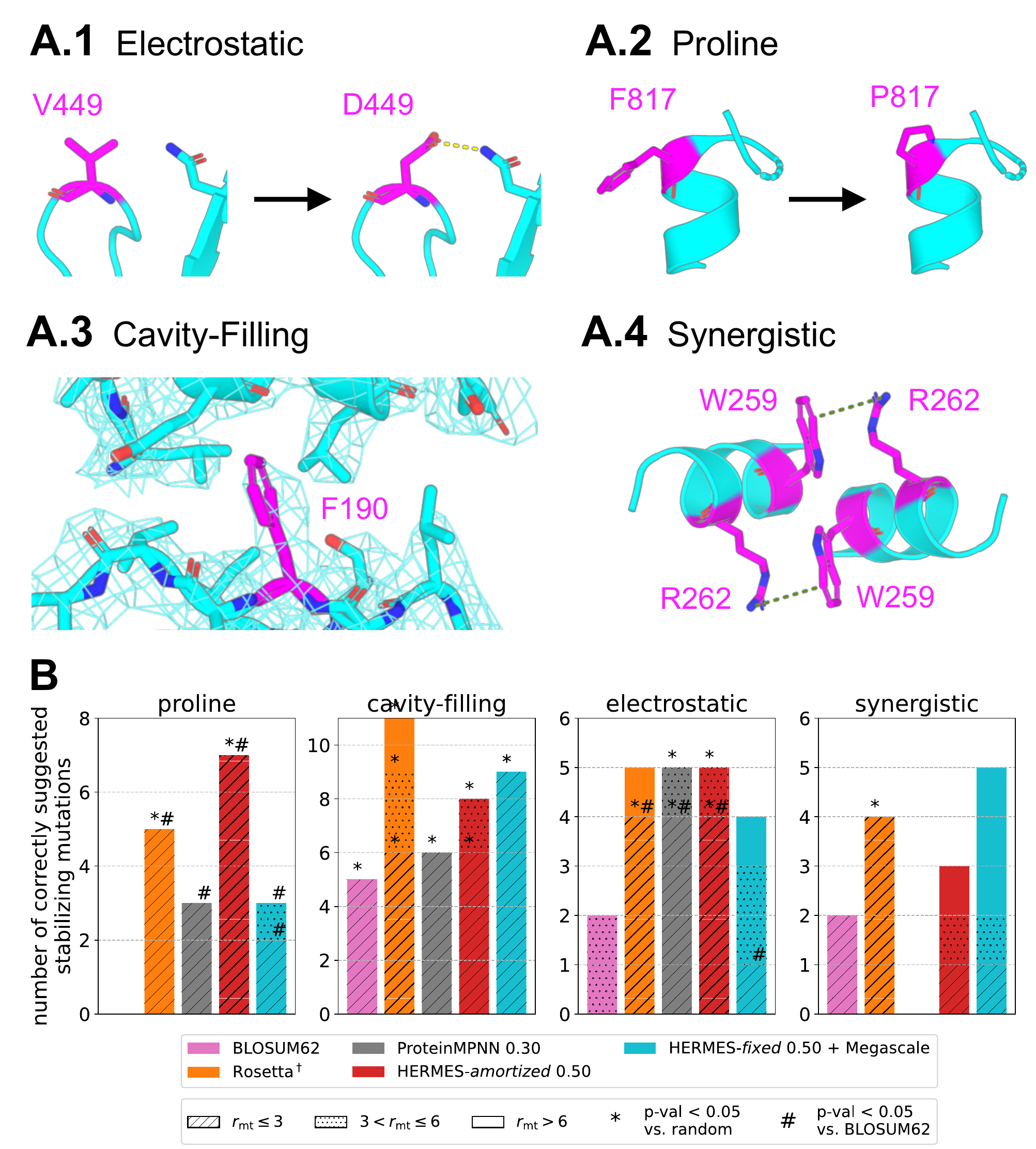}
    \caption{\textbf{Model ability to recover different mutation patterns that stabilize antigens.}
    \textbf{(A)} Representative examples of mutation types commonly seen in  antigen stabilization~\citep{byrne_principles_2022}, and analyzed in this study. The mutated residue(s) are shown in magenta.
    \textbf{(A.1)} {\em Electrostatic mutation} in hMPV-F: wild-type structure (PDB ID 5WB0, left), in-silico mutant {generated with PyMOL's mutagenesis wizard};  right. The introduced Aspartic acid forms an electrostatic interaction  with a nearby Asparagine (dashed yellow line).
    \textbf{(A.2)} {\em Proline mutation} at the N-terminal of a $\alpha$-helix cap in SARS-CoV-2 spike: wild-type structure (PDB ID 6VSB; left), in-silico mutant {generated with PyMOL's mutagenesis wizard}; right.
    \textbf{(A.3)} {\em Cavity-filling mutation} in RSV-F: mutant structure (PDB ID 4MMS). A bulky hydrophobic substitution packs a previously underfilled region: the $2Fo$-$Fc$ electron density map is shown as a thin mesh. 
    \textbf{(A.4)} {\em Synergistic dimer-stabilizing mutations} in DENV-E: A259W and T262R in the dimer mutant structure (PDB ID 6WY1). introduced in both chains, create a stabilizing cation-$\pi$ interaction (dashed green lines).
    \textbf{(B)} Number of stabilizing mutations recovered by each model, stratified by mutation class, reported as in Fig.~\ref{fig:antigen_figure_main_results}B.
    }
    \label{fig:antigen_figure_of_different_types_of_mutations}
\end{figure*}

Figs.~\ref{fig:antigen_figure_main_results},~\ref{fig:antigen_figure_of_different_types_of_mutations} {and Table~\ref{table:antigen_results}} summarize predictions across the 33 stabilizing mutations we studied. Among the machine-learning methods, HERMES-\emph{amortized} performs best: it assigns a better (lower) rank than wild-type to 24 stabilizing mutations, including 19 classified as strongly suggested ($r_\mt \leq 3$). Notably, stability fine-tuning does not consistently improve performance over the corresponding zero-shot ProteinMPNN and HERMES-{\em amortized} on this task (Table~\ref{table:antigen_results}).

For comparison, we also scored each mutant with Rosetta~\citep{chaudhury_pyrosetta_2010} (Methods). Rosetta recovers stabilizing mutations on par with HERMES-\emph{amortized} ({Fig.~\ref{fig:antigen_figure_main_results} and} Table~\ref{table:antigen_results}); however, 17 variants in this benchmark were originally selected (by the references that discovered the variants) using Rosetta-based screening, which partially biases this baseline in Rosetta's favor. Moreover, Rosetta inference is substantially more computationally intensive than the machine learning  models, limiting its practicality for predicting the stability effects in high-throughput saturation mutagenesis (Table~\ref{table:speed_on_antigens}).

To assess statistical enrichment of model predictions over chance, we compared the predicted number of stabilizing mutations in the strong/moderate categories to a random-ranking baseline using a binomial test (Methods); p-values are reported in Fig.~\ref{fig:antigen_pvalues_vs_random}. All models except for HERMES-{\em fixed} and ThermoMPNN significantly outperform random ranking in recovering stabilizing mutations with strong or moderate confidence ({Table~\ref{table:antigen_results}}).
As an additional baseline, we tested whether the BLOSUM62 (B62) substitution matrix can prioritize stabilizing mutations. BLOSUM62 recovers significantly fewer stabilizing mutations in the strong/moderate categories than HERMES-\emph{amortized} (defined as $r_\mt^{B62} \leq 3$ and $r_\mt^{B62} \leq 6$, noting that $r_\wt^{B62} = 1$ always; {results shown in Fig.~\ref{fig:antigen_figure_main_results},} and binomial test p-values reported in Fig.~\ref{fig:antigen_pvalues_vs_blosum62}), underscoring the value of incorporating structural context when predicting stabilizing substitutions. 

Practitioners commonly classify mutations in different types based upon the mechanism by which they stabilize pre-fusion antigens~\citep{byrne_principles_2022}.
We consider four types, examples of which are shown in Fig.~\ref{fig:antigen_figure_of_different_types_of_mutations}A.
(i) {\em Electrostatic mutations}, which  introduce polar or charged amino acids in environments where they can engage in stabilizing electrostatic interactions (Fig.~\ref{fig:antigen_figure_of_different_types_of_mutations}A.1);
(ii) {\em Proline mutations}, which are often used to impede post-fusion helix formation and stabilize pre-fusion conformations, due to proline's unique lack of amide hydrogens which prevents it from taking part in stable $\alpha$-helices, except at the N-terminal end of a helix cap (Fig.~\ref{fig:antigen_figure_of_different_types_of_mutations}A.2);
(iii) {\em Cavity-filling mutations}, which stabilize a particular conformation by filling cavities located within the hydrophobic core of an antigen by inserting a larger hydrophobic side-chain (Fig.~\ref{fig:antigen_figure_of_different_types_of_mutations}A.3);
(iv) {\em Synergistic mutations}, which  involve applying multiple mutations so that they positively interact through a variety of mechanisms (Fig.~\ref{fig:antigen_figure_of_different_types_of_mutations}A.4).

Based on these categories, we manually annotated the 33 antigen-stabilizing mutations  (Table~\ref{table:mutation_types}), and stratified model performance by mutational types (Fig.~\ref{fig:antigen_figure_of_different_types_of_mutations}B).
Among the models tested, HERMES-\emph{amortized} most reliably recovers stabilizing proline mutations (7 out of 8; Fig.~\ref{fig:antigen_figure_of_different_types_of_mutations}B), highlighting its potential utility for proposing pre-fusion stabilizing prolines. More broadly, all models recover proline and electrostatic mutations at higher rates than expected under BLOSUM62, whereas cavity-filling mutations show weaker gains. A plausible explanation is that cavity-filling changes typically replace a smaller hydrophobic residue with a larger one that preserves core hydrophobicity. Because such substitutions are common in natural sequence variation, they are partly captured by a sequence-derived matrix like BLOSUM62. In contrast, proline and charged substitutions are strongly context-dependent, with effects that hinge on local geometry and environment, and therefore benefit more from structure-aware modeling; see SI for a more extensive discussion.

Lastly, the synergistic RSV-F TriC mutations are difficult to recover for almost all methods (Fig.~\ref{fig:antigen_figure_main_results}). This is expected for the three ``acid patch neutralization" substitutions D486H, D489H, and E487Q~\citep{mclellan_structure-based_2013}, which were experimentally stabilizing only in combination (i.e., synergy) with F488W~\citep{mclellan_structure-based_2013}. These four residues are tightly clustered in both intra- and inter-chain space, consistent with a synergistic (epistatic) mechanism in which the substitutions act synergistically to stabilize the trimer. A second example of synergy is the DENV-E pair A259W and T262R, which together introduce a favorable cation–$\pi$ interaction (Fig.~\ref{fig:antigen_figure_of_different_types_of_mutations}A.4). Because our evaluation scheme scores mutations in a site-independent manner, the models are not able to capture such multi-residue dependencies. Specifically,  HERMES scores mutations at a single site under the assumption that all other amino-acid identities remain fixed, and therefore, cannot recover stabilization mechanisms that arise only from specific combinations of mutations. ProteinMPNN, ThermoMPNN, and Rosetta are likewise evaluated here under the same site-independent protocol for a fair comparison, although they can in principle  be applied to score multi-residue variants jointly; see SI for a more extensive discussion.

Beyond the inability to capture epistatic effects, several additional limitations of our approach should be noted. First, to emulate a practitioner's workflow we evaluate mutations using only their relative ranks; however, the absolute magnitudes of model scores carry additional information and could enable more robust decision-making when prioritizing antigen-stabilizing mutations.  Second, because the original studies rarely tested comprehensive alternative substitutions at the same sites, we cannot evaluate the models' ability to propose different stabilizing mutations that were not assayed experimentally. Third, experimental validation frequently involved multi-mutation constructs, complicating attribution of observed stabilization to any single residue. Finally, the modest number of curated stabilizing mutations limits statistical power and the strength of conclusions in our analyses. Despite these caveats, our results suggest that HERMES can serve as a practical screening tool for rational library design, prioritizing candidate antigen-stabilizing mutations for more efficient experimental validations.

\begin{table}[t!]
\centering
\begin{tabular}{l | c c | c c}
\toprule
 & \textbf{Per-Struct.} & \textbf{Per-Struct.} & \textbf{Overall} & \textbf{Overall} \\
\textbf{Method} & \textbf{Pearson} & \textbf{Spearman} & \textbf{Pearson} & \textbf{Spearman} \\
\midrule
Rosetta$^*$~\citep{park_simultaneous_2016} & 0.328 & 0.299 & 0.311 & 0.347 \\
FoldX$^*$~\citep{delgado_foldx_2019} & 0.391 & 0.364 & 0.356 & 0.351 \\
\midrule
DDGPred$^*$~\citep{shan_deep_2022} & 0.371 & 0.343 & 0.652 & 0.439 \\
ESM-IF$^*$~\citep{hsu_learning_2022} & 0.231 & 0.209 & 0.296 & 0.287 \\
MIF-Net.$^*$~\citep{luo_rotamer_2022} & 0.395 & 0.348 & 0.667 & 0.480 \\
RDE-Net.$^*$~\citep{luo_rotamer_2022} & 0.469 & 0.433 & 0.642 & 0.527 \\
\midrule
Vanilla Pythia PPI$^\dagger$~\citep{tao_reliable_2025} & 0.478 & 0.449 & 0.709 & 0.537 \\
Pythia-PPI$^\dagger$~\citep{tao_reliable_2025} & 0.565 & 0.527 & 0.785 & 0.637 \\
\midrule
\midrule
\proteinmpnn & 0.281 & 0.282 & 0.331 & 0.315 \\
\proteinmpnnNoise & 0.270 & 0.255 & 0.334 & 0.289 \\
\HERMESPy & 0.306 & 0.287 & 0.285 & 0.272 \\
\HERMESPyNoise & 0.317 & 0.308 & 0.291 & 0.286 \\
\HERMESPyFtRelaxed & 0.231 & 0.241 & 0.290 & 0.242 \\
\HERMESPyNoiseFtRelaxed & 0.222 & 0.239 & 0.276 & 0.222 \\
\HERMESPyFtCdna & 0.347 & 0.331 & 0.380 & 0.342 \\
\HERMESPyNoiseFtCdna & 0.305 & 0.294 & 0.344 & 0.288 \\
\midrule
\HERMESPyFtSkempiEasy & 0.471 & 0.433 & 0.578 & 0.476 \\
\HERMESPyNoiseFtSkempiEasy & 0.430 & 0.389 & 0.512 & 0.420 \\
\HERMESPyFtSkempiMedium & 0.472 & 0.430 & 0.576 & 0.466 \\
\HERMESPyNoiseFtSkempiMedium & 0.407 & 0.368 & 0.497 & 0.403 \\
\HERMESPyFtSkempiHard & 0.435 & 0.398 & 0.395 & 0.380 \\
\HERMESPyNoiseFtSkempiHard & 0.399 & 0.359 & 0.328 & 0.322 \\
\bottomrule
\end{tabular}
\caption{\textbf{Benchmark on predicting mutational effects on protein-protein binding in SKEMPI.}
The Spearman/Pearson correlations between model-predicted effects of single point mutations and experimental values from the SKEMPI v2.0 dataset~\citep{jankauskaite_skempi_2019} are reported both across mutations within each structure individually (``per-structure" correlations), and over mutations pooled across all complexes (``overall" correlations). Entries marked with $*$ are taken from ref.~\citep{luo_rotamer_2022} and, for machine learning-based methods (all except Rosetta and FoldX), were obtained using 3-fold cross-validation over SKEMPI complexes.
Entries marked with $\dagger$ are taken from ref.~\citep{tao_reliable_2025}, and were obtained via 5-fold coss validation on the SKEMPI complex structures.
We evaluate HERMES using three train/test splits defined from SKEMPI metadata, with increasing difficulty: (i) {\em Easy}, a random split; (ii) {\em Medium}, which groups sites with similar binding sites (hold-out proteins) into the same split; (iii) {\em Difficult}, which groups sites from the same held-out protein types (functional classes) into the same split (see Methods for details).
The HERMES models with most comparable training procedure to the other machine learning models are those fine-tuned on the \textit{Easy} split, for which we used 3-fold cross-validation with splits defined by PDB complex.
Note that, similar to HERMES, the machine learning models in ref.~\citep{luo_rotamer_2022} were fine-tuned on SKEMPI $\Delta\Delta G_{\text{binding}}$ labels only, whereas Pythia-PPI \citep{tao_reliable_2025} was trained on a mixture of SKEMPI binding labels and FireProtDB stability labels \citep{stourac_fireprotdb_2021} and further refined via self-distillation.
}
\label{table:skempi_spm}
\end{table}

\subsection{Predicting {binding} effect of mutations}
Predicting the effects of mutations on binding affinity is a central step in target-specific protein design. In prior work, we demonstrated the utility of HERMES for predicting how mutations in short peptide antigens alter binding to T-cell receptors in the context of MHC complexes~\citep{visani_t_2025}. We further leveraged this capability for de novo design of peptide antigens intended to elicit specific T-cell responses~\citep{visani_t_2025}. Here, we evaluate HERMES on the single-point mutation effects from the SKEMPI v2.0 dataset~\citep{jankauskaite_skempi_2019}, which spans a substantially broader range of protein–protein interactions and binding-affinity perturbations.

In a zero-shot setting, HERMES exhibits measurable predictive signal (HERMES-\emph{fixed} 0.50; Spearman's $\rho = 0.286$). ProteinMPNN achieves comparable overall performance, whereas physics-based approaches (Rosetta~\citep{chaudhury_pyrosetta_2010} and FoldX~\citep{schymkowitz_foldx_2005}) perform modestly better (Spearman's $\rho \approx 0.35$). In contrast to our results for stability prediction, HERMES-\emph{fixed} models correlate more strongly with binding effects than HERMES-\emph{amortized} models. Interestingly, fine-tuning for stability (HERMES-\emph{fixed} 0.00+ cDNA117k) yields a slight improvement in binding-effect prediction. A detailed comparison across models is provided in Table~\ref{table:skempi_spm}.

Next, we fine-tuned HERMES directly on SKEMPI using 3-fold cross-validation. To control for information leakage under structural similarity, we introduced three homology-aware splitting strategies of increasing difficulty; the most stringent split prevents complexes from the same interaction ``class" (e.g., antibody–antigen or TCR–pMHC) from appearing in different folds; see Methods for details. Fine-tuned HERMES models achieve substantially stronger correlations, including under the {\em Difficult} split (Table~\ref{table:skempi_spm}).

Finally, we compared HERMES to recent state-of-the-art methods, RDE-Network and MIF-Network~\citep{luo_rotamer_2022}, DDGPred~\citep{shan_deep_2022}, ESM-IF~\citep{hsu_learning_2022} and Pythia-PPI~\citep{tao_reliable_2025}. These baselines were also trained on SKEMPI using 3- or 5-fold cross-validation under protocols analogous, but not identical, to our {\em easy} split. HERMES fine-tuned on SKEMPI-Easy performs competitively with RDE-Network and MIF-Network, but trails Pythia-PPI (Table~\ref{table:skempi_spm}). Notably, RDE-Network and MIF-Network are fine-tuned on SKEMPI $\Delta\Delta G$ labels in a manner similar to our approach, whereas Pythia-PPI incorporates additional training procedures that further boost performance (Methods). These procedures are, in principle, applicable to HERMES as well, but we leave a systematic evaluation of such extensions, and of HERMES as a general framework for binding-affinity prediction, to future work.

\section{Discussion}
We introduced HERMES, a family of fast, structure-based machine learning models for predicting the effects of mutations on protein function. HERMES leverages SO(3)-equivariant neural networks operating on local atomic neighborhoods in a protein structure to predict amino acid propensities, whose differences can be used to predict  mutational effects. This formulation yields a computationally efficient predictor that is naturally suited to high-throughput settings. HERMES captures mutational impacts on diverse phenotypes, including thermodynamic stability and protein–protein binding affinity, and it provides a practical tool for antigen stabilization in vaccine design.

A central finding is that encoding packing flexibility is essential for accurate predictions across mutations involving amino acids of different sizes. HERMES-{\em fixed}, a lightweight protocol that evaluates mutations in the rigid wild-type structure, exhibits strong bias toward size-conserving substitutions. Explicitly modeling relaxation via Rosetta~\citep{chaudhury_pyrosetta_2010}, (HERMES-{\em relaxed} protocol) resolves this bias but incurs substantial computational cost. Our amortized approach offers an effective compromise: by fine-tuning on relaxed predictions for just 0.5\% of pre-training data, HERMES-{\em amortized} learns implicit packing flexibility that can be leveraged for predictions at HERMES-{\em fixed} speed. This capability proves particularly valuable for antigen stabilization, where HERMES-{\em amortized} recovers over half of the verified stabilizing mutations in our benchmark and shows strong performance on proline substitutions (87.5\% recovery), which stabilize proteins by restricting backbone conformational freedom.

Beyond zero-shot use, HERMES can be fine-tuned directly on experimental labels using a simple end-to-end procedure. Notably, both the native and the fine-tuned models use the  difference of amino acid scores  for predicting mutational effects, which yields an inherent reversibility with respect to mutation order--a thermodynamic consistency property that is often enforced via explicit 19$\times$ data augmentation~\citep{diaz_stability_2024}. 

With fine-tuning, HERMES achieves competitive accuracy on thermodynamic stability benchmarks when trained on the same data as prior methods~\citep{blaabjerg_rapid_2023, diaz_stability_2024, dieckhaus_transfer_2024}. Interestingly, stability fine-tuning does not consistently transfer to antigen stabilization and can even degrade performance. One plausible explanation is dataset mismatch: the high-throughput stability datasets used for fine-tuning~\citep{tsuboyama_mega-scale_2023} are enriched for relatively small, compact domains, whereas vaccine antigens are often larger, multi-domain, and conformationally heterogeneous, with stabilizing mutations frequently targeting quaternary contacts, glycoprotein-specific features, or prefusion-state constraints. Resolving this discrepancy will likely require broader supervision that better reflects antigen structure and design objectives, or task-aligned fine-tuning data.

We also demonstrate that pre-training on wild-type amino acid classification remains necessary for strong performance; models trained only on currently-available experimental stability data fail. Notably, pre-trained models exhibit ``wild-type preference," assigning elevated probabilities to wild-type residues in wild-type structural contexts. This bias correlates with sequence similarity to pre-training proteins, suggesting partial memorization rather than purely generalizable learning. Fine-tuning partially ameliorates but does not eliminate this effect, representing a fundamental trade-off in current training paradigms. More robust pre-training objectives, stronger regularization, or explicit debiasing strategies may be required to fully address this issue.

A key application we highlight is structure-based vaccine design, where practitioners seek to stabilize substitutions that preserve a desired prefusion conformation {of} a vaccine antigen to elicit immune responses targeting relevant epitopes. For this task, a useful model should propose stabilizing mutation candidates across different sites of an antigen. On the limited available data comprising 33 verified antigen-stabilizing mutations across five viruses, HERMES performs well in recovering these mutations. Moreover, the strict locality of the model makes it straightforward {and fast} to apply to large antigens and assemblies that can pose challenges for  architectures modeling global interactions in proteins. Our results show that local all-atom environments modeled by HERMES can encode multiple stabilization mechanisms, including cavity filling, backbone rigidification via proline, and favorable electrostatic remodeling, while also underscoring limitations for mechanisms that depend on long-range coupling, or multi-site epistasis.

We further evaluated HERMES on predicting changes in protein--protein binding affinity using the SKEMPI dataset~\citep{jankauskaite_skempi_2019}. In the zero-shot setting, HERMES shows measurable predictive signal, and supervised fine-tuning on SKEMPI {substantially} improves performance.
Recent state-of-the-art approaches for binding prediction like Pythia-PPI~\citep{tao_reliable_2025} indicate that design choices such as self-distillation can substantially improve performance~\citep{tao_reliable_2025}; we expect HERMES would similarly benefit from such approaches, representing a clear next step.

Several directions could extend HERMES' capabilities. Developing larger, more diverse training datasets, particularly for antigen stabilization and binding, would likely improve performance, as our results indicate that training data dominates architectural differences in determining benchmark performance. The locality of HERMES appears to be a double-edged sword: local neighborhoods reduce input complexity, remove constraints on protein size, and improve scalability, but necessarily limit the model's ability to capture long-range epistasis and allosteric coupling. We view HERMES as a hypothesis-generation tool whose predictions are most reliable when mutational effects are driven primarily by short-range interactions and local packing. Productive directions for future work include better characterizing when locality suffices, further reducing pre-training-induced biases, and developing multi-scale approaches that efficiently combine local all-atom predictors with complementary global or multi-site models to capture mechanisms beyond the local neighborhood.

\section{Acknowledgement}
This work has been supported by the National Institutes of Health MIRA award R35~GM142795,  the CAREER award from the National Science Foundation grant 2045054,  the Royalty Research Fund from the University of Washington no. A153352,   the Allen School Computer Science \& Engineering Research Fellowship from the Paul G. Allen School of Computer Science \& Engineering at the University of Washington (GV), and the Microsoft Azure award from the eScience institute at the University of Washington. This work is also supported, in part, through the Departments of Physics and Computer Science and Engineering, and the College of Arts and Sciences at the University of Washington. U.W. acknowledges the REU program at the Department of Physics at the University of Washington, which is supported by the National Science Foundation grant 2243362.
 The numerical analyses in this work  were completed on Hyak, the University of Washington's high performance computing cluster, which is  funded by the University of Washington student technology fee.

\clearpage{}
\newpage{}

\section{Methods}

\subsection{HERMES architecture}
The HERMES architecture closely follows our recent developments of SO(3)-equivariant neural network models for protein  structures~\citep{pun_learning_2024,visani_holographic-vae_2024,visani_h-packer_2024}. For completeness, we summarize here the main methodological components and refer the reader to those works for additional details.

The input to HERMES is a point cloud of atoms in a protein structure, which we term a \textit{neighborhood}. Each neighborhood is centered at the C-$\alpha$ atom of a focal residue and includes all atoms within a 10~\AA~radius of this center. Optionally, we add Gaussian noise with standard deviation 0.50~\AA~to the atomic coordinates to achieve further model robustness. The output of HERMES is a 20-dimensional representation of the neighborhood that is invariant to 3D rotations about its center (i.e., SO(3)-invariant). To obtain SO(3) invariance, HERMES is constructed from SO(3)-equivariant layers that progressively compute higher-level SO(3)-invariant features, which are then passed to a final multilayer perceptron (MLP) with a 20-dimensional output. Conceptually, symmetry awareness in HERMES is achieved through two main components: (i) a (holographic) encoding of the neighborhood into a basis suitable for SO(3)-equivariant operations, and (ii) processing of the input via a stack of SO(3)-equivariant neural network layers to learn an expressive and SO(3)-invariant representation of the neighborhood.\\

\noindent{\bf Holographic encoding of atomic protein structure neighborhoods.} We first represent the atomic point cloud of each structural neighborhood as a density function obtained by superposing (weighted) Dirac-$\delta$ functions, indicating the presence of atoms at a given position in space: $\rho(\mathbf{r}) = \sum_{i\in \text{points}} \omega_i \delta(\mathbf{r}_{i} - \mathbf{r})$; here, $\omega_i$ indicates the weight associated with point $i$ at position $\mathbf{r}_i$, and $\mathbf{r}_i$ can be decomposed in its constituents spherical components $(r_i, \theta_i, \varphi_i)$. 
We then use 3D Zernike Fourier Transform (ZFT)~\citep{pun_learning_2024} of the density function to encode the neighborhood into a convenient SO(3) equivariant basis, 
\begin{equation}
    \hat{Z}_{\ell m}^{n} = \sum_{i\in \text{points}} \omega_i R_{n}^{\ell}(r_i) Y_{\ell m}(\theta_i, \varphi_i)
   \label{eqn:spherical_and_radial_with_dirac_ft}
\end{equation}
where $Y_{\ell m}(\theta,\phi) $ is the spherical harmonics of integer degree $\ell \geq 0$ and integer order $m \leq |\ell|$, and $R^{n}_{\ell}(r)$ is the radial Zernike polynomial in 3D with integer radial frequency $n \geq 0$ and degree $\ell$. $R^{n}_{\ell}(r)$ is non-zero only when $n - \ell$ is even and $\geq 0$. We keep coefficients of up to and including $\ell = 5$, and, for every $\ell$, we keep the first 11 non-zero radial frequencies.

Notably, the spherical harmonics that describe the angular component of ZFT arise from the irreducible representations of the 3D rotation group SO(3), and form a convenient basis under rotation in 3D (see the Appendix of~\citep{visani_h-packer_2024} for a formal mathematical introduction). Zernike projections in spherical Fourier space can be understood as a superposition of spherical holograms of an input point cloud, and thus, we term this operation an {\em holographic encoding} of the data~\cite{visani_holographic-vae_2024, pun_learning_2024}. The resulting holograms are primarily indexed by the degree $\ell$: different values of $\ell$ encode components that transform in specific ways under 3D rotations. For example, the $\ell = 0$ component is rotation-invariant.

Following~\citep{pun_learning_2024,visani_holographic-vae_2024}, we incorporate atom-level features by partitioning the holographic encoding into multiple \textit{channels} (Fig.~\ref{fig:HERMES}A). Specifically, we use separate channels for C, N, O, S, computationally added hydrogens, partial charge, and solvent-accessible surface area (SASA). Partial charge and SASA are defined for all atoms, and their values are incorporated through the weights $\omega_i$.\\

\noindent{\bf SO(3)-equivariant neural network architecture.}
The neighborhood holograms are then processed by a stack of SO(3)-equivariant layers (Fig.~\ref{fig:HERMES}A). The final SO(3)-invariant representation is obtained by reading out the $\ell = 0$ component of the last layer and passing it through an MLP to produce a 20-dimensional embedding. We employ three types of SO(3)-equivariant building blocks, inspired by the H-CNN architecture~\citep{pun_learning_2024} and also detailed in ref.~\citep{visani_h-packer_2024}: (i) SO(3)-equivariant linear layers (Lin), which mix channels while preserving the SO(3) transformation properties; (ii) tensor-product nonlinearities (TP), which couple different irreducible components; and (iii) SO(3)-equivariant layer normalization (LN).

All SO(3)-equivariant components are implemented using \texttt{e3nn} primitives~\citep{geiger_e3nn_2022}. Within the same framework, we re-implemented the H-CNN architecture described in~\citep{pun_learning_2024} for comparison. HERMES achieves a forward pass that is approximately $2.75\times$ faster than H-CNN, while using a similar number of parameters ($\sim 3.5$M).

\subsection{Pre-processing of protein structure data}
To pre-process protein structure data, we developed two distinct pipelines based on either (i) PyRosetta~\citep{chaudhury_pyrosetta_2010} or (ii) Biopython~\citep{cock_biopython_2009} together with other open-source tools, with code adapted from~\citep{blaabjerg_rapid_2023}. The PyRosetta-based pipeline is considerably faster but requires a license, whereas the Biopython-based pipeline is fully open source. We train models using data generated by both pipelines, with the constraint that the Pre-processing pipeline used at inference must match that used during training. Performance differences between the two pipelines are minor (Fig.~\ref{fig:rasp_exp} and~\ref{fig:t2837_py_vs_bp}, and Table~\ref{table:aa_wt_cls}); unless otherwise stated, we report results obtained with the PyRosetta pipeline, which yields slightly better performance and has faster Pre-processing runtime.

For the PyRosetta workflow, we use PyRosetta functionalities for all Pre-processing steps: repairing PDB files by adding missing residues, adding hydrogen atoms, assigning partial atomic charges, computing solvent-accessible surface areas (SASA); we ignore non-canonical amino acids.
For the open-source Pre-processing workflow (``Biopython" pipeline), we proceed as follows. First, we use OpenMM~\citep{eastman_openmm_2013} to repair PDB files by adding missing residues and substituting non-canonical residues with their canonical counterparts. Hydrogen atoms are then added using the \texttt{reduce} program~\citep{word_asparagine_1999}. Partial atomic charges are assigned from the AMBER99sb force field~\citep{ponder_force_2003}, and SASA are computed using Biopython~\citep{cock_biopython_2009}.

Both pre-processing pipelines retain atoms belonging to non-protein residues and ions, in contrast to the RaSP pre-processing procedure~\citep{blaabjerg_rapid_2023}. Notably, the PyRosetta-based pipeline does not replace non-canonical residues.

\subsection{HERMES pre-training}
We pre-trained HERMES using an inverse folding objective, in which the model predicts the identity of a masked focal amino acid (i.e., the native residue in the structure) given its surrounding atomic neighborhood. This training task is analogous to that used for H-CNN~\citep{pun_learning_2024}. We adopted the same data splits as in H-CNN: 10,957 structures for training, 2,730 for validation, and 212 for testing. These splits are derived from ProteinNet's~\citep{alquraishi_proteinnet_2019} 30\% sequence-identity clustering of PDB structures available at the time of CASP12.

Model parameters were optimized for 10 epochs using the Adam optimizer~\citep{kingma_adam_2015} with a learning rate of $10^{-3}$. We selected the model checkpoint with the lowest validation loss at the end of each epoch. A single HERMES model is implemented as an ensemble of 10 independently trained neural network instances, whose predictions are averaged at inference time. Pre-training a single network instance required approximately 40 minutes per epoch on a single NVIDIA A40 GPU.

\subsection{Rosetta relaxation of mutant structures for the HERMES-{\em relaxed} protocol}
For the HERMES-{\em relaxed} protocol, we use PyRosetta~\citep{chaudhury_pyrosetta_2010} to generate and locally refine mutant protein structures. Specifically, the focal residue is first substituted with the desired mutant amino acid, after which all side-chain atoms within 12~\AA~of the focal residue's C-$\alpha$ atom are relaxed using the FastRelax protocol with the \texttt{ref2015\_cart} energy function and a single relaxation cycle.

We performed a targeted ablation study to select the relaxation parameters. These experiments indicate that allowing backbone atoms to relax slightly degrades performance, and that ensembling over multiple independent relaxation runs does not improve accuracy (Fig.~\ref{fig:relaxations_ablation}A), while substantially increasing computational cost (Fig.~\ref{fig:relaxations_ablation}B). Thus, all results for the HERMES-{\em relaxed} protocol are reported using side-chain–only relaxation with a single FastRelax cycle.

\subsection{Training HERMES-\emph{amortized} to implicitly learn about relaxation}
We fine-tuned HERMES to align HERMES-\emph{fixed} predictions to those of HERMES-\emph{relaxed}, on a subset of the pre-training protein sites. Specifically, we uniformly sampled 10\% of the proteins pre-training  training set (1,284), and from those we uniformly sampled 5\% of the sites. We similarly sampled 1\% of the sites from the proteins of the pre-training validation set, and 5\% of the pre-training testing set.

\subsection{HERMES fine-tuning for downstream tasks}
\label{sec:finetuning_details}
In all analyses, we fine-tuned HERMES models under a Huber Loss objective with hyperparameter $\delta = 1.0$, for 15 epochs using the Adam optimizer~\citep{kingma_adam_2015} with a learning rate of $10^{-3}$ and a batch size of 128 mutations, selecting the model checkpoint with the lowest validation loss at the end of each epoch. The only exception is the models without pre-training, which are trained for 25 epochs.

By convention, we define model outputs such that higher predicted values correspond to more favorable mutation effects. Accordingly, when fine-tuning on experimental $\Delta\Delta G$ measurements--where lower values indicate greater stability--we trained the model to predict  $-\Delta\Delta G$. This sign convention ensures consistent interpretation of model outputs across tasks and is fixed throughout all reported analyses.

To speed-up convergence during fine-tuning, we first  rescale the weight matrix and bias vector of the network's output layer so that the mean and variance of the output logits match those of  the validation scores. This initialization step requires a forward pass through the validation data to estimate the mean and variance, but it makes the model outputs immediately on the same scale as the experimental  scores. Thus, fine-tuning avoids spending early epochs merely adjusting output magnitudes.

Overall, fine-tuning is computationally efficient. Training a single neural network instance requires approximately 2.5 minutes per epoch on the cDNA117k dataset ($\sim$117,000 mutations) and about 4 minutes per epoch on the Megascale training dataset ($\sim$217,000 mutations) on a single NVIDIA A40 GPU.

\subsection{Fine-tuning datasets}
\label{sec:datasets}

\noindent {\bf Stability effect prediction.} For protein stability prediction, we fine-tuned and evaluated the performance of HERMES on the same datasets of three recent structure-based predictors of mutational stability effects--RaSP~\citep{blaabjerg_rapid_2023}, Stability-Oracle~\citep{diaz_stability_2024}, and ThermoMPNN~\citep{dieckhaus_transfer_2024}--so results are comparable. The data from these models are used in following way:

\begin{itemize}
\item {\em RaSP dataset.} We used the RaSP dataset~\citep{blaabjerg_rapid_2023} as provided by the authors on their github repository (\url{https://github.com/KULL-Centre/_2022_ML-ddG-Blaabjerg}). As in the original RaSP paper, the target values are Rosetta-computed stability changes ($\Delta\Delta G$), which are known to be reliable primarily within the range [-7, 1] kcal/mol. To account for this, RaSP applies a sigmoid transformation to the raw $\Delta\Delta G$ values prior to training, effectively saturating the targets outside this interval.

We adopt the same sigmoid transformation, with one modification: we center the transformed values such that $\Delta\Delta G = 0$ maps to zero after transformation. This centering is required by the HERMES output parameterization, in which the predicted stability change for a mutation to the same amino acid is constrained to be zero (i.e., $\Delta\Delta G_{aa_i \rightarrow aa_i} = 0$). While this property holds for physical $\Delta\Delta G$ values, it is not preserved by the uncentered sigmoid transform used in RaSP. Our centered transformation therefore ensures consistency between the model's output space and the physical interpretation of stability changes, and it follows,

\begin{equation}
    F(\Delta\Delta G) = \frac{1}{1 + e^{-\beta(\Delta\Delta G - \alpha)}} - \frac{1}{1 + e^{\beta \alpha}}
    \label{eq:centered_fermi_transform}
\end{equation}\\

\item {\em Stability-Oracle datasets.} Stability-Oracle introduces two curated datasets that we use in this work~\citep{diaz_stability_2024}:\\

(i) {cDNA117k} (training set): derived from the Megascale cDNA display proteolysis dataset \# 1~\citep{tsuboyama_mega-scale_2023}. The original Megascale dataset reports approximately 850,000 thermodynamic folding stability measurements ($\Delta G$) across 354 natural and 188 de novo mini-protein domains (40–72 amino acids in length). Following the Stability-Oracle protocol, mutations are filtered to enforce at most 30\% sequence identity with the test set, yielding a reduced set of approximately 117,000 mutation-induced stability changes ($\Delta\Delta G$), referred to as cDNA117k.\\

(ii) {T2837} (test set): assembled by combining several commonly used benchmarking datasets of experimentally measured $\Delta\Delta G$ values, including S669~\citep{pancotti_predicting_2022}, myoglobin~\citep{li_predicting_2020}, ssym~\citep{pucci_quantification_2018}, and p53~\citep{caldararu_base_2021}. The resulting test set comprises 2,837 mutations.

We obtained  the cDNA117k and T2837 datasets from the Stability-Oracle~\citep{diaz_stability_2024} github repository (\url{https://github.com/danny305/StabilityOracle/tree/master}). At the time of access, the residue indices provided in the dataset did not correspond to the residue numbering in the original PDB files, but instead to an intermediate, post-processed representation that was not documented in sufficient detail to allow straightforward recovery of the original numbering. To ensure consistency with structural data, we manually corrected the residue numbers in the CSV files to match those in the corresponding PDB structures. The corrected versions of these datasets are included in our repository.\\

\item {\em ThermoMPNN dataset.} 
ThermoMPNN also leveraged  Megascale~\citep{tsuboyama_mega-scale_2023}, but used datasets \# 2 and \# 3, curated and split by the authors at a 25\% sequence-identity cutoff into 216k training and 28k test mutation effects~\citep{dieckhaus_transfer_2024}. We used the train, valid, and test splits of the Megascale dataset as curated in ThermoMPNN, and provided on their github repository (\url{https://github.com/Kuhlman-Lab/ThermoMPNN}). Throughout the text, we refer to these datasets as the Megascale training (train + valid) and testing sets. We downloaded the structures' \texttt{.pdb} files from \url{https://zenodo.org/records/7992926}. We note that the Megascale training set was not controlled for maximum similarity with the T2837 dataset, and six of the T2837 proteins ($\sim 5\%$)  have sequence homologs in the Megascale training set with sequence similarity above 90\%.
\\\\
\end{itemize}

\noindent {\bf Binding effect prediction.} For predicting mutational effects on binding affinity, we  fine-tuned HERMES on the SKEMPI v2.0 dataset~\citep{jankauskaite_skempi_2019}, using 3-fold cross-validation.

\begin{itemize} 
\item{\em SKEMPI dataset:}
After removing duplicate experimental entries, SKEMPI v2.0 contains 5,713 measurements of binding free-energy changes ($\Delta\Delta G^{\text{binding}}$) spanning 331 protein–protein complex structures. We further restrict the dataset to complexes with at least 10 annotated mutations, resulting in 116 structures and 5,025 mutations. Finally, we retain only single-point mutations, yielding 93 structures and 3,485 mutations.

SKEMPI provides metadata explicitly designed to support leakage-aware evaluation via two fields: \emph{hold-out type} and \emph{hold-out proteins} (see SKEMPI documentation: \url{https://life.bsc.es/pid/skempi2/info/faq_and_help}). Briefly, {\em hold-out type} groups complexes into broad categories (e.g., protease-inhibitor, antibody-antigen, and pMHC-TCR), while {\em hold-out proteins} lists PDB identifiers and/or hold-out types that should be co-held-out to avoid training on closely related binding sites. Using this information, we define three split strategies:\\

(i) \textit{Easy:} random splitting without using hold-out metadata.\\

(ii) \textit{Medium:} we enforce that all entries linked via a mutation's \emph{hold-out proteins} annotation are assigned to the same split (but we do not additionally group by \emph{hold-out type}).\\

(iii) \textit{Difficult:} we enforce that all complexes sharing the same \emph{hold-out type} are assigned to the same split, producing a stringent evaluation of generalization across interaction classes.\\

In cases where a complex is associated with multiple hold-out types, we assign it to a single type by randomly selecting one of the available labels.

\end{itemize}

\subsection{Baseline models}
Here, we describe the baseline models used to benchmark HERMES on mutational effect prediction.\\

\noindent \textbf{ProteinMPNN \citep{dauparas_robust_2022}.} ProteinMPNN is an inverse-folding model that samples amino-acid sequences conditioned on a protein backbone (optionally with a partial sequence fixed). Because it also outputs per-site amino-acid probabilities, we used it to score mutational effects via the log-likelihood ratio in Eq.~\ref{eq:log_ratio_wt_and_mt}. As for HERMES, we evaluate ProteinMPNN models trained with two noise levels: 0.02 \AA (virtually no noise) and 0.30 \AA. We used, and provide, scripts to infer mutation effects built upon a public fork of the ProteinMPNN repository (\url{https://github.com/gvisani/ProteinMPNN-copy}).\\
\\
\noindent\textbf{ThermoMPNN~\citep{dieckhaus_transfer_2024}.} ThermoMPNN is a thermodynamic stability predictor built on top of ProteinMPNN. For a given structure, it extracts the final ProteinMPNN residue embedding at the site of interest and feeds it to a separate head that predicts per–amino-acid $\Delta G$ values, from which $\Delta\Delta G$ is computed. Similar to HERMES, this formulation enforces the permutation symmetry of mutational effects by construction, without requiring data augmentation. For our experiments, we used native functionalities in the ThermoMPNN repository (\url{https://github.com/Kuhlman-Lab/ThermoMPNN}). \\
\\
\noindent \textbf{Stability-Oracle~\citep{diaz_stability_2024}.} Similar to HERMES, Stability-Oracle is trained in two stages. First, a graph-attention network is pre-trained to predict masked amino acids from their local atomic environment (``neighborhood"). Next, the embeddings from the pre-trained model is used to regress over mutation-effect. Specifically,  For a target site on a structure, the masked-neighborhood embedding $h$ is extracted from the pre-trained network and concatenated separately with embeddings of the ``from" and ``to" amino acids. Each concatenated input is passed through a transformer to produce amino-acid–specific embeddings $e_{aa_{\from}}$ and $e_{aa_{\tto}}$, whose difference $(e_{aa_{\tto}}-e_{aa_{\from}})$ is fed to a two-layer MLP to predict $\Delta\Delta G_{aa_{\from}\rightarrow aa_{\tto}}$. This construction is permutation-symmetric up to the final MLP, since each $e_{aa}$ is computed independently; symmetry is broken only by the MLP, and would have been preserved by a bias-free linear layer. The original method enforces symmetry during training via $19\times$ data augmentation. We report performance scores calculated using predictions provided in the Stability-Oracale's repository (\url{https://github.com/danny305/StabilityOracle}).\\
\\
\noindent \textbf{RaSP \citep{blaabjerg_rapid_2023}.} Similar to HERMES, RaSP is trained in two steps. First, a 3D CNN is pre-trained to predict masked amino acids from their local atomic environment (``neighborhood"). Then, a small fully-connected neural network with a single output is trained to regress over mutation effects, using as input neighborhoods' embeddings from the 3DCNN, the one-hot encodings of wildtype and mutant amino-acids, and the wildtype and mutant amino-acids' frequencies in the pre-training data.
RaSP is fine-tuned on the stability effect of mutations $\Delta\Delta G$, computationally determined with Rosetta~\citep{chaudhury_pyrosetta_2010}. We report performance scores calculated using predictions provided in the authors' repository (\url{https://github.com/KULL-Centre/_2022_ML-ddG-Blaabjerg}).\\
\\
\textbf{RDE-Network~\citep{luo_rotamer_2022}.} RDE stands for Rotamer Density Estimator. This model consists of first a graph neural network encoder that is trained via a normalizing flow objective to predict distributions of side-chain conformations. Then, a prediction head is added, and it is trained on $\Delta\Delta G_{\text{binding}}$ effects from the SKEMPI dataset. We report performance scores as provided in ref.~\citep{luo_rotamer_2022}.
\\
\\
\textbf{MIF-Network~\citep{luo_rotamer_2022}.} This model's architecture is the same as the encoder in RDE-Network. It is first pre-trained on the task of wild-type amino acid classification. Then, a prediction head is added to the encoder, and it is trained on $\Delta\Delta G_{\text{binding}}$ effects from the SKEMPI dataset~\citep{jankauskaite_skempi_2019} in the same manner as RDE-Network. We report performance scores as provided in Ref.~\citep{luo_rotamer_2022}.\\
\\
\textbf{Pythia-PPI~\citep{tao_reliable_2025}.} Pythia-PPI uses a graph neural network encoder pre-trained for wild-type amino acid classification, followed by a $\Delta\Delta G$ prediction module with two heads: one for $\Delta\Delta G_{\text{stability}}$ and one for $\Delta\Delta G_{\text{binding}}$. The model is fine-tuned jointly on stability labels from FireProtDB~\citep{stourac_fireprotdb_2021} and binding labels from SKEMPI~\citep{jankauskaite_skempi_2019}, using a validation selected 20:80 stability:binding mixing ratio. The resulting checkpoint (``Vanilla Pythia-PPI") is then further trained via self-distillation by fine-tuning on its own predictions over all SKEMPI complex structures to obtain the final model (Pythia-PPI). We should note that these two procedures (i.e., joint fine-tuning on stability and binding with a validation selected mixing ratio and self-distillation) could also be applied to HERMES to potentially improve performance. We report performance scores as provided in ref.~\citep{tao_reliable_2025}.

\subsection{ESMFold for computational modeling of protein structures}
For the analysis in Fig.~\ref{fig:radial_plots__esmfold}, we used the ESM Metagenomic Atlas API to fold each sequence individually (\url{https://esmatlas.com/resources?action=fold}).

\subsection{Structure-conditioned reversibility analysis}

In Figs.~\ref{fig:ssym_figure}~and~\ref{fig:ssym_antisymmetry_reversibility_score_results}, we characterized structure-conditioned reversibility on the SSym dataset~\cite{pucci_quantification_2018}, which contains wild-type and single-mutant structures for 352 mutations across 19 proteins~\citep{pancotti_predicting_2022}. 

We assess structure-conditioned reversibility by whether the effects of forward $\alpha\to \beta$ and reverse $\beta\to \alpha$ mutations are equal, using the outgoing structural context to compute mutational effects. Specifically, the forward mutational effect  $\Delta \hat F^{(\text{model})}_{\alpha\to \beta}$ is computed by conditioning on the structure $X_\alpha$, and  the reverse effect $\Delta \hat F^{(\text{model})}_{\beta\to \alpha}$ is computed using $X_\beta$, 
\begin{equation}
\begin{aligned}
    \Delta \hat F^{(\text{model})}_{\alpha\to \beta} = \log P(\beta | X_{\alpha}) - \log P(\alpha |X_{\alpha}) \\
    \Delta \hat F^{(\text{model})}_{\beta\to \alpha} = \log P(\alpha | X_{\beta}) - \log P(\beta |X_{\beta})
\end{aligned}
\end{equation}
With this definition, we do not expect reversibility, as forward and reverse transitions are conditioned on different structures.\\

\noindent{\bf Reversibility score.} For the analysis in Figs.~\ref{fig:ssym_figure}~and~\ref{fig:ssym_antisymmetry_reversibility_score_results} we construct a reversibility score as a mean squared error normalized to be between -1 and 1,
\EQ
\text{rev. score} = 1 - \frac{\text{mean}((\Delta\log p_{fwd}+\Delta\log p_{rev})^2)}{\text{mean}({\Delta\log p_{fwd}}^2)+\text{mean}({\Delta\log p_{rev}}^2)}
\label{eq:rev_score}
\EE
where $\Delta \log p_{fwd} = \log p(\mt | X_\wt) - \log p(\wt | X_\wt)$ is the forward mutational effect computed by  conditioning on the wild type structure, and $\Delta \log p_{rev} = \log p(\wt | X_\mt) - \log p(\mt | X_\mt)$ is the reverse mutational effect computed by  conditioning  on the mutants structure. The averages (mean) are computed over the 352 mutations in the SSym dataset, or stratified as needed by the desired criteria.
With this definition, a larger value of {\bf rev. score} implies more degree of reversibility between forward and reverse mutations. \\

\noindent{\bf Sequence similarity calculation between the Ssym dataset and pre-training proteins.} We stratified the data based on their similarity to the training data. To do so, we used BLASTp via NCBI BLAST+~\cite{camacho_blast_2009} with individual chains in the SSym structures as queries~\cite{pucci_quantification_2018}, and individual chains in the pre-training set as the database; for each SSym chain, we then considered its maximum similarity to any sequence in the pre-training set. We found that 9 Ssym proteins have similarity below 70\% to the pre-training proteins (only 5 are below 50\%, none are below 40\%), and 10 proteins have similarity above 70\% (comprising the two panels in Fig.~\ref{fig:ssym_figure}B).\\

\subsection{Statistical comparison of model performances on classification metrics}

\noindent{\bf Comparing models' performances on the same tasks.} Figures~\ref{fig:radial_plots__noise}~and~\ref{fig:megascale_precision_recall_f1_by_size_cutoff_0p0} report classification performance (precision, recall, F1, etc.) for predicting the stability effects of mutations across models. Figures~\ref{fig:t2837_and_megascale_pairwise_pvalues_permutation}~and~\ref{fig:pairwise_pvalues_permutation_bucketed_by_sizereduced_precision_recall_f1} report pairwise significance tests (p-values) for differences in these metrics using permutation tests, with the null hypothesis that two models' predictions are exchangeable. To compute these p-values, for each model pair, we generated permuted prediction sets by swapping paired prediction between models with probability 0.5, and repeat this procedure for 1000 random seeds. The p-value is the fraction of permutations in which the metric difference between the permuted sets is at least as large ($\geq$) as the observed difference. We correct the computed p-values for multiple comparisons using Holm–Bonferroni across all model pairs within each metric (i.e., within each panel in Figs.~\ref{fig:t2837_and_megascale_pairwise_pvalues_permutation},~\ref{fig:pairwise_pvalues_permutation_bucketed_by_sizereduced_precision_recall_f1}).\\

\noindent{\bf Comparing a model's performance across tasks.} Figure~\ref{fig:megascale_precision_recall_f1_by_size_cutoff_0p0} reports classification performance (precision, recall, F1, etc.) for predicting mutation stability effects across mutation size categories (tasks). Figure~\ref{fig:pvalues__bucketed_by_size__between_small_large_buckets__vertical} reports, for each model, pairwise significance tests for differences in these metrics between tasks. We estimate p-values via bootstrap resampling: for each task pair, we repeatedly (1000 seeds) draw bootstrap samples of predictions within each task, recompute the metric difference, and define the p-value as the fraction of bootstrap replicates in which the resampled metric difference is at least as large ($\geq$) as the observed metric difference. We apply Holm–Bonferroni correction across all task pairs within each metric.\\

\noindent{\bf Significance of the number of retrieved antigen-stabilizing mutations by different models.}
In Figures~\ref{fig:antigen_figure_main_results},~\ref{fig:antigen_figure_of_different_types_of_mutations} and Table~\ref{table:antigen_results}, we report, for each model, the number of antigen-stabilizing mutations retrieved ($x$) out of a total of $N=33$ experimentally verified such mutations. Retrieval is based on amino-acid ranking: a mutation is counted as retrieved if (i) the mutant amino acid is ranked better than the wild type ($r_\mt<r_\wt$), and (ii) that its rank is  below or equal to a certain threshold $R$, with $R=3$ for ``strongly suggested" and $R=6$ for ``moderately suggested." We assess significance in the number of retrieved antigen-stabilizing mutations by a given model with a one-sided binomial test: for each model, the p-value is the probability under a null model of recovering at least $x$ successes in $N$ trials with per-mutation success probability $p_{\text{null}}$, i.e., $p = 1-\text{BinomCDF}(x-1, N, p_{\text{null}})$. We adjust p-values for multiple testing using Holm–Bonferroni, and report the resulting p-values in Figs.~\ref{fig:antigen_pvalues_vs_random}~and~\ref{fig:antigen_pvalues_vs_blosum62}.\\
To compute the p-values, we consider two null models:
\begin{enumerate}[(i)]
    \item \textbf{Random null.} For each mutation, we sample ranks for the mutant and wild type uniformly without replacement: $r_{\mathrm{mt}}\sim \mathrm{Unif}\{1,\dots,20\}$ and $r_{\mathrm{wt}}\sim \mathrm{Unif}\{1,\dots,20\}\setminus{r_{\mathrm{mt}}}$. A mutation is a ``success" if $r_{\mathrm{mt}}<r_{\mathrm{wt}}$ and $r_{\mathrm{mt}}\leq R$, which occurs with probability $p_{null} = \frac{1}{19 \times 20}\sum_{i=1}^{R} (20 - i)$.\\
    \item \textbf{BLOSUM62 null.} We simply set $p_{null} = x_{B62} / N$, where $x_{B62}$ is the number of mutations with BLOSUM62 rank  $r^{B62}_\mt \leq R$. We note that the rank of the wild-type is always 1 for BLOSUM62, so we omit the condition that the rank of the mutant has to be lower than the rank of the wild-type, which is instead applied to all other models, including the random null model.
\end{enumerate}

\subsection{Using Rosetta to score antigen-stabilizing mutations}
In Figures~\ref{fig:antigen_figure_main_results},~\ref{fig:antigen_figure_of_different_types_of_mutations},~\ref{fig:antigen_heatmap} and Tables~\ref{table:antigen_results},~\ref{table:speed_on_antigens}, we reported the Rosetta scores for the verified antigen-stabilizing mutations. To compute these scores, we used the PyRosetta software~\citep{chaudhury_pyrosetta_2010} to model protein structures, with wild-type structures serving as template. 
For each target sequence containing a single point mutation, we threaded the mutant sequence onto all chains of the oligomeric template. Each resulting model was then subjected to a high-resolution refinement protocol using Rosetta's FastRelax application. This protocol involved five cycles of side-chain rotamer repacking followed by gradient-based energy minimization of backbone ($\phi$, $\psi$) and side-chain ($\chi$) torsion angles, guided by the \texttt{ref2015\_cart} all-atom energy function. To improve conformational sampling, we repeated the threading-and-relaxation procedure 20 times per mutation. For each mutant, we report the mean Rosetta Energy Unit (REU) score of the five lowest-energy (most favorable) relaxed models.

% \bibliographystyle{unsrt}
% \bibliography{protholo.bib, armita_bib_safe.bib, tcr_antigen_design_gm_safe.bib}

\begin{thebibliography}{10}
\expandafter\ifx\csname url\endcsname\relax
  \def\url#1{\burl{#1}}\fi
\expandafter\ifx\csname urlprefix\endcsname\relax\def\urlprefix{URL }\fi
\providecommand{\bibinfo}[2]{#2}
\providecommand{\eprint}[2][]{\url{#2}}
\providecommand{\doi}[1]{\url{https://doi.org/#1}}
\bibcommenthead

\bibitem{gerasimavicius_identification_2020}
\bibinfo{author}{Gerasimavicius, L.}, \bibinfo{author}{Liu, X.} \&
  \bibinfo{author}{Marsh, J.~A.}
\newblock \bibinfo{title}{Identification of pathogenic missense mutations using
  protein stability predictors}.
\newblock \emph{\bibinfo{journal}{Scientific Reports}}
  \textbf{\bibinfo{volume}{10}}, \bibinfo{pages}{15387} (\bibinfo{year}{2020}).
\newblock \urlprefix\url{https://www.nature.com/articles/s41598-020-72404-w}.
\newblock \bibinfo{note}{Publisher: Nature Publishing Group}.

\bibitem{blaabjerg_rapid_2023}
\bibinfo{author}{Blaabjerg, L.~M.} \emph{et~al.}
\newblock \bibinfo{title}{Rapid protein stability prediction using deep
  learning representations}.
\newblock \emph{\bibinfo{journal}{eLife}} \textbf{\bibinfo{volume}{12}},
  \bibinfo{pages}{e82593} (\bibinfo{year}{2023}).
\newblock \urlprefix\url{https://doi.org/10.7554/eLife.82593}.
\newblock \bibinfo{note}{Publisher: eLife Sciences Publications, Ltd}.

\bibitem{ishida_effects_2010}
\bibinfo{author}{Ishida, T.}
\newblock \bibinfo{title}{Effects of {Point} {Mutation} on {Enzymatic}
  {Activity}: {Correlation} between {Protein} {Electronic} {Structure} and
  {Motion} in {Chorismate} {Mutase} {Reaction}}.
\newblock \emph{\bibinfo{journal}{Journal of the American Chemical Society}}
  \textbf{\bibinfo{volume}{132}}, \bibinfo{pages}{7104--7118}
  (\bibinfo{year}{2010}).
\newblock \urlprefix\url{https://doi.org/10.1021/ja100744h}.
\newblock \bibinfo{note}{Publisher: American Chemical Society}.

\bibitem{wang_d3distalmutation_2021}
\bibinfo{author}{Wang, X.} \emph{et~al.}
\newblock \bibinfo{title}{{D3DistalMutation}: a {Database} to {Explore} the
  {Effect} of {Distal} {Mutations} on {Enzyme} {Activity}}.
\newblock \emph{\bibinfo{journal}{Journal of Chemical Information and
  Modeling}} \textbf{\bibinfo{volume}{61}}, \bibinfo{pages}{2499--2508}
  (\bibinfo{year}{2021}).
\newblock \urlprefix\url{https://doi.org/10.1021/acs.jcim.1c00318}.
\newblock \bibinfo{note}{Publisher: American Chemical Society}.

\bibitem{thadani_learning_2023}
\bibinfo{author}{Thadani, N.~N.} \emph{et~al.}
\newblock \bibinfo{title}{Learning from prepandemic data to forecast viral
  escape}.
\newblock \emph{\bibinfo{journal}{Nature}} \textbf{\bibinfo{volume}{622}},
  \bibinfo{pages}{818--825} (\bibinfo{year}{2023}).
\newblock \urlprefix\url{https://www.nature.com/articles/s41586-023-06617-0}.
\newblock \bibinfo{note}{Publisher: Nature Publishing Group}.

\bibitem{Luksza2014-pg}
\bibinfo{author}{Łuksza, M.} \& \bibinfo{author}{L\"{a}ssig, M.}
\newblock \bibinfo{title}{A predictive fitness model for influenza}.
\newblock \emph{\bibinfo{journal}{Nature}} \textbf{\bibinfo{volume}{507}},
  \bibinfo{pages}{57--61} (\bibinfo{year}{2014}).

\bibitem{Neher2014-pf}
\bibinfo{author}{Neher, R.~A.}, \bibinfo{author}{Russell, C.~A.} \&
  \bibinfo{author}{Shraiman, B.~I.}
\newblock \bibinfo{title}{Predicting evolution from the shape of genealogical
  trees}.
\newblock \emph{\bibinfo{journal}{Elife}} \textbf{\bibinfo{volume}{3}}
  (\bibinfo{year}{2014}).

\bibitem{hie_efficient_2024}
\bibinfo{author}{Hie, B.~L.} \emph{et~al.}
\newblock \bibinfo{title}{Efficient evolution of human antibodies from general
  protein language models}.
\newblock \emph{\bibinfo{journal}{Nature Biotechnology}}
  \textbf{\bibinfo{volume}{42}}, \bibinfo{pages}{275--283}
  (\bibinfo{year}{2024}).
\newblock \urlprefix\url{https://www.nature.com/articles/s41587-023-01763-2}.
\newblock \bibinfo{note}{Publisher: Nature Publishing Group}.

\bibitem{mclellan_structure-based_2013}
\bibinfo{author}{McLellan, J.~S.} \emph{et~al.}
\newblock \bibinfo{title}{Structure-{Based} {Design} of a {Fusion}
  {Glycoprotein} {Vaccine} for {Respiratory} {Syncytial} {Virus}}.
\newblock \emph{\bibinfo{journal}{Science}} \textbf{\bibinfo{volume}{342}},
  \bibinfo{pages}{592--598} (\bibinfo{year}{2013}).
\newblock \urlprefix\url{https://www.science.org/doi/10.1126/science.1243283}.
\newblock \bibinfo{note}{Publisher: American Association for the Advancement of
  Science}.

\bibitem{gonzalez_general_2024}
\bibinfo{author}{Gonzalez, K.~J.} \emph{et~al.}
\newblock \bibinfo{title}{A general computational design strategy for
  stabilizing viral class {I} fusion proteins}.
\newblock \emph{\bibinfo{journal}{Nature Communications}}
  \textbf{\bibinfo{volume}{15}}, \bibinfo{pages}{1335} (\bibinfo{year}{2024}).
\newblock \urlprefix\url{https://www.nature.com/articles/s41467-024-45480-z}.
\newblock \bibinfo{note}{Publisher: Nature Publishing Group}.

\bibitem{bakkers_efficacious_2024}
\bibinfo{author}{Bakkers, M. J.~G.} \emph{et~al.}
\newblock \bibinfo{title}{Efficacious human metapneumovirus vaccine based on
  {AI}-guided engineering of a closed prefusion trimer}.
\newblock \emph{\bibinfo{journal}{Nature Communications}}
  \textbf{\bibinfo{volume}{15}}, \bibinfo{pages}{6270} (\bibinfo{year}{2024}).
\newblock \urlprefix\url{https://www.nature.com/articles/s41467-024-50659-5}.
\newblock \bibinfo{note}{Publisher: Nature Publishing Group}.

\bibitem{milder_universal_2022}
\bibinfo{author}{Milder, F.~J.} \emph{et~al.}
\newblock \bibinfo{title}{Universal stabilization of the influenza
  hemagglutinin by structure-based redesign of the {pH} switch regions}.
\newblock \emph{\bibinfo{journal}{Proceedings of the National Academy of
  Sciences}} \textbf{\bibinfo{volume}{119}}, \bibinfo{pages}{e2115379119}
  (\bibinfo{year}{2022}).
\newblock
  \urlprefix\url{https://www.pnas.org/doi/full/10.1073/pnas.2115379119}.
\newblock \bibinfo{note}{Publisher: Proceedings of the National Academy of
  Sciences}.

\bibitem{phan_conserved_2022}
\bibinfo{author}{Phan, T. T.~N.} \emph{et~al.}
\newblock \bibinfo{title}{A conserved set of mutations for stabilizing soluble
  envelope protein dimers from dengue and {Zika} viruses to advance the
  development of subunit vaccines}.
\newblock \emph{\bibinfo{journal}{Journal of Biological Chemistry}}
  \textbf{\bibinfo{volume}{298}}, \bibinfo{pages}{102079}
  (\bibinfo{year}{2022}).
\newblock
  \urlprefix\url{https://www.sciencedirect.com/science/article/pii/S0021925822005191}.

\bibitem{rocklin_global_2017}
\bibinfo{author}{Rocklin, G.~J.} \emph{et~al.}
\newblock \bibinfo{title}{Global analysis of protein folding using massively
  parallel design, synthesis, and testing}.
\newblock \emph{\bibinfo{journal}{Science}} \textbf{\bibinfo{volume}{357}},
  \bibinfo{pages}{168--175} (\bibinfo{year}{2017}).
\newblock \urlprefix\url{https://www.science.org/doi/10.1126/science.aan0693}.
\newblock \bibinfo{note}{Publisher: American Association for the Advancement of
  Science}.

\bibitem{lindorff-larsen_linking_2021}
\bibinfo{author}{Lindorff-Larsen, K.} \& \bibinfo{author}{Teilum, K.}
\newblock \bibinfo{title}{Linking thermodynamics and measurements of protein
  stability}.
\newblock \emph{\bibinfo{journal}{Protein Engineering, Design and Selection}}
  \textbf{\bibinfo{volume}{34}}, \bibinfo{pages}{gzab002}
  (\bibinfo{year}{2021}).
\newblock \urlprefix\url{https://doi.org/10.1093/protein/gzab002}.

\bibitem{karlsson_spr_2004}
\bibinfo{author}{Karlsson, R.}
\newblock \bibinfo{title}{{SPR} for molecular interaction analysis: a review of
  emerging application areas}.
\newblock \emph{\bibinfo{journal}{Journal of Molecular Recognition}}
  \textbf{\bibinfo{volume}{17}}, \bibinfo{pages}{151--161}
  (\bibinfo{year}{2004}).
\newblock
  \urlprefix\url{https://onlinelibrary.wiley.com/doi/abs/10.1002/jmr.660}.
\newblock \bibinfo{note}{\_eprint:
  https://onlinelibrary.wiley.com/doi/pdf/10.1002/jmr.660}.

\bibitem{fowler_deep_2014}
\bibinfo{author}{Fowler, D.~M.} \& \bibinfo{author}{Fields, S.}
\newblock \bibinfo{title}{Deep mutational scanning: a new style of protein
  science}.
\newblock \emph{\bibinfo{journal}{Nature Methods}}
  \textbf{\bibinfo{volume}{11}}, \bibinfo{pages}{801--807}
  (\bibinfo{year}{2014}).
\newblock \urlprefix\url{https://www.nature.com/articles/nmeth.3027}.
\newblock \bibinfo{note}{Publisher: Nature Publishing Group}.

\bibitem{Kinney2019-wc}
\bibinfo{author}{Kinney, J.~B.} \& \bibinfo{author}{McCandlish, D.~M.}
\newblock \bibinfo{title}{Massively parallel assays and quantitative
  {Sequence–Function} relationships}.
\newblock \emph{\bibinfo{journal}{Annu. Rev. Genomics Hum. Genet.}}
  \textbf{\bibinfo{volume}{20}}, \bibinfo{pages}{99--127}
  (\bibinfo{year}{2019}).

\bibitem{starr_deep_2020}
\bibinfo{author}{Starr, T.~N.} \emph{et~al.}
\newblock \bibinfo{title}{Deep {Mutational} {Scanning} of {SARS}-{CoV}-2
  {Receptor} {Binding} {Domain} {Reveals} {Constraints} on {Folding} and {ACE2}
  {Binding}}.
\newblock \emph{\bibinfo{journal}{Cell}} \textbf{\bibinfo{volume}{182}},
  \bibinfo{pages}{1295--1310.e20} (\bibinfo{year}{2020}).
\newblock
  \urlprefix\url{https://www.ncbi.nlm.nih.gov/pmc/articles/PMC7418704/}.

\bibitem{tsuboyama_mega-scale_2023}
\bibinfo{author}{Tsuboyama, K.} \emph{et~al.}
\newblock \bibinfo{title}{Mega-scale experimental analysis of protein folding
  stability in biology and design}.
\newblock \emph{\bibinfo{journal}{Nature}} \textbf{\bibinfo{volume}{620}},
  \bibinfo{pages}{434--444} (\bibinfo{year}{2023}).
\newblock \urlprefix\url{https://www.nature.com/articles/s41586-023-06328-6}.
\newblock \bibinfo{note}{Publisher: Nature Publishing Group}.

\bibitem{gapsys_accurate_2016}
\bibinfo{author}{Gapsys, V.}, \bibinfo{author}{Michielssens, S.},
  \bibinfo{author}{Seeliger, D.} \& \bibinfo{author}{de~Groot, B.~L.}
\newblock \bibinfo{title}{Accurate and {Rigorous} {Prediction} of the {Changes}
  in {Protein} {Free} {Energies} in a {Large}-{Scale} {Mutation} {Scan}}.
\newblock \emph{\bibinfo{journal}{Angewandte Chemie International Edition}}
  \textbf{\bibinfo{volume}{55}}, \bibinfo{pages}{7364--7368}
  (\bibinfo{year}{2016}).
\newblock
  \urlprefix\url{https://onlinelibrary.wiley.com/doi/abs/10.1002/anie.201510054}.
\newblock \bibinfo{note}{\_eprint:
  https://onlinelibrary.wiley.com/doi/pdf/10.1002/anie.201510054}.

\bibitem{schymkowitz_foldx_2005}
\bibinfo{author}{Schymkowitz, J.} \emph{et~al.}
\newblock \bibinfo{title}{The {FoldX} web server: an online force field}.
\newblock \emph{\bibinfo{journal}{Nucleic Acids Research}}
  \textbf{\bibinfo{volume}{33}}, \bibinfo{pages}{W382--W388}
  (\bibinfo{year}{2005}).
\newblock \urlprefix\url{https://doi.org/10.1093/nar/gki387}.

\bibitem{kellogg_role_2011}
\bibinfo{author}{Kellogg, E.~H.}, \bibinfo{author}{Leaver-Fay, A.} \&
  \bibinfo{author}{Baker, D.}
\newblock \bibinfo{title}{Role of conformational sampling in computing
  mutation-induced changes in protein structure and stability}.
\newblock \emph{\bibinfo{journal}{Proteins: Structure, Function, and
  Bioinformatics}} \textbf{\bibinfo{volume}{79}}, \bibinfo{pages}{830--838}
  (\bibinfo{year}{2011}).
\newblock
  \urlprefix\url{https://onlinelibrary.wiley.com/doi/abs/10.1002/prot.22921}.
\newblock \bibinfo{note}{\_eprint:
  https://onlinelibrary.wiley.com/doi/pdf/10.1002/prot.22921}.

\bibitem{riesselman_deep_2018}
\bibinfo{author}{Riesselman, A.~J.}, \bibinfo{author}{Ingraham, J.~B.} \&
  \bibinfo{author}{Marks, D.~S.}
\newblock \bibinfo{title}{Deep generative models of genetic variation capture
  the effects of mutations}.
\newblock \emph{\bibinfo{journal}{Nature Methods}}
  \textbf{\bibinfo{volume}{15}}, \bibinfo{pages}{816--822}
  (\bibinfo{year}{2018}).
\newblock \urlprefix\url{https://www.nature.com/articles/s41592-018-0138-4}.
\newblock \bibinfo{note}{Publisher: Nature Publishing Group}.

\bibitem{meier_language_2021}
\bibinfo{author}{Meier, J.} \emph{et~al.}
\newblock \emph{\bibinfo{title}{Language models enable zero-shot prediction of
  the effects of mutations on protein function}}, Vol.~\bibinfo{volume}{34},
  \bibinfo{pages}{29287--29303} (\bibinfo{publisher}{Curran Associates, Inc.},
  \bibinfo{year}{2021}).
\newblock
  \urlprefix\url{https://proceedings.neurips.cc/paper/2021/hash/f51338d736f95dd42427296047067694-Abstract.html}.

\bibitem{pun_learning_2024}
\bibinfo{author}{Pun, M.~N.} \emph{et~al.}
\newblock \bibinfo{title}{Learning the shape of protein microenvironments with
  a holographic convolutional neural network}.
\newblock \emph{\bibinfo{journal}{Proceedings of the National Academy of
  Sciences}} \textbf{\bibinfo{volume}{121}} (\bibinfo{year}{2024}).
\newblock \urlprefix\url{https://www.pnas.org/doi/10.1073/pnas.2300838121}.
\newblock \bibinfo{note}{Publisher: Proceedings of the National Academy of
  Sciences}.

\bibitem{diaz_stability_2024}
\bibinfo{author}{Diaz, D.~J.} \emph{et~al.}
\newblock \bibinfo{title}{Stability {Oracle}: a structure-based
  graph-transformer framework for identifying stabilizing mutations}.
\newblock \emph{\bibinfo{journal}{Nature Communications}}
  \textbf{\bibinfo{volume}{15}}, \bibinfo{pages}{6170} (\bibinfo{year}{2024}).
\newblock \urlprefix\url{https://www.nature.com/articles/s41467-024-49780-2}.
\newblock \bibinfo{note}{Publisher: Nature Publishing Group}.

\bibitem{li_predicting_2020}
\bibinfo{author}{Li, B.}, \bibinfo{author}{Yang, Y.~T.},
  \bibinfo{author}{Capra, J.~A.} \& \bibinfo{author}{Gerstein, M.~B.}
\newblock \bibinfo{title}{Predicting changes in protein thermodynamic stability
  upon point mutation with deep {3D} convolutional neural networks}.
\newblock \emph{\bibinfo{journal}{PLOS Computational Biology}}
  \textbf{\bibinfo{volume}{16}}, \bibinfo{pages}{e1008291}
  (\bibinfo{year}{2020}).
\newblock
  \urlprefix\url{https://journals.plos.org/ploscompbiol/article?id=10.1371/journal.pcbi.1008291}.
\newblock \bibinfo{note}{Publisher: Public Library of Science}.

\bibitem{benevenuta_antisymmetric_2021}
\bibinfo{author}{Benevenuta, S.}, \bibinfo{author}{Pancotti, C.},
  \bibinfo{author}{Fariselli, P.}, \bibinfo{author}{Birolo, G.} \&
  \bibinfo{author}{Sanavia, T.}
\newblock \bibinfo{title}{An antisymmetric neural network to predict free
  energy changes in protein variants}.
\newblock \emph{\bibinfo{journal}{Journal of Physics D: Applied Physics}}
  \textbf{\bibinfo{volume}{54}}, \bibinfo{pages}{245403}
  (\bibinfo{year}{2021}).
\newblock \urlprefix\url{https://dx.doi.org/10.1088/1361-6463/abedfb}.
\newblock \bibinfo{note}{Publisher: IOP Publishing}.

\bibitem{dieckhaus_transfer_2024}
\bibinfo{author}{Dieckhaus, H.}, \bibinfo{author}{Brocidiacono, M.},
  \bibinfo{author}{Randolph, N.~Z.} \& \bibinfo{author}{Kuhlman, B.}
\newblock \bibinfo{title}{Transfer learning to leverage larger datasets for
  improved prediction of protein stability changes}.
\newblock \emph{\bibinfo{journal}{Proceedings of the National Academy of
  Sciences}} \textbf{\bibinfo{volume}{121}}, \bibinfo{pages}{e2314853121}
  (\bibinfo{year}{2024}).
\newblock \urlprefix\url{https://www.pnas.org/doi/10.1073/pnas.2314853121}.
\newblock \bibinfo{note}{Publisher: Proceedings of the National Academy of
  Sciences}.

\bibitem{gordon_protein_2024}
\bibinfo{author}{Gordon, C.~W.}, \bibinfo{author}{Lu, A.~X.} \&
  \bibinfo{author}{Abbeel, P.}
\newblock \emph{\bibinfo{title}{Protein {Language} {Model} {Fitness} is a
  {Matter} of {Preference}}} (\bibinfo{year}{2024}).
\newblock \urlprefix\url{https://openreview.net/forum?id=UvPdpa4LuV}.

\bibitem{gong_evolution-inspired_2024}
\bibinfo{author}{Gong, C.} \emph{et~al.}
\newblock \emph{\bibinfo{title}{Evolution-{Inspired} {Loss} {Functions} for
  {Protein} {Representation} {Learning}}} (\bibinfo{year}{2024}).
\newblock
  \urlprefix\url{https://openreview.net/forum?id=y5L8W0KRUX&referrer=\%5Bthe\%20profile\%20of\%20Chengyue\%20Gong\%5D(\%2Fprofile\%3Fid\%3D~Chengyue_Gong1)}.

\bibitem{luo_rotamer_2022}
\bibinfo{author}{Luo, S.} \emph{et~al.}
\newblock \emph{\bibinfo{title}{Rotamer {Density} {Estimator} is an
  {Unsupervised} {Learner} of the {Effect} of {Mutations} on
  {Protein}-{Protein} {Interaction}}} (\bibinfo{year}{2022}).
\newblock \urlprefix\url{https://openreview.net/forum?id=_X9Yl1K2mD}.

\bibitem{chaudhury_pyrosetta_2010}
\bibinfo{author}{Chaudhury, S.}, \bibinfo{author}{Lyskov, S.} \&
  \bibinfo{author}{Gray, J.~J.}
\newblock \bibinfo{title}{{PyRosetta}: a script-based interface for
  implementing molecular modeling algorithms using {Rosetta}}.
\newblock \emph{\bibinfo{journal}{Bioinformatics (Oxford, England)}}
  \textbf{\bibinfo{volume}{26}}, \bibinfo{pages}{689--691}
  (\bibinfo{year}{2010}).

\bibitem{dauparas_robust_2022}
\bibinfo{author}{Dauparas, J.} \emph{et~al.}
\newblock \bibinfo{title}{Robust deep learning–based protein sequence design
  using {ProteinMPNN}}.
\newblock \emph{\bibinfo{journal}{Science}} \textbf{\bibinfo{volume}{378}},
  \bibinfo{pages}{49--56} (\bibinfo{year}{2022}).
\newblock \urlprefix\url{https://www.science.org/doi/10.1126/science.add2187}.
\newblock \bibinfo{note}{Publisher: American Association for the Advancement of
  Science}.

\bibitem{visani_holographic-vae_2024}
\bibinfo{author}{Visani, G.~M.}, \bibinfo{author}{Pun, M.~N.},
  \bibinfo{author}{Angaji, A.} \& \bibinfo{author}{Nourmohammad, A.}
\newblock \bibinfo{title}{Holographic-({V}){AE}: {An} end-to-end
  {SO}(3)-equivariant (variational) autoencoder in {Fourier} space}.
\newblock \emph{\bibinfo{journal}{Physical Review Research}}
  \textbf{\bibinfo{volume}{6}}, \bibinfo{pages}{023006} (\bibinfo{year}{2024}).
\newblock
  \urlprefix\url{https://link.aps.org/doi/10.1103/PhysRevResearch.6.023006}.
\newblock \bibinfo{note}{Publisher: American Physical Society}.

\bibitem{visani_h-packer_2024}
\bibinfo{author}{Visani, G.~M.}, \bibinfo{author}{Galvin, W.},
  \bibinfo{author}{Pun, M.} \& \bibinfo{author}{Nourmohammad, A.}
\newblock \emph{\bibinfo{title}{H-{Packer}: {Holographic} {Rotationally}
  {Equivariant} {Convolutional} {Neural} {Network} for {Protein} {Side}-{Chain}
  {Packing}}}, \bibinfo{pages}{230--249} (\bibinfo{publisher}{PMLR},
  \bibinfo{year}{2024}).
\newblock \urlprefix\url{https://proceedings.mlr.press/v240/visani24a.html}.
\newblock \bibinfo{note}{ISSN: 2640-3498}.

\bibitem{alquraishi_proteinnet_2019}
\bibinfo{author}{AlQuraishi, M.}
\newblock \bibinfo{title}{{ProteinNet}: a standardized data set for machine
  learning of protein structure}.
\newblock \emph{\bibinfo{journal}{BMC Bioinformatics}}
  \textbf{\bibinfo{volume}{20}}, \bibinfo{pages}{311} (\bibinfo{year}{2019}).
\newblock \urlprefix\url{https://doi.org/10.1186/s12859-019-2932-0}.

\bibitem{mei_new_2005}
\bibinfo{author}{Mei, H.}, \bibinfo{author}{Liao, Z.~H.},
  \bibinfo{author}{Zhou, Y.} \& \bibinfo{author}{Li, S.~Z.}
\newblock \bibinfo{title}{A new set of amino acid descriptors and its
  application in peptide {QSARs}}.
\newblock \emph{\bibinfo{journal}{Peptide Science}}
  \textbf{\bibinfo{volume}{80}}, \bibinfo{pages}{775--786}
  (\bibinfo{year}{2005}).
\newblock
  \urlprefix\url{https://onlinelibrary.wiley.com/doi/abs/10.1002/bip.20296}.
\newblock \bibinfo{note}{\_eprint:
  https://onlinelibrary.wiley.com/doi/pdf/10.1002/bip.20296}.

\bibitem{lin_evolutionary-scale_2023}
\bibinfo{author}{Lin, Z.} \emph{et~al.}
\newblock \bibinfo{title}{Evolutionary-scale prediction of atomic-level protein
  structure with a language model}.
\newblock \emph{\bibinfo{journal}{Science}} \textbf{\bibinfo{volume}{379}},
  \bibinfo{pages}{1123--1130} (\bibinfo{year}{2023}).
\newblock \urlprefix\url{https://www.science.org/doi/10.1126/science.ade2574}.
\newblock \bibinfo{note}{Publisher: American Association for the Advancement of
  Science}.

\bibitem{pyo_data-driven_2025}
\bibinfo{author}{Pyo, A. G.~T.} \emph{et~al.}
\newblock \bibinfo{title}{Data-{Driven} {Discovery} of {Biophysical} {T} {Cell}
  {Receptor} {Cospecificity} {Rules}}.
\newblock \emph{\bibinfo{journal}{PRX Life}} \textbf{\bibinfo{volume}{3}},
  \bibinfo{pages}{033005} (\bibinfo{year}{2025}).
\newblock \urlprefix\url{https://link.aps.org/doi/10.1103/14j1-wrh5}.
\newblock \bibinfo{note}{Publisher: American Physical Society}.

\bibitem{pancotti_predicting_2022}
\bibinfo{author}{Pancotti, C.} \emph{et~al.}
\newblock \bibinfo{title}{Predicting protein stability changes upon
  single-point mutation: a thorough comparison of the available tools on a new
  dataset}.
\newblock \emph{\bibinfo{journal}{Briefings in Bioinformatics}}
  \textbf{\bibinfo{volume}{23}}, \bibinfo{pages}{bbab555}
  (\bibinfo{year}{2022}).
\newblock \urlprefix\url{https://doi.org/10.1093/bib/bbab555}.

\bibitem{lee_rational_2024}
\bibinfo{author}{Lee, Y.-Z.} \emph{et~al.}
\newblock \bibinfo{title}{Rational design of uncleaved prefusion-closed trimer
  vaccines for human respiratory syncytial virus and metapneumovirus}.
\newblock \emph{\bibinfo{journal}{Nature Communications}}
  \textbf{\bibinfo{volume}{15}}, \bibinfo{pages}{9939} (\bibinfo{year}{2024}).
\newblock \urlprefix\url{https://www.nature.com/articles/s41467-024-54287-x}.
\newblock \bibinfo{note}{Publisher: Nature Publishing Group}.

\bibitem{hsieh_structure-based_2020}
\bibinfo{author}{Hsieh, C.-L.} \emph{et~al.}
\newblock \bibinfo{title}{Structure-based design of prefusion-stabilized
  {SARS}-{CoV}-2 spikes}.
\newblock \emph{\bibinfo{journal}{Science}} \textbf{\bibinfo{volume}{369}},
  \bibinfo{pages}{1501--1505} (\bibinfo{year}{2020}).
\newblock \urlprefix\url{https://www.science.org/doi/10.1126/science.abd0826}.
\newblock \bibinfo{note}{Publisher: American Association for the Advancement of
  Science}.

\bibitem{byrne_principles_2022}
\bibinfo{author}{Byrne, P.~O.} \& \bibinfo{author}{McLellan, J.~S.}
\newblock \bibinfo{title}{Principles and practical applications of
  structure-based vaccine design}.
\newblock \emph{\bibinfo{journal}{Current Opinion in Immunology}}
  \textbf{\bibinfo{volume}{77}}, \bibinfo{pages}{102209}
  (\bibinfo{year}{2022}).
\newblock
  \urlprefix\url{https://www.sciencedirect.com/science/article/pii/S0952791522000565}.

\bibitem{park_simultaneous_2016}
\bibinfo{author}{Park, H.} \emph{et~al.}
\newblock \bibinfo{title}{Simultaneous {Optimization} of {Biomolecular}
  {Energy} {Functions} on {Features} from {Small} {Molecules} and
  {Macromolecules}}.
\newblock \emph{\bibinfo{journal}{Journal of Chemical Theory and Computation}}
  \textbf{\bibinfo{volume}{12}}, \bibinfo{pages}{6201--6212}
  (\bibinfo{year}{2016}).
\newblock \urlprefix\url{https://doi.org/10.1021/acs.jctc.6b00819}.
\newblock \bibinfo{note}{Publisher: American Chemical Society}.

\bibitem{delgado_foldx_2019}
\bibinfo{author}{Delgado, J.}, \bibinfo{author}{Radusky, L.~G.},
  \bibinfo{author}{Cianferoni, D.} \& \bibinfo{author}{Serrano, L.}
\newblock \bibinfo{title}{{FoldX} 5.0: working with {RNA}, small molecules and
  a new graphical interface}.
\newblock \emph{\bibinfo{journal}{Bioinformatics}}
  \textbf{\bibinfo{volume}{35}}, \bibinfo{pages}{4168--4169}
  (\bibinfo{year}{2019}).
\newblock \urlprefix\url{https://doi.org/10.1093/bioinformatics/btz184}.

\bibitem{shan_deep_2022}
\bibinfo{author}{Shan, S.} \emph{et~al.}
\newblock \bibinfo{title}{Deep learning guided optimization of human antibody
  against {SARS}-{CoV}-2 variants with broad neutralization}.
\newblock \emph{\bibinfo{journal}{Proceedings of the National Academy of
  Sciences}} \textbf{\bibinfo{volume}{119}}, \bibinfo{pages}{e2122954119}
  (\bibinfo{year}{2022}).
\newblock
  \urlprefix\url{https://www.pnas.org/doi/full/10.1073/pnas.2122954119}.
\newblock \bibinfo{note}{Publisher: Proceedings of the National Academy of
  Sciences}.

\bibitem{hsu_learning_2022}
\bibinfo{author}{Hsu, C.} \emph{et~al.}
\newblock \emph{\bibinfo{title}{Learning inverse folding from millions of
  predicted structures}}, \bibinfo{pages}{8946--8970}
  (\bibinfo{publisher}{PMLR}, \bibinfo{year}{2022}).
\newblock \urlprefix\url{https://proceedings.mlr.press/v162/hsu22a.html}.
\newblock \bibinfo{note}{ISSN: 2640-3498}.

\bibitem{tao_reliable_2025}
\bibinfo{author}{Tao, F.}, \bibinfo{author}{Sun, J.}, \bibinfo{author}{Gao,
  P.}, \bibinfo{author}{Gao, G.~F.} \& \bibinfo{author}{Wu, B.}
\newblock \bibinfo{title}{Reliable prediction of protein–protein binding
  affinity changes upon mutations with {Pythia}-{PPI}}.
\newblock \emph{\bibinfo{journal}{National Science Review}}
  \textbf{\bibinfo{volume}{12}}, \bibinfo{pages}{nwaf231}
  (\bibinfo{year}{2025}).
\newblock \urlprefix\url{https://doi.org/10.1093/nsr/nwaf231}.

\bibitem{jankauskaite_skempi_2019}
\bibinfo{author}{Jankauskaitė, J.}, \bibinfo{author}{Jiménez-García, B.},
  \bibinfo{author}{Dapkūnas, J.}, \bibinfo{author}{Fernández-Recio, J.} \&
  \bibinfo{author}{Moal, I.~H.}
\newblock \bibinfo{title}{{SKEMPI} 2.0: an updated benchmark of changes in
  protein–protein binding energy, kinetics and thermodynamics upon mutation}.
\newblock \emph{\bibinfo{journal}{Bioinformatics}}
  \textbf{\bibinfo{volume}{35}}, \bibinfo{pages}{462--469}
  (\bibinfo{year}{2019}).
\newblock \urlprefix\url{https://doi.org/10.1093/bioinformatics/bty635}.

\bibitem{stourac_fireprotdb_2021}
\bibinfo{author}{Stourac, J.} \emph{et~al.}
\newblock \bibinfo{title}{{FireProtDB}: database of manually curated protein
  stability data}.
\newblock \emph{\bibinfo{journal}{Nucleic Acids Research}}
  \textbf{\bibinfo{volume}{49}}, \bibinfo{pages}{D319--D324}
  (\bibinfo{year}{2021}).
\newblock \urlprefix\url{https://doi.org/10.1093/nar/gkaa981}.

\bibitem{visani_t_2025}
\bibinfo{author}{Visani, G.~M.} \emph{et~al.}
\newblock \bibinfo{title}{T cell receptor specificity landscape revealed
  through de novo peptide design}.
\newblock \emph{\bibinfo{journal}{Proceedings of the National Academy of
  Sciences}} \textbf{\bibinfo{volume}{122}}, \bibinfo{pages}{e2504783122}
  (\bibinfo{year}{2025}).
\newblock \urlprefix\url{https://www.pnas.org/doi/10.1073/pnas.2504783122}.
\newblock \bibinfo{note}{Publisher: Proceedings of the National Academy of
  Sciences}.

\bibitem{geiger_e3nn_2022}
\bibinfo{author}{Geiger, M.} \& \bibinfo{author}{Smidt, T.}
\newblock \bibinfo{title}{e3nn: {Euclidean} {Neural} {Networks}}
  (\bibinfo{year}{2022}).
\newblock \urlprefix\url{http://arxiv.org/abs/2207.09453}.
\newblock \bibinfo{note}{ArXiv:2207.09453 [cs]}.

\bibitem{cock_biopython_2009}
\bibinfo{author}{Cock, P. J.~A.} \emph{et~al.}
\newblock \bibinfo{title}{Biopython: freely available {Python} tools for
  computational molecular biology and bioinformatics}.
\newblock \emph{\bibinfo{journal}{Bioinformatics}}
  \textbf{\bibinfo{volume}{25}}, \bibinfo{pages}{1422--1423}
  (\bibinfo{year}{2009}).
\newblock \urlprefix\url{https://doi.org/10.1093/bioinformatics/btp163}.

\bibitem{eastman_openmm_2013}
\bibinfo{author}{Eastman, P.} \emph{et~al.}
\newblock \bibinfo{title}{{OpenMM} 4: {A} {Reusable}, {Extensible}, {Hardware}
  {Independent} {Library} for {High} {Performance} {Molecular} {Simulation}}.
\newblock \emph{\bibinfo{journal}{Journal of Chemical Theory and Computation}}
  \textbf{\bibinfo{volume}{9}}, \bibinfo{pages}{461--469}
  (\bibinfo{year}{2013}).
\newblock \urlprefix\url{https://doi.org/10.1021/ct300857j}.
\newblock \bibinfo{note}{Publisher: American Chemical Society}.

\bibitem{word_asparagine_1999}
\bibinfo{author}{Word, J.~M.}, \bibinfo{author}{Lovell, S.~C.},
  \bibinfo{author}{Richardson, J.~S.} \& \bibinfo{author}{Richardson, D.~C.}
\newblock \bibinfo{title}{Asparagine and glutamine: using hydrogen atom
  contacts in the choice of side-chain amide orientation1}.
\newblock \emph{\bibinfo{journal}{Journal of Molecular Biology}}
  \textbf{\bibinfo{volume}{285}}, \bibinfo{pages}{1735--1747}
  (\bibinfo{year}{1999}).
\newblock
  \urlprefix\url{https://www.sciencedirect.com/science/article/pii/S0022283698924019}.

\bibitem{ponder_force_2003}
\bibinfo{author}{Ponder, J.~W.} \& \bibinfo{author}{Case, D.~A.}
\newblock \bibinfo{title}{ in \textit{Force {Fields} for {Protein}
  {Simulations}}} , Vol.~\bibinfo{volume}{66} of \emph{\bibinfo{series}{Protein
  {Simulations}}} \bibinfo{pages}{27--85} (\bibinfo{publisher}{Academic Press},
  \bibinfo{year}{2003}).
\newblock
  \urlprefix\url{https://www.sciencedirect.com/science/article/pii/S006532330366002X}.

\bibitem{kingma_adam_2015}
\bibinfo{author}{Kingma, D.~P.} \& \bibinfo{author}{Ba, J.}
\newblock \bibinfo{editor}{Bengio, Y.} \& \bibinfo{editor}{LeCun, Y.} (eds)
  \emph{\bibinfo{title}{Adam: {A} {Method} for {Stochastic} {Optimization}}}.
\newblock (eds \bibinfo{editor}{Bengio, Y.} \& \bibinfo{editor}{LeCun, Y.})
  \emph{\bibinfo{booktitle}{3rd {International} {Conference} on {Learning}
  {Representations}, {ICLR} 2015, {San} {Diego}, {CA}, {USA}, {May} 7-9, 2015,
  {Conference} {Track} {Proceedings}}} (\bibinfo{year}{2015}).
\newblock \urlprefix\url{http://arxiv.org/abs/1412.6980}.

\bibitem{pucci_quantification_2018}
\bibinfo{author}{Pucci, F.}, \bibinfo{author}{Bernaerts, K.~V.},
  \bibinfo{author}{Kwasigroch, J.~M.} \& \bibinfo{author}{Rooman, M.}
\newblock \bibinfo{title}{Quantification of biases in predictions of protein
  stability changes upon mutations}.
\newblock \emph{\bibinfo{journal}{Bioinformatics}}
  \textbf{\bibinfo{volume}{34}}, \bibinfo{pages}{3659--3665}
  (\bibinfo{year}{2018}).
\newblock \urlprefix\url{https://doi.org/10.1093/bioinformatics/bty348}.

\bibitem{caldararu_base_2021}
\bibinfo{author}{Caldararu, O.}, \bibinfo{author}{Blundell, T.~L.} \&
  \bibinfo{author}{Kepp, K.~P.}
\newblock \bibinfo{title}{A base measure of precision for protein stability
  predictors: structural sensitivity}.
\newblock \emph{\bibinfo{journal}{BMC Bioinformatics}}
  \textbf{\bibinfo{volume}{22}}, \bibinfo{pages}{88} (\bibinfo{year}{2021}).
\newblock \urlprefix\url{https://doi.org/10.1186/s12859-021-04030-w}.

\bibitem{camacho_blast_2009}
\bibinfo{author}{Camacho, C.} \emph{et~al.}
\newblock \bibinfo{title}{{BLAST}+: architecture and applications}.
\newblock \emph{\bibinfo{journal}{BMC bioinformatics}}
  \textbf{\bibinfo{volume}{10}}, \bibinfo{pages}{421} (\bibinfo{year}{2009}).

\bibitem{modis_ligand-binding_2003}
\bibinfo{author}{Modis, Y.}, \bibinfo{author}{Ogata, S.},
  \bibinfo{author}{Clements, D.} \& \bibinfo{author}{Harrison, S.~C.}
\newblock \bibinfo{title}{A ligand-binding pocket in the dengue virus envelope
  glycoprotein}.
\newblock \emph{\bibinfo{journal}{Proceedings of the National Academy of
  Sciences}} \textbf{\bibinfo{volume}{100}}, \bibinfo{pages}{6986--6991}
  (\bibinfo{year}{2003}).
\newblock
  \urlprefix\url{https://www.pnas.org/doi/full/10.1073/pnas.0832193100}.
\newblock \bibinfo{note}{Publisher: Proceedings of the National Academy of
  Sciences}.

\bibitem{kudlacek_designed_2021}
\bibinfo{author}{Kudlacek, S.~T.} \emph{et~al.}
\newblock \bibinfo{title}{Designed, highly expressing, thermostable dengue
  virus 2 envelope protein dimers elicit quaternary epitope antibodies}.
\newblock \emph{\bibinfo{journal}{Science Advances}}
  \textbf{\bibinfo{volume}{7}}, \bibinfo{pages}{eabg4084}
  (\bibinfo{year}{2021}).
\newblock \urlprefix\url{https://www.science.org/doi/10.1126/sciadv.abg4084}.
\newblock \bibinfo{note}{Publisher: American Association for the Advancement of
  Science}.

\end{thebibliography}

\clearpage{}
\newpage{}

\appendix

\setcounter{secnumdepth}{2} 

\counterwithin{table}{section}
\setcounter{table}{0}
\renewcommand{\thetable}{S\arabic{table}}

\counterwithin{figure}{section}
\setcounter{figure}{0}
\renewcommand{\thefigure}{S\arabic{figure}}

\counterwithin{equation}{section}
\setcounter{equation}{0}
\renewcommand{\theequation}{S\arabic{equation}}

\section*{Supplementary Information}

\subsection*{Extended analysis of antigen stabilization}

\begin{figure*}[ht!]
    \centering
    \includegraphics[width=1.0\textwidth]{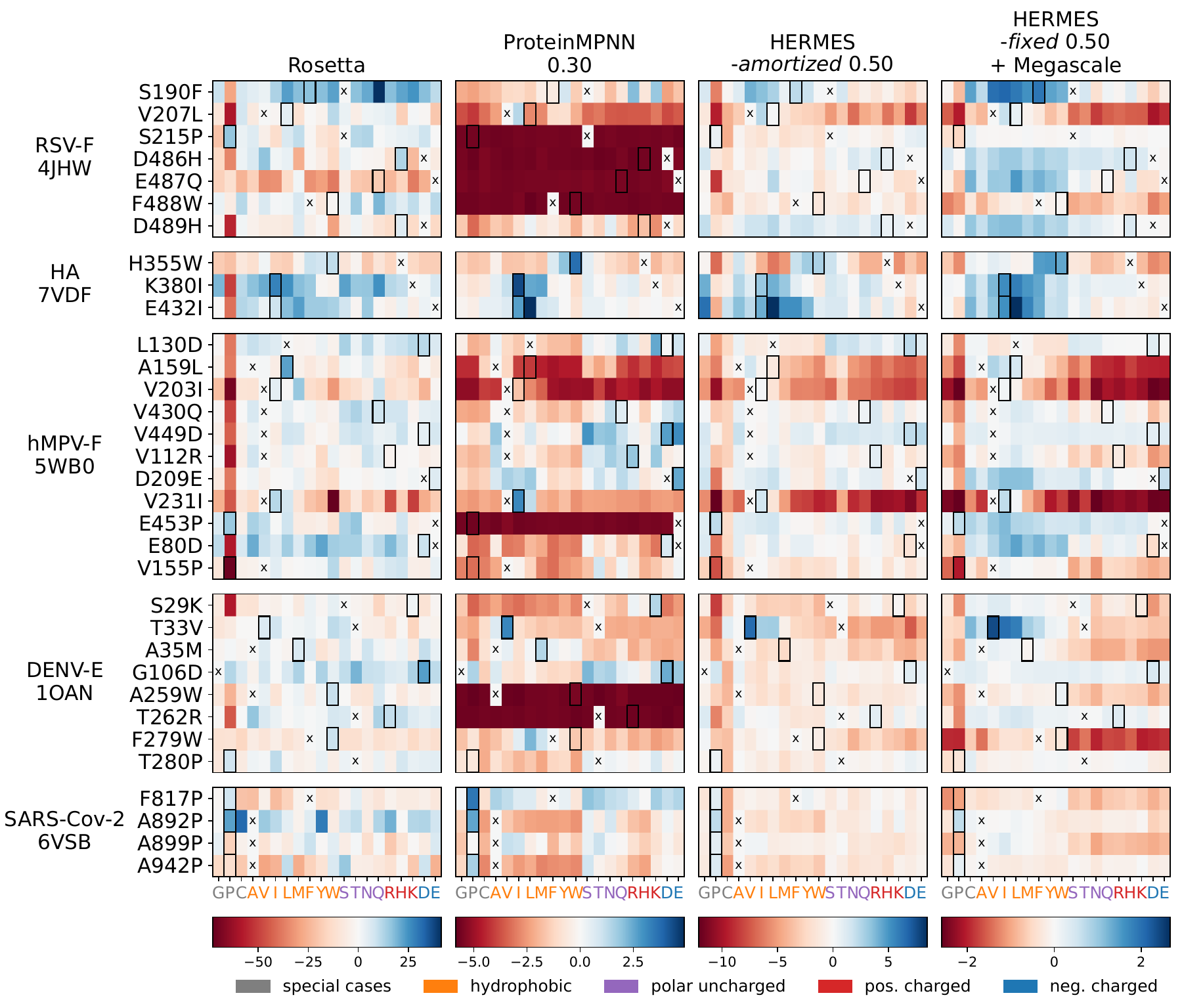}
    \caption{\textbf{Model predictions for amino acid preferences at sites with known antigen-stabilizing mutations.} Predictions from different models (columns) are shown for sites with known antigen-stabilizing mutations specified in Fig.~\ref{fig:antigen_figure_main_results} and Table~\ref{table:antigen_results}. For each antigen (rows), predictions are computed using the protein structure corresponding to the PDB ID indicated on the left. Predictions are reported as changes in Rosetta Energy Units (REU) for Rosetta, and $\Delta \log p$ for ProteinMPNN and HERMES models.
    Wild-type amino acids are marked with centered crosses, while stabilizing mutant amino acids are indicated by dark borders. Amino acids are grouped by broad biochemical class (see legend at the bottom) and, within each class, ordered by increasing size (number of atoms).
    }
   \label{fig:antigen_heatmap} 
\end{figure*}

Here, we present a more detailed analysis of our antigen stabilization predictions from Figures~\ref{fig:antigen_figure_main_results},~\ref{fig:antigen_figure_of_different_types_of_mutations} and Table~\ref{table:antigen_results}. For each antigen, we contextualize the candidate mutations by their local structural environment and examine the characteristics and physico-chemical features of the HERMES-predicted top-ranked antigen-stabilizing  substitutions (Fig.~\ref{fig:antigen_heatmap}). This analysis is intended to help practitioners incorporate these models into design workflows and to motivate quantitative, context-dependent success criteria for real-world stabilization tasks. As demonstrated in the main text (Figs.~\ref{fig:antigen_figure_main_results},~\ref{fig:antigen_figure_of_different_types_of_mutations}), we find that comparing the rank of the wild-type residue to that of putative stabilizing substitutions is an informative diagnostic of model behavior and performance.\\\\
\noindent\textbf{RSV-F.} 
RSV-F Cav1 stabilitizing mutations S190F and V207L are well-predicted by different HERMES models (the stability-fine-tuned models trained on +cDNA117k and +Megascale, and HERMES-\textit{amortized}), whereas Rosetta and ProteinMPNN struggle to identify these mutations. We therefore focus on these two residues to analyze, post hoc, the structural context underlying HERMES' predictions, with the goal of clarifying which classes of stabilizing mutations HERMES is best suited to characterize.

\begin{figure*}
    \centering
    \includegraphics[width=1.0\textwidth]{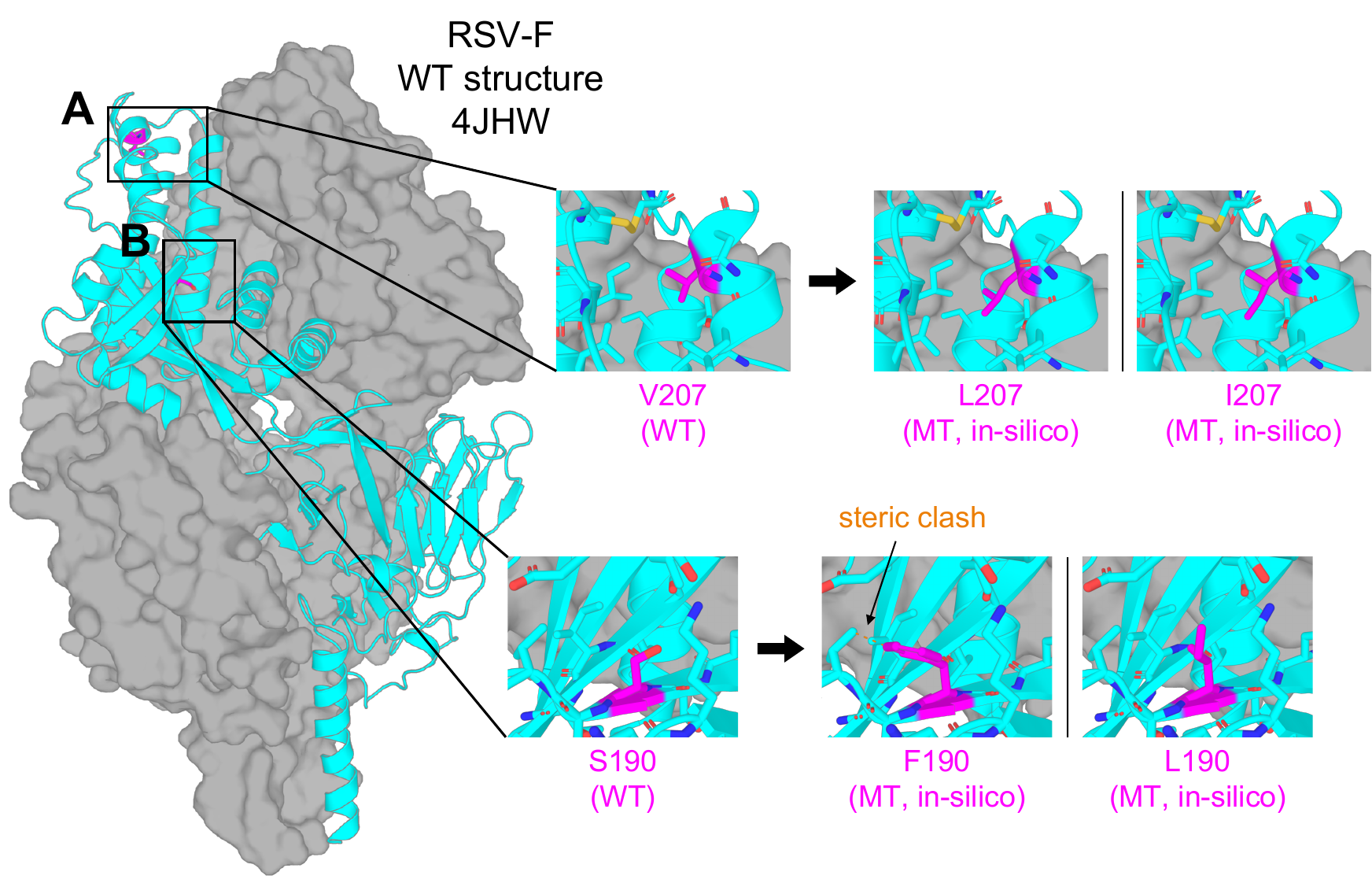}
    \caption{\textbf{Structure of the wild-type pre-fusion RSV-F antigen highlighting observed and candidate antigen-stabilizing mutations.}
The wild-type structure is shown on the left (PDB ID: 4JHW). A single protomer of the trimer is displayed as a cyan cartoon, with the remaining protomers shown as a grey surface. WT denotes wild type and MT denotes mutant. Mutations labeled as ``in silico" were introduced using PyMOL's Mutagenesis Wizard starting from the wild-type structure. Steric clashes (orange dashed lines) were identified using PyMOL's ``find clashes" command, and polar contacts (yellow dashed lines) were identified using the corresponding PyMOL command.
    \textbf{(A)} The L207 mutant has been experimentally shown to stabilize the pre-fusion conformation~\citep{mclellan_structure-based_2013} and appears to enhance intraprotomer packing. We speculate that the I207 mutant, which HERMES-\textit{amortized} predicts to have a comparable ranking to L207 in Fig.~\ref{fig:antigen_heatmap}, would pack similarly and may therefore represent an additional stabilizing mutation worth screening.
\textbf{(B)} Mutant F190 is observed to be stabilizing~\citep{mclellan_structure-based_2013}, though it appears to slightly over-pack the region. We speculate that L190, which HERMES-\textit{amortized} predicts to have a comparable ranking to F190 in Fig.~\ref{fig:antigen_heatmap}, would provide a similarly stabilizing effect without over-packing the region.
    }
    \label{fig:rsvf_si_figure}
\end{figure*}

Both S190F and V207L enhance hydrophobic packing in an underpacked region near the trimer apex (Fig.~\ref{fig:rsvf_si_figure}). V207L is a significantly more anticipated mutation than S190F, indicated by a positive BLOSUM62 score; this is reflected in HERMES assigning a higher rank to the wild-type Val207 relative to the mutant Leu (Fig.~\ref{fig:antigen_heatmap} and Table~\ref{table:antigen_results}). Notably, the highest-ranking residue predicted by HERMES models at position 207 is Ile (Fig.~\ref{fig:antigen_heatmap}). Inspection  of the native wild-type structure (PDB ID: 4JHW~\citep{mclellan_structure-based_2013}) suggests that an Ile mutation would pack exceptionally well in the hydrophobic pocket surrounding residue 207, suggesting V207I might outperform the identified V207L mutation (Fig.~\ref{fig:rsvf_si_figure}A).

At position 190, the top-ranked substitutions from the wild-type Ser are Val or Ile. Structural examination suggests both residues can be accommodated within the existing 4JHW structure pocket without requiring the repacking of surrounding residues (Fig.~\ref{fig:rsvf_si_figure}B). Conversely, the known stabilizing mutation, Phe, appears to slightly overpack the region in the 4JHW crystal structure (Fig.~\ref{fig:rsvf_si_figure}B). While the HERMES-\textit{fixed} model does not strongly prioritize  Phe, both HERMES stability-fine-tuned models rank Phe third, behind the smaller Val and Ile.
 This indicates that fine-tuning on $\Delta\Delta G$ datasets enhances implicit reasoning regarding local repacking and structural relation possibilities, particularly for larger residues. The HERMES-\textit{amortized} model also outperforms the HERMES-\textit{fixed} model, likely due to the model's de-emphasis on strict steric constraints (Table~\ref{table:antigen_results}).\\
\\
\textbf{Universal-HA.} 
We observe robust performance in recovering Universal-HA pH switch mutations across all tested models. All three mutations carry ``unanticipated" BLOSUM62 scores.

\begin{figure*}
    \centering
    \includegraphics[width=1.0\textwidth]{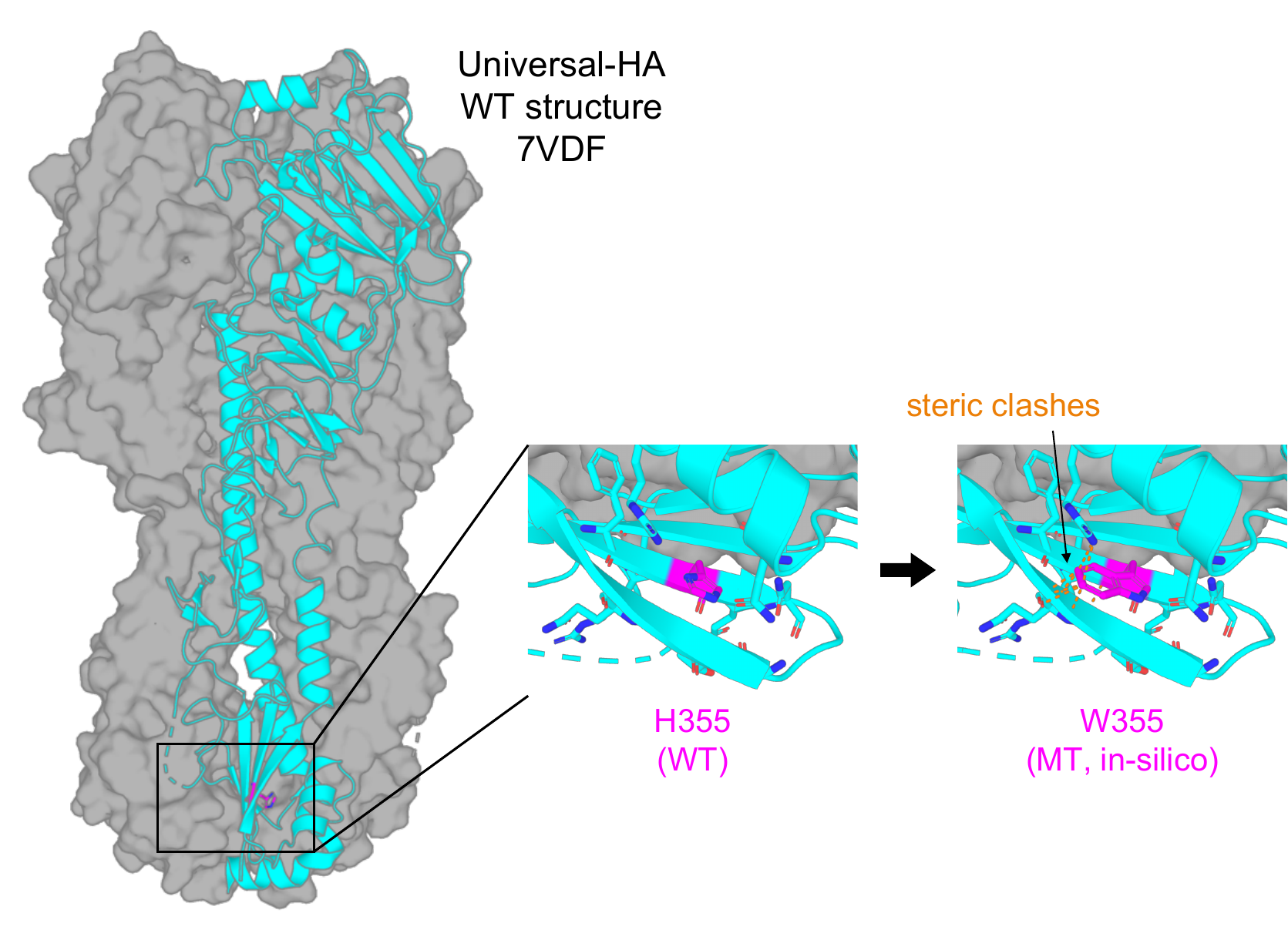}
    \caption{\textbf{Structure of the wild-type pre-fusion Universal-HA antigen, with highlighted observed and candidate antigen-stabilizing mutations.}
    The wild-type structure is shown on the left (PDB ID 7VDF).
    A single protomer of the trimer is shown as cyan cartoon; the other copies are shown as grey surface.
    WT denotes wild type and MT denotes mutant. Mutations labeled as ``in silico" were introduced using PyMOL's Mutagenesis Wizard starting from the wild-type structure. Steric clashes (orange dashed lines) were identified using PyMOL's ``find clashes" command, and polar contacts (yellow dashed lines) were identified using the corresponding PyMOL command. 
    We highlight the H355W mutation which, despite stabilizing the pre-fusion conformation, exhibits substantial steric clashes in the wild-type structural context.
    }
    \label{fig:ha_si_figure}
\end{figure*}

\begin{figure*}
    \centering
    \includegraphics[width=1.0\textwidth]{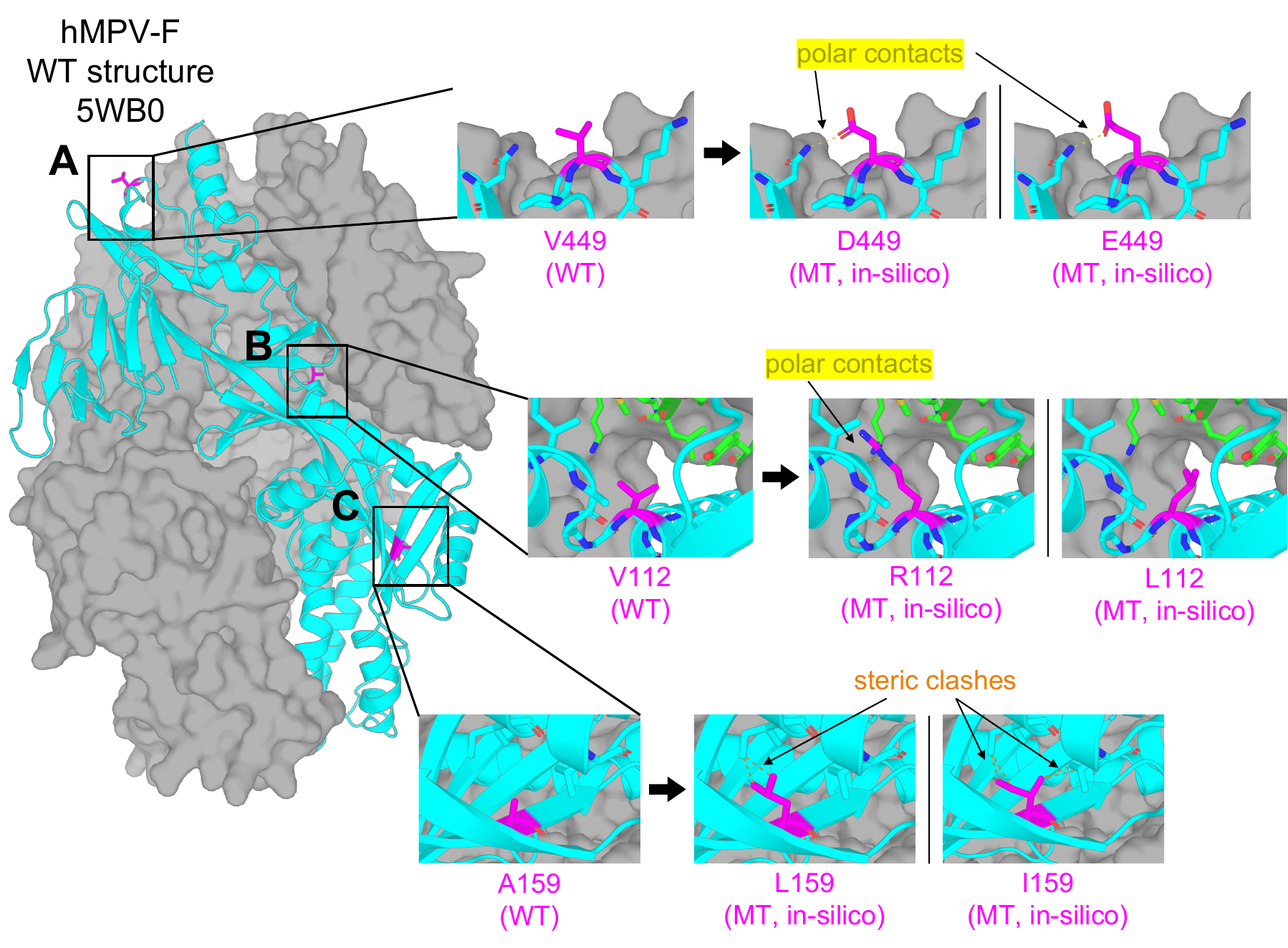}
    \caption{\textbf{Structure of the wild-type pre-fusion hMPV-F antigen, with highlighted observed and candidate antigen-stabilizing mutations.}
    The wild-type structure is shown on the left (PDB ID 5WB0).
 A single protomer of the trimer is shown as cyan cartoon; the other copies are shown as grey surface.
    WT denotes wild type and MT denotes mutant. Mutations labeled as ``in silico" were introduced using PyMOL's Mutagenesis Wizard starting from the wild-type structure. Steric clashes (orange dashed lines) were identified using PyMOL's ``find clashes" command, and polar contacts (yellow dashed lines) were identified using the corresponding PyMOL command. 
    \textbf{(A)} The proposed V449D mutation is highlighted, which likely stabilizes the complex by removing a surface-exposed hydrophobic residue and introducing an intraprotomer polar contact. We speculate that substitution with glutamic acid, which HERMES-\textit{amortized} predicts to have a comparable ranking to D449 in Fig.~\ref{fig:antigen_heatmap}, would produce a similar stabilizing effect.
     \textbf{(B)} Introduction of an Arginine in place of a Valine at position 112 likely introduces an intraprotomer polar contact, as well as packing against the adjacent protomer (shown in green); we believe L112, which HERMES-\textit{amortized} predicts to have a comparable ranking to R112 in Fig.~\ref{fig:antigen_heatmap}, would also pack against the adjacent protomer better than Valine.
    \textbf{(C)} A159L stabilizes the complex likely due to a cavity-filling effect, albeit necessitating nearby I137 to adopt a different rotamer; we believe I159, which HERMES-\textit{amortized} predicts to have a comparable ranking to L159 in Fig.~\ref{fig:antigen_heatmap}, would also fulfill a similar role on the condition of L141 also adopting a different rotamer.
    }
    \label{fig:hmpvf_si_figure}
\end{figure*}

The mutational-effect heatmaps (Fig.~\ref{fig:antigen_heatmap}) show that, for HA, ProteinMPNN's preferences are narrowly concentrated, strongly favoring the reported stabilizing substitutions, whereas HERMES-\textit{amortized}, HERMES-\textit{fixed} (0.50 + Megascale), and Rosetta yield substantially broader preference profiles.  For H355W, all HERMES models correctly indicate that the wild-type His is disfavored, but they rank other bulky aromatics (Tyr or Phe) above the stabilizing Trp. 
 Despite being a known stabilizing mutation, Trp does not appear to fit in the 7VDF crystal structure without clashes (Fig.~\ref{fig:ha_si_figure}). This discrepancy may stem from the limitations of rigid-body mutagenesis (e.g., PyMOL's wizard), as subtle backbone movements are likely required to accommodate the mutation. This finding cautions against  over-interpreting apparent clashes in a single experimentally determined static crystal  structure, especially for bulky substitutions, and motivates follow-up evaluation with relaxation-aware structure prediction models or physics-based repacking/scoring tools to capture local flexibility and repacking that may render seemingly sterically forbidden mutations feasible.\\\\
\noindent \textbf{hMPV-F.} 
We first analyze stabilizing mutations in the M-104 hMPV variant~\citep{gonzalez_general_2024}.
The A159L substitution requires the nearby Ile-137 to adopt an alternative rotamer~\citep{gonzalez_general_2024}. The two $\Delta\Delta G$-fine-tuned HERMES models prioritize Val (V) and Ile (I) over Leu (L) at position 159. We note that mutation to either Val or Ile at position 159 would require both Ile-137 and Leu-141 to adopt different rotamers to accommodate their beta-branched side chains (Fig.~\ref{fig:hmpvf_si_figure}C). Nonetheless,  A159V and A159I remain  plausible stabilizing mutations and therefore, warrant experimental screening.
ProteinMPNN identifies Ala as the most favorable residue at this position, indicating limited sensitivity to larger hydrophobic substitutions. By comparison, the $\Delta\Delta G$-fine-tuned HERMES models rank the wild-type Ala substantially lower (ordinal rank 5--7), while consistently favoring larger hydrophobic residues. Low rankings for buried hydrophobic positions may therefore indicate suboptimal core packing, particularly when alternative residues with greater side-chain volume are predicted. Although Leu is not the top-ranked substitution, the $\Delta\Delta G$-fine-tuned HERMES models correctly capture the preference for a residue larger than Ala to improve packing within this pocket.

The V203I substitution was recovered by nearly all models.  A Val to Ile substitution is highly anticipated, with a BLOSUM62 score of 3.  Notably, the $\Delta\Delta G$-fine-tuned HERMES models rank Ile as more favorable than the wild-type Val, whereas ProteinMPNN, ThermoMPNN, HERMES-\textit{fixed}, and HERMES-\textit{amortized} prefer the wild-type Val. In accordance with this mutation being more highly anticipated, it is likely that the pocket occupied by V203 is less underpacked relative to A159L in the M-104 variant, with the wild-type residue still being quite highly-favored. The subtle nature of Val to Ile mutations may result in less signal overall when comparing amino acid ranks as we do in this work.

The V449D substitution replaces a surface-exposed hydrophobic residue with a polar side chain and introduces hydrogen-bonding contacts to Asn298 (Fig.~\ref{fig:hmpvf_si_figure}A). ProteinMPNN, ThermoMPNN, and all HERMES variants instead rank Glu as the top substitution; structural inspection suggests that Glu could form similar hydrogen bonds, and may therefore also be stabilizing (Fig.~\ref{fig:hmpvf_si_figure}A). Overall, the models performed well in identifying mutations that enable additional hydrogen bonding.

For V430Q, the predicted $\Delta\Delta G$ reported in the original study is small in magnitude (albeit negative), suggesting that it should not be classified as strongly stabilizing, as suggested in the original study~\citep{gonzalez_general_2024}. Consistent with this, neither of the $\Delta\Delta G$–fine-tuned HERMES models prioritize V430Q. Interestingly, the only model that ranks V430Q as the top substitution is HERMES-\textit{fixed}.

In the MPV-2c variant~\citep{bakkers_efficacious_2024}, recovery of V112R depends on providing the model with the trimeric PDB. The introduced Arg forms interprotomer van der Waals contacts and intraprotomer hydrogen bonds that stabilize the prefusion trimer (Fig.~\ref{fig:hmpvf_si_figure}B). Because this mechanism is inherently multimeric and not captured by a strictly local energetic proxy, it may explain why the $\Delta\Delta G$-fine-tuned HERMES models do not prioritize V112R. In contrast, HERMES-\textit{amortized} ranks Arg second, slightly favoring the more conservative Leu, which could plausibly adjust hydrophobic interprotomer contacts.

D209E is correctly predicted by ProteinMPNN, HERMES-\textit{fixed}, and HERMES-\textit{amortized}, whereas the $\Delta\Delta G$-fine-tuned HERMES models perform poorly in proposing this mutation. In particular, the $\Delta\Delta G$-fine-tuned models rank bulky hydrophobics (Leu, Met, Ile, Val) ahead of the charge-conserving Glu. Although the charged substitution D209E plausibly stabilizes the prefusion state via favorable polar interactions, the stabilizing potential of these alternative hydrophobic substitutions remains untested.

\begin{figure*}
    \centering
    \includegraphics[width=1.0\textwidth]{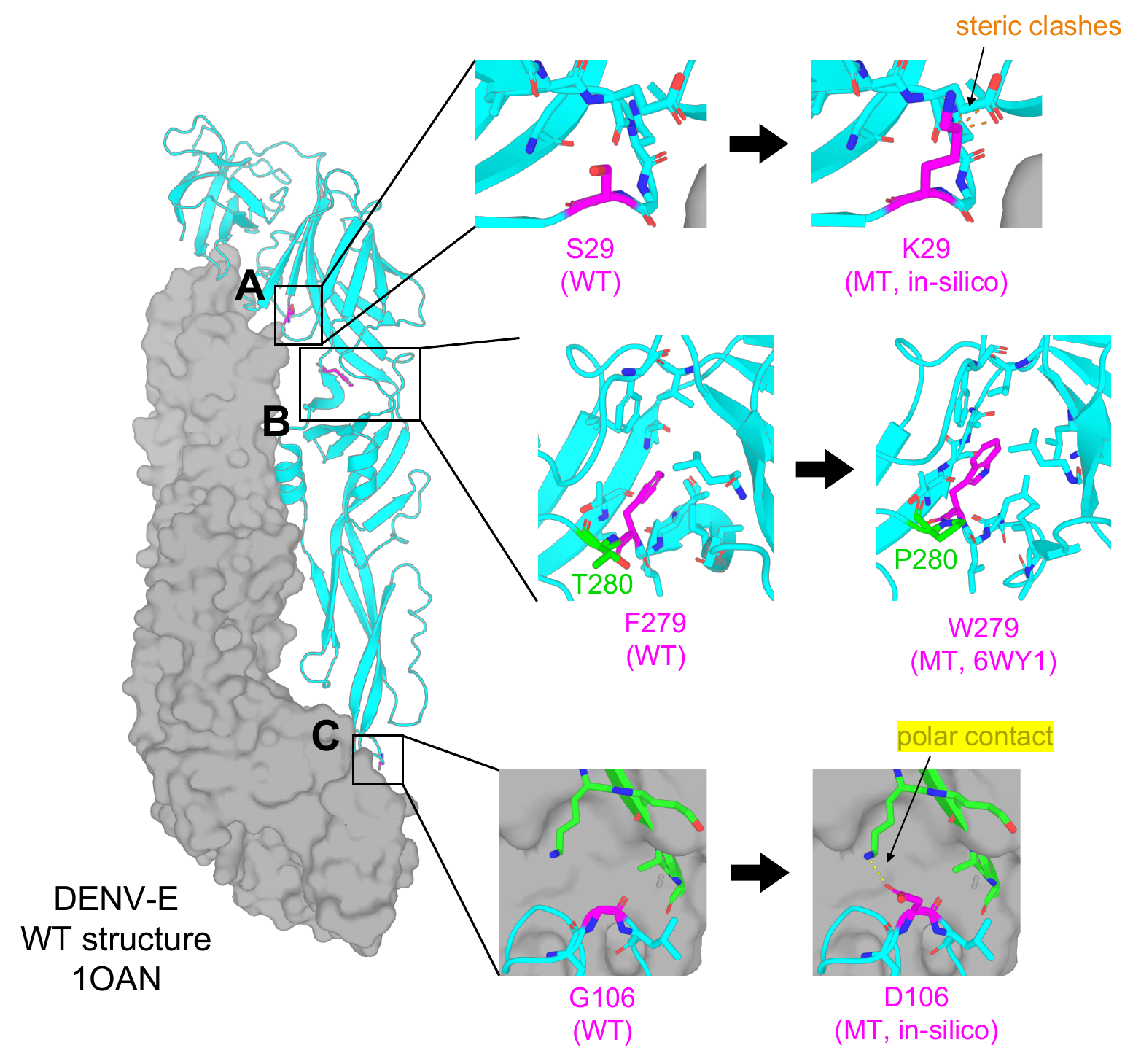}
    \caption{\textbf{Structure of wild-type pre-fusion DENV-E antigen, with highlighted observed and candidate mutations.}
    The wild-type structure is shown on the left (PDB ID 1OAN).
   A single protomer of the trimer is shown as cyan cartoon; the other copies are shown as grey surface.
    WT denotes wild type and MT denotes mutant. Mutations labeled as ``in silico" were introduced using PyMOL's Mutagenesis Wizard starting from the wild-type structure. Steric clashes (orange dashed lines) were identified using PyMOL's ``find clashes" command, and polar contacts (yellow dashed lines) were identified using the corresponding PyMOL command. 
   \textbf{(A)} The S29K mutation is experimentally observed to be stabilizing~\citep{phan_conserved_2022}, despite appearing to introduce substantial steric clashes when inspected in the wild-type structure, suggesting that stabilization requires a shift in backbone conformation.
    \textbf{(B)} The F279W mutation is experimentally observed to be stabilizing~\citep{phan_conserved_2022}, likely by filling an under-packed cavity. A substantial backbone rearrangement is also observed in the mutant structure (residues 269–281), potentially facilitated by the  T280P mutation.
    \textbf{(C)} The G106D mutation is experimentally observed to be stabilizing~\citep{phan_conserved_2022}, likely through the introduction of a polar contact with the adjacent protomer (shown in green) and/or interactions with surrounding water molecules.
    }
    \label{fig:denve_si_figure}
\end{figure*}

E453P replaces a glutamate with proline, a substitution that can be strongly stabilizing by rigidifying locally flexible regions. Among the evaluated methods, only HERMES-\textit{amortized} predicts this mutation. ProteinMPNN ranks the wild-type Glu as most favored and Pro as most disfavored (rank 1 vs. 20), while ThermoMPNN partially recovers the substitution but still favors the native Glu. Consistent with these mixed signals, the original study~\citep{bakkers_efficacious_2024} reports that E453P improves stability but reduces expression, and that E453Q yields weaker stabilization with improved expression. This highlights the poorly understood coupling between stability and expression and suggests that some models may systematically downweigh proline substitutions, potentially because prolines are underrepresented in natural sequences, and they appropriately learned to rarely recommend them.
The fact that HERMES-\textit{amortized}—but not HERMES-\textit{fixed}—recovers E453P suggests that modeling relaxed neighborhoods helps identify sites that can accommodate Pro's constrained Ramachandran geometry. Finally,  we suspect that the utility and stabilizing ability of proline substitutions in antigen design is not fully captured by HERMES when fine-tuned with the cDNA117k or Megascale datasets, as the proteins represented in those datasets are much smaller, and lack the interplay of conformational switching between prefusion and post-fusion.\\\\
\textbf{DENV-E.} 
We observe mixed performance on DENV-E stabilizing mutations. S29K is recovered only by ProteinMPNN and ThermoMPNN. Although S29K, T33V, and A35M (``PM4" mutations as a group) are spatially proximal and may act synergistically, they were originally identified by Rosetta site-saturation mutagenesis, suggesting they should be accessible to single point-mutation scoring schemes~\citep{phan_conserved_2022}. In the 1OAN structure~\citep{modis_ligand-binding_2003}, introducing Lys at position 29 appears to create substantial steric clashes across rotamers (Fig.~\ref{fig:denve_si_figure}A). S29K and A35M are not anticipated by BLOSUM62, whereas T33V is neutral according to BLOSUM62 (score of $0$, Table~\ref{table:antigen_results}) but is strongly preferred by all models tested (Fig.~\ref{fig:antigen_figure_main_results} and Table~\ref{table:antigen_results}). Structurally, T33 lines a hydrophobic pocket, making the isosteric Val substitution easier to predict. Notably, the PM4 mutations are not present in the best-available SC12 structure (PDB: 6WY1)~\citep{kudlacek_designed_2021}.

A major challenge in stabilizing DENV-E is strengthening homodimer interactions. A259W and T262R were mutations made at the dimer interface with the intention of enhancing the strength of the homodimer. A259W is highly unanticipated (BLOSUM62 = $-3$) but, together with T262R, can form a favorable cation--$\pi$ interaction while improving packing against the neighboring protomer (Fig.~\ref{fig:antigen_figure_of_different_types_of_mutations}A.4). Notably, the HERMES model fine-tuned on Megascale favors A259W despite its apparent synergy with T262R; this could reflect sensitivity to the interprotomer packing geometry, although a chance effect cannot be excluded. We also expect the generally weak recovery of T262R across models to improve when Trp is present at position 259, consistent with the coupled nature of this interface motif.

Like E453P in hMPV-F, the proline substitution T280P is recovered only by HERMES-\textit{amortized}, further highlighting the model's ability to reason about proline accommodation. F279W fills an underpacked region. However, comparison of the stabilized 6WY1 structure with the native 1OAN crystal structure suggests that backbone movement (residues 269--281) is required to optimally fit Trp, potentially aided by T280P (Fig.~\ref{fig:denve_si_figure}B). ThermoMPNN correctly reasons regarding the Trp introduction, and HERMES-\textit{fixed} $+$ Megascale ranks Trp second. Other stability-fine-tuned HERMES models prefer Phe, Ile, or Leu, likely adhering more strictly to the steric limitations of the 1OAN backbone.

We observe strong reasoning across almost all HERMES models, ProteinMPNN, and Rosetta in identifying G106D, which introduces stabilizing polar and electrostatic contacts (Fig.~\ref{fig:denve_si_figure}C). Curiously, ThermoMPNN struggles with this specific mutation.\\
\\
\textbf{SARS-CoV-2 Spike.} 
The spike protein of SARS-CoV-2 was initially stabilized through the introduction of two proline mutations (``S-2P"), and subsequent work identified four additional prolines that further improve stability and expression of the full-length spike~\citep{hsieh_structure-based_2020}. Here we evaluate recovery of these four additional proline sites (excluding the original S-2P mutations). Consistent with its strong performance on proline substitutions, HERMES-\textit{amortized} ranks Pro as the top choice at all four positions. ProteinMPNN also performs well, recovering Pro at 3/4 sites. \\\\
\textbf{Summary.} 
Our analysis across multiple antigens reveals several consistent themes regarding model performance and decision-making logic.

First, we observe distinct behaviors regarding hydrophobic packing. The HERMES models fine-tuned on stability datasets consistently demonstrate enhanced reasoning for hydrophobic core packing. These models frequently suggest larger hydrophobic residues (e.g., Ile or Phe) to fill underpacked cavities where models without explicit stability training might prefer the native residue or smaller conservative substitutions. However, this sensitivity can sometimes lead to ``confusion" among similar hydrophobic amino acids (e.g., Val vs. Ile vs. Leu), where the models correctly identify the chemical property needed but may rank several hydrophobic options similarly.

Second, the interplay between structural rigidity and model flexibility is evident in the prediction of proline mutations. HERMES-\textit{amortized} consistently outperforms other models in identifying stabilizing proline substitutions. This suggests that the model's training, which incorporates relaxed neighborhoods, allows it to better identify backbone locations capable of accommodating the steric constraints of proline, whereas other methods more often penalize these stabilizing mutations due to perceived steric incompatibilities of the provided protein structure.

Third, we note differences in the mutational landscape profiles predicted by different architectures (Fig.~\ref{fig:antigen_heatmap}. ProteinMPNN tends to produce narrow substitution profiles, often assigning very low probabilities to non-native residues unless the signal is overwhelmingly strong. In contrast, HERMES models--particularly HERMES-\textit{amortized} and those fine-tuned on Megascale stability data--exhibit broader mutational profiles. This broader landscape may be more advantageous for design applications, as it provides a richer set of plausible candidate hypotheses for experimental validation and may guide a practitioner's intuition regarding the nature or predicted effects of various substitutions in a more granular manner.

Finally, while ProteinMPNN and Rosetta remain powerful tools for sequence recovery, the HERMES-\textit{amortized} and stability-fine-tuned models demonstrate a unique capacity to prioritize mutations that improve local packing density even when such mutations appear sterically challenging in the provided structure. This highlights the utility of using an ensemble of models to capture different modes of stabilization, from electrostatic optimization to hydrophobic core repacking and backbone rigidification.

\clearpage{}
\newpage{}

\begin{table}[h!]
\centering
\resizebox{0.9\textwidth}{!}{
\begin{tabular}{l|c|c}
\toprule
\multirow{2}{*} {\bf Model} &{\bf Accuracy}& {\bf Accuracy}\\
& Pyrosetta  pre-processing& {Biopython pre-processing} \\
\midrule
\HERMESPy & 0.73 & 0.75 \\
\HERMESPyNoise & 0.64 & 0.65 \\
\HERMESPyFtRelaxed & 0.55 & 0.47 \\
\HERMESPyNoiseFtRelaxed & 0.50 & 0.44 \\
\midrule
\HERMESPyFtRos & 0.41 & 0.40 \\
\HERMESPyNoiseFtRos & 0.38 & 0.37 \\
\HERMESPyFtCdna & 0.47 & 0.45 \\
\HERMESPyNoiseFtCdna & 0.39 & 0.38 \\
\HERMESPyFtRelaxedFtCdna & 0.37 & - \\
\HERMESPyNoiseFtRelaxedFtCdna & 0.34 & - \\
\HERMESPyFtCdnaESM & 0.46 & 0.49 \\
\HERMESPyNoiseFtCdnaESM & 0.40 & 0.40 \\
\midrule
\HERMESPyUntrainedFtCdna & 0.09 & - \\
\HERMESPyNoiseUntrainedFtCdna & 0.08 & - \\
\bottomrule
\end{tabular}}
\caption{\textbf{Accuracy of HERMES models on wildtype amino-acid classification on all sites across 40 CASP12 test proteins.}
Accuracy is defined as proportion of sites for which the wild-type amino acid is predicted with the highest probability among all 20 canonical amino acids.
Model names indicate the architecture, the coordinate-noise amplitude used, and when applicable, the fine-tuning dataset (listed after ``$+$"); \textit{Untr.} is short for \textit{Untrained}, indicating models that had no pre-training and were instead only trained on stability effects. Accuracy is reported for the two pre-processing schemes (with PyRosetta and Biopython) used in HERMES.
}
\label{table:aa_wt_cls}
\end{table}

\begin{table}[h!]
\centering
\resizebox{0.4\textwidth}{!}{%
\begin{tabular}{l | c}
\toprule
\textbf{Model} & \textbf{hh:mm:ss} \\
\midrule
\HERMESPyNoise & 00:11:23 \\
\HERMESPyNoiseFtRelaxed & 00:11:23 \\
\HERMESPyNoiseRelaxed & 12:13:20 \\
\bottomrule
\end{tabular}}
\caption{\textbf{Inference speed of HERMES models on the T2837 dataset.} Runtimes are reported in hours (hh), minutes (mm), and seconds (ss) for inference on the T2837 dataset, which comprises 2837 mutation effects across 129 proteins. For HERMES-{\em fixed} and HERMES-{\em amortized}, the script `mutation\_effect\_prediction\_with\_hermes.py` was used; for HERMES-{\em relaxed}, the script `mutation\_effect\_prediction\_with\_hermes\_with\_relaxation.py` was used. Both scripts, along with the dataset `csv` file, are available in our GitHub repository. All models were executed using a single CPU and a single A40 GPU.}
\label{table:inference_speed}
\end{table}

\begin{table}[h!]
\centering
\footnotesize
\setlength{\tabcolsep}{6pt}
\begin{tabularx}{\textwidth}{>{\raggedright\arraybackslash}p{0.27\linewidth} >{\raggedright\arraybackslash}X}
\toprule
\textbf{Category} & \textbf{Description} \\
\midrule

\multirow{18}{*}{Hydrophobic Property}
& 1.\ Retention coefficient in TFA \\
& 2.\ Free energy of solution in water \\
& 3.\ Solvation free energy \\
& 4.\ Melting point \\
& 5.\ Number of hydrogen-bond donors \\
& 6.\ Number of full nonbonding orbitals \\
& 7.\ Partition energy \\
& 8.\ Hydration number \\
& 9.\ Retention coefficient in high performance liquid chromatography (HPLC), pH 7.4 \\
& 10.\ Retention coefficient in HPLC, pH 2.1 \\
& 11.\ Partition coefficient in thin-layer chromatography \\
& 12.\ Retention coefficient at pH 2 \\
& 13.\ $R_f$ for 1-N-(4-nitrobenzofurazono)-amino acids in ethyl acetate/pyridine/water \\
& 14.\ $\Delta G$ of transfer from organic solvent to water \\
& 15.\ Hydration potential or free energy of transfer from vapor phase to water \\
& 16.\ $R_f$, salt chromatography \\
& 17.\ $\log D$, partition coefficient at pH 7.1 for acetamide derivatives of amino acids in octanol/water \\
& 18.\ $\Delta G = RT \log f$, $f$ = fraction buried/accessible amino acids in 22 proteins \\
\midrule

\multirow{17}{*}{Steric Property}
& 19.\ Average volume of buried residue \\
& 20.\ Residue accessible surface area in tripeptide \\
& 21.\ Graph shape index \\
& 22.\ Normalized van der Waals volume \\
& 23.\ STERMIMOL length of the side chain \\
& 24.\ STERMIMOL minimum width of the side chain \\
& 25.\ STERMIMOL maximum width of the side chain \\
& 26.\ Average accessible surface area \\
& 27.\ Distance between $C_{\alpha}$ and centroid of side chain \\
& 28.\ Side-chain angle $\theta$ \\
& 29.\ Side-chain torsion angle $\phi$ \\
& 30.\ Radius of gyration of side chain \\
& 31.\ Van der Waals parameter $R_0$ \\
& 32.\ Van der Waals parameter $\varepsilon$ \\
& 33.\ Refractivity \\
& 34.\ Value of $\theta$ (i) \\
& 35.\ Substituent van der Waals volume \\
\midrule

\multirow{15}{*}{Electronic Property}
& 36.\ $\alpha$CH chemical shifts \\
& 37.\ $\alpha$NH chemical shifts \\
& 38.\ A parameter of charge transfer capability \\
& 39.\ A parameter of charge transfer donor capability \\
& 40.\ Nuclear magnetic resonance (NMR) chemical shift of $\alpha$ carbon \\
& 41.\ Localized electrical effect \\
& 42.\ Positive charge \\
& 43.\ Negative charge \\
& 44.\ Polarity \\
& 45.\ Net charge \\
& 46.\ Amphipathicity index \\
& 47.\ Isoelectric point \\
& 48.\ Electron-ion interaction potential values \\
& 49.\ $\mathrm{pK}_{\mathrm{NH}_2}$ (NH$_2$ on $C_{\alpha}$) \\
& 50.\ $\mathrm{pK}_{\mathrm{COOH}}$ (COOH on $C_{\alpha}$) \\
\bottomrule
\end{tabularx}
\caption{\textbf{Table of amino-acid properties used for comparison with substitution matrices.} Values associated with each amino acid are listed in ref.~\cite{mei_new_2005}.}
\label{table:aa_properties}
\end{table}

\begin{table}[h!]
    \centering
    \resizebox{\textwidth}{!}{
        \begin{tabular}{c c c c | c c c c c c c}
            \toprule
            \multirow{2}{*}{\makecell{antigen\\PDB id}} &
            \multirow{2}{*}{\makecell{stabilized\\name}} &
            \multirow{2}{*}{mutation} &
            \multirow{2}{*}{\makecell{BLOSUM62\\score ; rank}} &
                \makecell{Rosetta} & \makecell{Protein\\MPNN\\0.30} & \makecell{Thermo\\MPNN\\+ Megascale} & \makecell{HERMES\\-{\em fixed} 0.50} & \makecell{HERMES\\-{\em amortized}\\0.50} & \makecell{HERMES\\-{\em fixed} 0.50\\+ cDNA117k} & \makecell{HERMES\\-{\em fixed} 0.50\\+ Megascale} \\
                \cmidrule(lr){5-11}
                & & & & $r_\wt \rightarrow r_\mt$ & $r_\wt \rightarrow r_\mt$ & $r_\wt \rightarrow r_\mt$ & $r_\wt \rightarrow r_\mt$ & $r_\wt \rightarrow r_\mt$ & $r_\wt \rightarrow r_\mt$ & $r_\wt \rightarrow r_\mt$ \\
                \midrule
            \multirow{7}{*}{\makecell{RSV-F\\4JHW}} & \multirow{2}{*}{Cav1 \citep{mclellan_structure-based_2013}} & S190F & -2 ; 16 & \yellowl 19 $\rightarrow$ 11 & 4 $\rightarrow$ 7 & \green 14 $\rightarrow$ 3 & 4 $\rightarrow$ 8 & \greenl 11 $\rightarrow$ 4 & \green 12 $\rightarrow$ 3 & \green 13 $\rightarrow$ 3 \\
             &  & V207L & 1 ; 3 & \greenl 9 $\rightarrow$ 4 & 2 $\rightarrow$ 7 & \greenl 6 $\rightarrow$ 5 &  2 $\rightarrow$ 3 & \green 3 $\rightarrow$ 2 & \green 3 $\rightarrow$ 2 & \green 3 $\rightarrow$ 2 \\
            \cmidrule(lr){2-11}
             & \multirow{1}{*}{Uncl. \citep{lee_rational_2024}} & S215P & -1 ; 13 & \green 13 $\rightarrow$ 1 & 1 $\rightarrow$ 14 & 10 $\rightarrow$ 20 & 10 $\rightarrow$ 15 & \green 6 $\rightarrow$ 3 & 16 $\rightarrow$ 19 & 4 $\rightarrow$ 20 \\
            \cmidrule(lr){2-11}
             & \multirow{4}{*}{TriC \citep{mclellan_structure-based_2013}} & D486H & -1 ; 7 & \green 9 $\rightarrow$ 2 & 1 $\rightarrow$ 10 & 1 $\rightarrow$ 8 & \greenl 5 $\rightarrow$ 4 & \greenl 10 $\rightarrow$ 4 & \yellowl 20 $\rightarrow$ 9 & \yellowl 18 $\rightarrow$ 9 \\
             &  & E487Q & 2 ; 3 & 2 $\rightarrow$ 9 & 1 $\rightarrow$ 13 & 1 $\rightarrow$ 12 &  2 $\rightarrow$ 3 & 4 $\rightarrow$ 5 & 11 $\rightarrow$ 12 & \yellowl 14 $\rightarrow$ 12 \\
             &  & F488W & 1 ; 3 & 15 $\rightarrow$ 17 & 1 $\rightarrow$ 14 & \green 3 $\rightarrow$ 1 & 2 $\rightarrow$ 6 & 1 $\rightarrow$ 13 &  1 $\rightarrow$ 3 &  1 $\rightarrow$ 2 \\
             &  & D489H & -1 ; 7 & \green 5 $\rightarrow$ 2 & 3 $\rightarrow$ 14 & 5 $\rightarrow$ 12 & \yellowl 11 $\rightarrow$ 8 & \yellowl 18 $\rightarrow$ 15 & \yellowl 20 $\rightarrow$ 11 & \yellowl 17 $\rightarrow$ 12 \\
            \midrule
            \multirow{3}{*}{\makecell{HA\\7VDF}} & \multirow{3}{*}{Universal-HA \citep{milder_universal_2022}} & H355W & -2 ; 12 & \green 5 $\rightarrow$ 1 & \green 5 $\rightarrow$ 1 & \green 4 $\rightarrow$ 2 & \green 4 $\rightarrow$ 3 & \green 7 $\rightarrow$ 2 & \green 5 $\rightarrow$ 3 & \green 8 $\rightarrow$ 3 \\
             &  & K380I & -3 ; 19 & \green 17 $\rightarrow$ 1 & \green 9 $\rightarrow$ 1 & \green 11 $\rightarrow$ 1 & \greenl 8 $\rightarrow$ 4 & \green 14 $\rightarrow$ 3 & \green 13 $\rightarrow$ 2 & \green 17 $\rightarrow$ 2 \\
             &  & E432I & -3 ; 19 & \yellowl 14 $\rightarrow$ 9 & \green 12 $\rightarrow$ 2 & \green 13 $\rightarrow$ 2 & \greenl 9 $\rightarrow$ 4 & \greenl 15 $\rightarrow$ 5 & \green 12 $\rightarrow$ 3 & \green 14 $\rightarrow$ 3 \\
            \midrule
            \multirow{11}{*}{\makecell{hMPV-F\\5WB0}} & \multirow{5}{*}{M104 \citep{gonzalez_general_2024}} & L130D & -4 ; 19 & \green 15 $\rightarrow$ 3$^\dagger$ & 8 $\rightarrow$ 9 & 4 $\rightarrow$ 10 & 5 $\rightarrow$ 9 & \green 12 $\rightarrow$ 2 & \green 8 $\rightarrow$ 2 & \greenl 10 $\rightarrow$ 4 \\
             &  & A159L & -1 ; 10 & \green 6 $\rightarrow$ 1$^\dagger$ & 1 $\rightarrow$ 12 & 3 $\rightarrow$ 5 & 1 $\rightarrow$ 7 & 2 $\rightarrow$ 4 & \green 7 $\rightarrow$ 3 & \green 5 $\rightarrow$ 3 \\
             &  & V203I & 3 ; 2 & \green 4 $\rightarrow$ 3$^\dagger$ &  1 $\rightarrow$ 2 &  1 $\rightarrow$ 2 &  1 $\rightarrow$ 2 & \green 2 $\rightarrow$ 1 & \green 3 $\rightarrow$ 1 & \green 2 $\rightarrow$ 1 \\
             &  & V430Q & -2 ; 10 & \green 12 $\rightarrow$ 3$^\dagger$ & \greenl 5 $\rightarrow$ 4 & \greenl 12 $\rightarrow$ 6 & \green 8 $\rightarrow$ 1 & \greenl 10 $\rightarrow$ 5 & 5 $\rightarrow$ 12 & \yellowl 10 $\rightarrow$ 8 \\
             &  & V449D & -3 ; 16 & \yellowl 13 $\rightarrow$ 8$^\dagger$ & \green 13 $\rightarrow$ 3 & \green 16 $\rightarrow$ 2 & \green 11 $\rightarrow$ 2 & \green 15 $\rightarrow$ 2 & \green 18 $\rightarrow$ 2 & \greenl 19 $\rightarrow$ 6 \\
            \cmidrule(lr){2-11}
             & \multirow{4}{*}{MPV-2cREKR \citep{bakkers_efficacious_2024}} & V112R & -3 ; 19 & 6 $\rightarrow$ 13$^\dagger$ & \green 8 $\rightarrow$ 1 & \green 2 $\rightarrow$ 1 & 2 $\rightarrow$ 9 & \green 8 $\rightarrow$ 3 & 4 $\rightarrow$ 11 & 5 $\rightarrow$ 10 \\
             &  & D209E & 2 ; 2 & \green 8 $\rightarrow$ 1$^\dagger$ & \green 11 $\rightarrow$ 1 & \greenl 12 $\rightarrow$ 4 & \green 3 $\rightarrow$ 1 & \green 14 $\rightarrow$ 3 & \greenl 15 $\rightarrow$ 6 & \yellowl 16 $\rightarrow$ 7 \\
             &  & V231I & 3 ; 2 & \green 3 $\rightarrow$ 1$^\dagger$ & \green 3 $\rightarrow$ 1 & \green 3 $\rightarrow$ 1 & \green 2 $\rightarrow$ 1 & \green 2 $\rightarrow$ 1 & \green 3 $\rightarrow$ 1 & \green 3 $\rightarrow$ 1 \\
             &  & E453P & -1 ; 10 & \green 10 $\rightarrow$ 2$^\dagger$ & 1 $\rightarrow$ 20 &  1 $\rightarrow$ 2 & 8 $\rightarrow$ 19 & \green 13 $\rightarrow$ 1 & \yellowl 17 $\rightarrow$ 16 & \greenl 19 $\rightarrow$ 6 \\
            \cmidrule(lr){2-11}
             & \multirow{2}{*}{Uncl. \citep{lee_rational_2024}} & E80D & 2 ; 2 & \yellowl 18 $\rightarrow$ 13 & \green 3 $\rightarrow$ 1 & 13 $\rightarrow$ 18 & 1 $\rightarrow$ 12 & 7 $\rightarrow$ 17 & 15 $\rightarrow$ 19 & 14 $\rightarrow$ 18 \\
             &  & V155P & -2 ; 13 & 10 $\rightarrow$ 20 & 4 $\rightarrow$ 20 & 1 $\rightarrow$ 20 & 1 $\rightarrow$ 19 & 1 $\rightarrow$ 20 & 2 $\rightarrow$ 20 & 4 $\rightarrow$ 20 \\
            \midrule
            \multirow{8}{*}{\makecell{DENV-E\\1OAN}} & \multirow{8}{*}{SC12 \citep{phan_conserved_2022}} & S29K & 0 ; 6 & \green 4 $\rightarrow$ 3$^\dagger$ & \green 3 $\rightarrow$ 1 & \green 7 $\rightarrow$ 2 & 1 $\rightarrow$ 12 & 2 $\rightarrow$ 5 & 6 $\rightarrow$ 12 & 9 $\rightarrow$ 14 \\
             &  & T33V & 0 ; 3 & \greenl 9 $\rightarrow$ 4$^\dagger$ & \green 3 $\rightarrow$ 1 & \green 6 $\rightarrow$ 1 & \green 3 $\rightarrow$ 1 & \green 5 $\rightarrow$ 1 & \green 10 $\rightarrow$ 1 & \green 10 $\rightarrow$ 1 \\
             &  & A35M & -1 ; 13 & \greenl 11 $\rightarrow$ 4$^\dagger$ & \green 6 $\rightarrow$ 1 & \yellowl 10 $\rightarrow$ 8 & 1 $\rightarrow$ 12 & 1 $\rightarrow$ 11 & 1 $\rightarrow$ 7 & 6 $\rightarrow$ 7 \\
             &  & G106D & -1 ; 5 & \green 18 $\rightarrow$ 1$^\dagger$ & \green 11 $\rightarrow$ 1 & 6 $\rightarrow$ 11 & \green 8 $\rightarrow$ 1 & \green 6 $\rightarrow$ 2 & \green 20 $\rightarrow$ 1 & \green 18 $\rightarrow$ 1 \\
             &  & A259W & -3 ; 20 & \green 7 $\rightarrow$ 1$^\dagger$ &  1 $\rightarrow$ 3 & 4 $\rightarrow$ 11 & 1 $\rightarrow$ 20 & 1 $\rightarrow$ 13 &  1 $\rightarrow$ 2 & \green 2 $\rightarrow$ 1 \\
             &  & T262R & -1 ; 11 & \green 15 $\rightarrow$ 3$^\dagger$ & 1 $\rightarrow$ 15 & \yellowl 18 $\rightarrow$ 7 & 3 $\rightarrow$ 14 & \green 10 $\rightarrow$ 3 & \green 16 $\rightarrow$ 3 & \greenl 17 $\rightarrow$ 4 \\
             &  & F279W & 1 ; 3 & \green 3 $\rightarrow$ 1$^\dagger$ & 4 $\rightarrow$ 11 & \green 4 $\rightarrow$ 1 & 1 $\rightarrow$ 6 & 3 $\rightarrow$ 7 & 1 $\rightarrow$ 4 &  1 $\rightarrow$ 2 \\
             &  & T280P & -1 ; 13 & \green 12 $\rightarrow$ 2$^\dagger$ & 7 $\rightarrow$ 10 & 9 $\rightarrow$ 14 & 4 $\rightarrow$ 19 & \green 6 $\rightarrow$ 2 & 2 $\rightarrow$ 20 & 8 $\rightarrow$ 19 \\
            \midrule
            \multirow{4}{*}{\makecell{SARS-Cov-2\\6VSB}} & \multirow{4}{*}{hexapro \citep{hsieh_structure-based_2020}} & F817P & -4 ; 20 & \green 3 $\rightarrow$ 1 & \green 16 $\rightarrow$ 1 & 1 $\rightarrow$ 12 & 2 $\rightarrow$ 16 & \green 3 $\rightarrow$ 1 & 2 $\rightarrow$ 18 & 2 $\rightarrow$ 19 \\
             &  & A892P & -1 ; 14 & \green 16 $\rightarrow$ 3 & \green 5 $\rightarrow$ 1 & 2 $\rightarrow$ 4 & \green 2 $\rightarrow$ 1 & \green 4 $\rightarrow$ 1 & \green 4 $\rightarrow$ 1 & \green 5 $\rightarrow$ 1 \\
             &  & A899P & -1 ; 14 & 8 $\rightarrow$ 19 & 6 $\rightarrow$ 8 & 7 $\rightarrow$ 18 & 2 $\rightarrow$ 6 & \green 4 $\rightarrow$ 1 & 8 $\rightarrow$ 15 & 5 $\rightarrow$ 17 \\
             &  & A942P & -1 ; 14 & 5 $\rightarrow$ 12 & \green 3 $\rightarrow$ 1 & 2 $\rightarrow$ 8 & 1 $\rightarrow$ 4 & \green 2 $\rightarrow$ 1 & \green 2 $\rightarrow$ 1 & \green 2 $\rightarrow$ 1 \\
            \midrule
            & \multicolumn{3}{c|}{\green \makecell{proportion of correctly and\\strongly suggested mutations}} & \textbf{20/33} & \textbf{15/33} & \textbf{11/33} & 8/33 & \textbf{19/33} & \textbf{15/33} & \textbf{13/33} \\
            \cmidrule(lr){2-11}
            & \multicolumn{3}{c|}{\greenl \makecell{proportion of correctly and\\at least moderately suggested mutations}} & \textbf{23/33} & \textbf{16/33} & 14/33 & 11/33 & \textbf{23/33} & \textbf{16/33} & \textbf{17/33} \\
            \cmidrule(lr){2-11}
            & \multicolumn{3}{c|}{\makecell{proportion of correctly and\\at least weakly suggested mutations}} & \textbf{27/33} & 16/33 & 16/33 & 12/33 & \textbf{24/33} & 19/33 & 22/33 \\

            \bottomrule
        \end{tabular}
    }
    \caption{
{\bf Predicting antigen-stabilizing mutations with HERMES: extended results.} Recall for different models (columns) is evaluated on 33 previously reported antigen-stabilizing mutations (rows) spanning five viral antigens. For each antigen, we list the PDB structure used for scoring and the publication(s) that originally reported the mutation. Mutations are specified as wild-type$\to$mutant substitutions at the annotated site. Seven models are compared (columns). We additionally report the BLOSUM62 substitution score for each mutation and the mutant's rank among the 20 possible amino-acid substitutions for the wild-type residue (per BLOSUM62). For each model and mutation, predicted ranks of the wild-type and mutant amino acids are shown as $r_\wt \to r_\mt$. Dagger symbols ($\dagger$) indicate mutations originally proposed as stabilizing by Rosetta-based pipelines in the source reference. When a model ranks the mutant better than the wild type ($r_\mt < r_\wt$), the cell is shaded by the prediction strength  based on the value of $r_\mt$: dark green, strongly suggested ($r_\mt \leq 3$); light green, moderately suggested ($4 \leq r_\mt \leq 6$); light yellow, weakly suggested ($r_\mt \geq 6$). Column summaries report counts of strongly, at least moderately, and at least weakly suggested mutations (out of 33); {\bf bold} indicates significance (p-value $<0.05$) for the number of recalled mutations relative to a random null model (see Fig.~\ref{fig:antigen_pvalues_vs_random} for p-values and Methods  for details.) All structures were scored in their native multimeric states, generating symmetric partners when needed. ThermoMPNN's native mode predicts mutation effects only for monomers, ignoring multimeric assemblies even when present in the input structure. ``Uncl." stands for uncleaved prefusion-closed state.
    }
    \label{table:antigen_results}
\end{table}

\begin{table}[h!]
    \centering
    \begin{tabular}{c c | c c | c c}
        \toprule
          & & \makecell{\textbf{HERMES GPU}\\{[seconds]}} & \makecell{\textbf{HERMES CPU}\\{[seconds]}} & \multicolumn{2}{|c}{\makecell{\textbf{Rosetta}\\{[CPU-hours]}}} \\
        pdbID & \makecell{\# of monomer sites} & \makecell{all sites} & \makecell{all sites} & \makecell{one site} & \makecell{all sites} \\
        \midrule
        4JHW & 449 & 57 & 112 & 150 & 67,350 \\
        7VDF & 485 & 64 & 118 & 164 & 79,540 \\
        5WB0 & 442 & 43 & 154 & 151 & 66,742 \\
        1OAN & 394 & 32 & 151 & 59 & 23,246 \\
        6VSB & 968 & 69 & 265 & 156 & 151,008 \\
        \bottomrule
    \end{tabular}
    \caption{
    \textbf{Execution times for saturation mutagenesis predictions on the viral antigens considered in this study.} Executions times (in seconds) of HERMES apply to HERMES-{\em fixed} and HERMES-{\em amortized} models, regardless of whether zero-shot or fine-tuned. Times were computed when running the script \texttt{run\_hermes\_on\_pdbfiles.py} providing as input the pdbfile as well as a single monomeric chain. A single CPU core with 64GB of memory was used, and a NVIDIA A40 GPU when applicable. For Rosetta, we computed times (in CPU-hours) for a single CPU core with 4 GBs of memory, and averaging 10 relaxation instances, which we consider the minimum number of instances for robust results. Times for all sites in the structure were extrapolated by multiplying the calculated average time for a single mutation by the number of monomeric sites.
    }
    \label{table:speed_on_antigens}
\end{table}

\begin{table}[h!]
    \centering
    \resizebox{\textwidth}{!}{
        \begin{tabular}{c c c | c c c}
            \toprule
            \makecell{antigen\\PDB id} &
            \makecell{stabilized\\name} &
            \makecell{mutation} &
            \makecell{mutation\\type} &
            \makecell{is synergistic} &
            \makecell{notes} \\

                \midrule
            \multirow{7}{*}{\makecell{RSV-F\\4JHW}} & \multirow{2}{*}{Cav1 \citep{mclellan_structure-based_2013}} & S190F & cavity-filling & False &   \\
             &  & V207L & cavity-filling & False &   \\
            \cmidrule(lr){2-6}
             & \multirow{1}{*}{Uncl. \citep{lee_rational_2024}} & S215P & proline & False &   \\
            \cmidrule(lr){2-6}
             & \multirow{4}{*}{TriC \citep{mclellan_structure-based_2013}} & D486H & electrostatic & True &   \\
             &  & E487Q & electrostatic & True &   \\
             &  & F488W & cavity-filling & True &   \\
             &  & D489H & electrostatic & True &   \\
            \midrule
            \multirow{3}{*}{\makecell{HA\\7VDF}} & \multirow{3}{*}{Universal-HA \citep{milder_universal_2022}} & H355W & cavity-filling & False &   \\
             &  & K380I & cavity-filling & False &   \\
             &  & E432I & cavity-filling & False &   \\
            \midrule
            \multirow{11}{*}{\makecell{hMPV-F\\5WB0}} & \multirow{5}{*}{M104 \citep{gonzalez_general_2024}} & L130D & electrostatic & False &   \\
             &  & A159L & cavity-filling & False &   \\
             &  & V203I & cavity-filling & False &   \\
             &  & V430Q & electrostatic & False &   \\
             &  & V449D & electrostatic & False &   \\
            \cmidrule(lr){2-6}
             & \multirow{4}{*}{MPV-2cREKR \citep{bakkers_efficacious_2024}} & V112R & electrostatic & False &   \\
             &  & D209E &  & False & same charge, slightly different size: unclear  \\
             &  & V231I & cavity-filling & False &   \\
             &  & E453P & proline & False &   \\
            \cmidrule(lr){2-6}
             & \multirow{2}{*}{Uncl. \citep{lee_rational_2024}} & E80D &  & False & same charge, slightly different size: unclear  \\
             &  & V155P & proline & False &   \\
            \midrule
            \multirow{8}{*}{\makecell{DENV-E\\1OAN}} & \multirow{8}{*}{SC12 \citep{phan_conserved_2022}} & S29K & electrostatic & False &   \\
             &  & T33V & cavity-filling & False &   \\
             &  & A35M & cavity-filling & False &   \\
             &  & G106D & electrostatic & False &   \\
             &  & A259W & cavity-filling & True &   \\
             &  & T262R & electrostatic & True &   \\
             &  & F279W & cavity-filling & False &   \\
             &  & T280P & proline & False &   \\
            \midrule
            \multirow{4}{*}{\makecell{SARS-Cov-2\\6VSB}} & \multirow{4}{*}{hexapro \citep{hsieh_structure-based_2020}} & F817P & proline & False &   \\
             &  & A892P & proline & False &   \\
             &  & A899P & proline & False &   \\
             &  & A942P & proline & False &   \\

            \bottomrule
        \end{tabular}
    }
    \caption{\textbf{Characteristics of antigen-stabilizing mutations.}
    Hand-curated mutation types are listed for antigen-stabilizing mutations reported in Fig.~\ref{fig:antigen_figure_main_results} and Table~\ref{table:antigen_results}. Cavity-filling mutations are defined as substitutions to hydrophobic residues that are larger than the wild-type when the wild-type is also hydrophobic. Electrostatic mutations are substitutions that change the residue's net charge. Proline mutations correspond to substitutions to proline. Synergistic mutations were identified through structural reasoning based on the spatial arrangement of mutations within the corresponding structure; see ref.~\citep{byrne_principles_2022} for a breakdown of mutation types considered in the structure-based vaccine design literature. ``Uncl." stands for ``Uncleaved Prefusion-Closed".
     }
    \label{table:mutation_types}
\end{table}

\clearpage

\begin{figure*}[h!]
    \centering
    \includegraphics[width=1.0\textwidth]{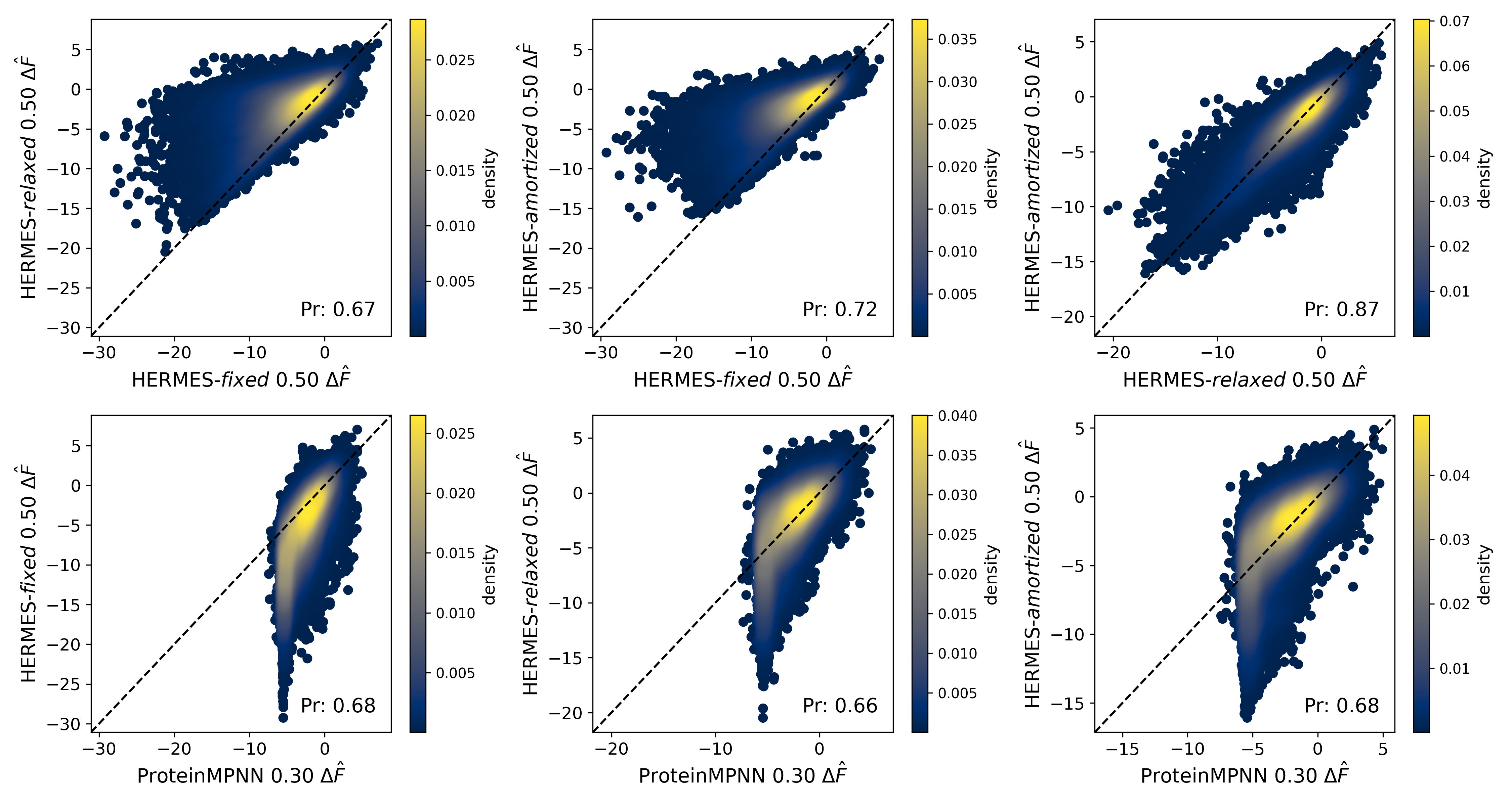}
    \caption{\textbf{Comparison of zero-shot model predictions on the Megascale test set.} For each substitution in the Megascale test set, the predicted change in amino acid propensity upon mutation ($\delta \log p$) is compared between two models in each panel. Color indicates local point density (blue denotes low density and yellow denotes high density). The reported “Pr” in each panel corresponds to the Pearson correlation coefficient between the predictions of the model pair. Model names indicate the architecture and the coordinate-noise amplitude used. $\Delta \hat{F}$ indicates the model's prediction, following Equations~\ref{eq:log_ratio_wt_and_mt}~and~\ref{eq:deltaF_variants}.} 
    \label{fig:relaxed_vs_ft_relaxed}
\end{figure*}

\begin{figure}[ht!]
    \centering
    \includegraphics[width=0.85\textwidth]{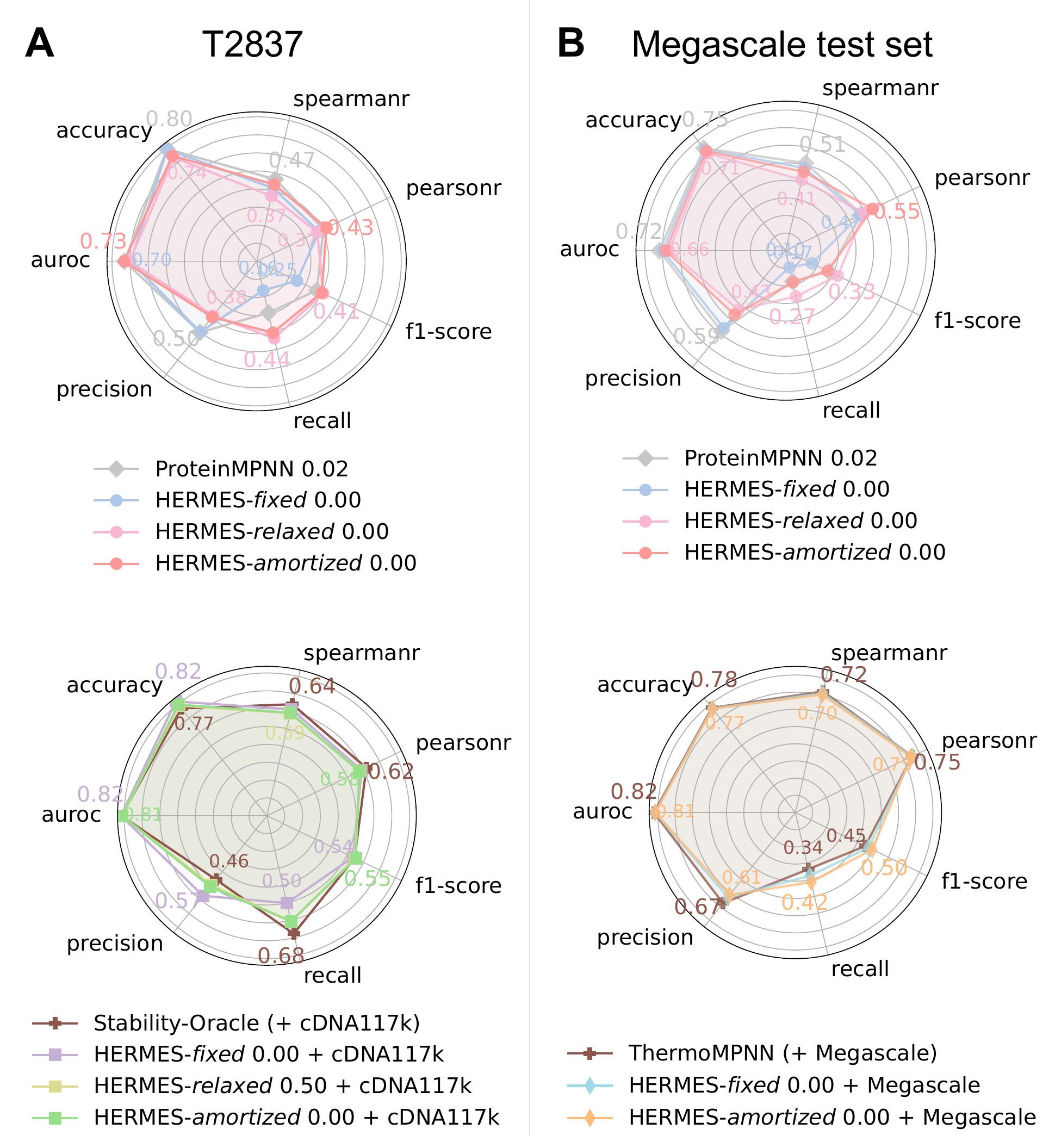}
    \caption{
    \textbf{Predicting mutational effects  on thermodynamic folding stability.} Stabilizing-versus-destabilizing classification metrics are computed using $\Delta\Delta G < 0$ (experimental) and $\Delta \log p > 0$ (predicted) as  cutoffs for  stabilizing mutations.
    \textbf{(A)} Evaluation on the T2837 results: zero-shot models (top) and models fine-tuned on cDNA117k (bottom). \textbf{(B)}  Evaluation on Megascale test set results: zero-shot models (top) and models fine-tuned on the Megascale training set (bottom). Model names indicate the architecture, the coordinate-noise amplitude used, and when applicable, the fine-tuning dataset (listed after ``$+$"); \textit{Untr.} is short for \textit{Untrained}, indicating models that had no pre-training and were instead only trained on stability effects. Only models trained without coordinate noise are shown; the noise amplitude is indicated within each model name.
    }
   \label{fig:radial_plots__no_noise} 
\end{figure}

\begin{figure}
    \centering
    \includegraphics[width=1.0\linewidth]{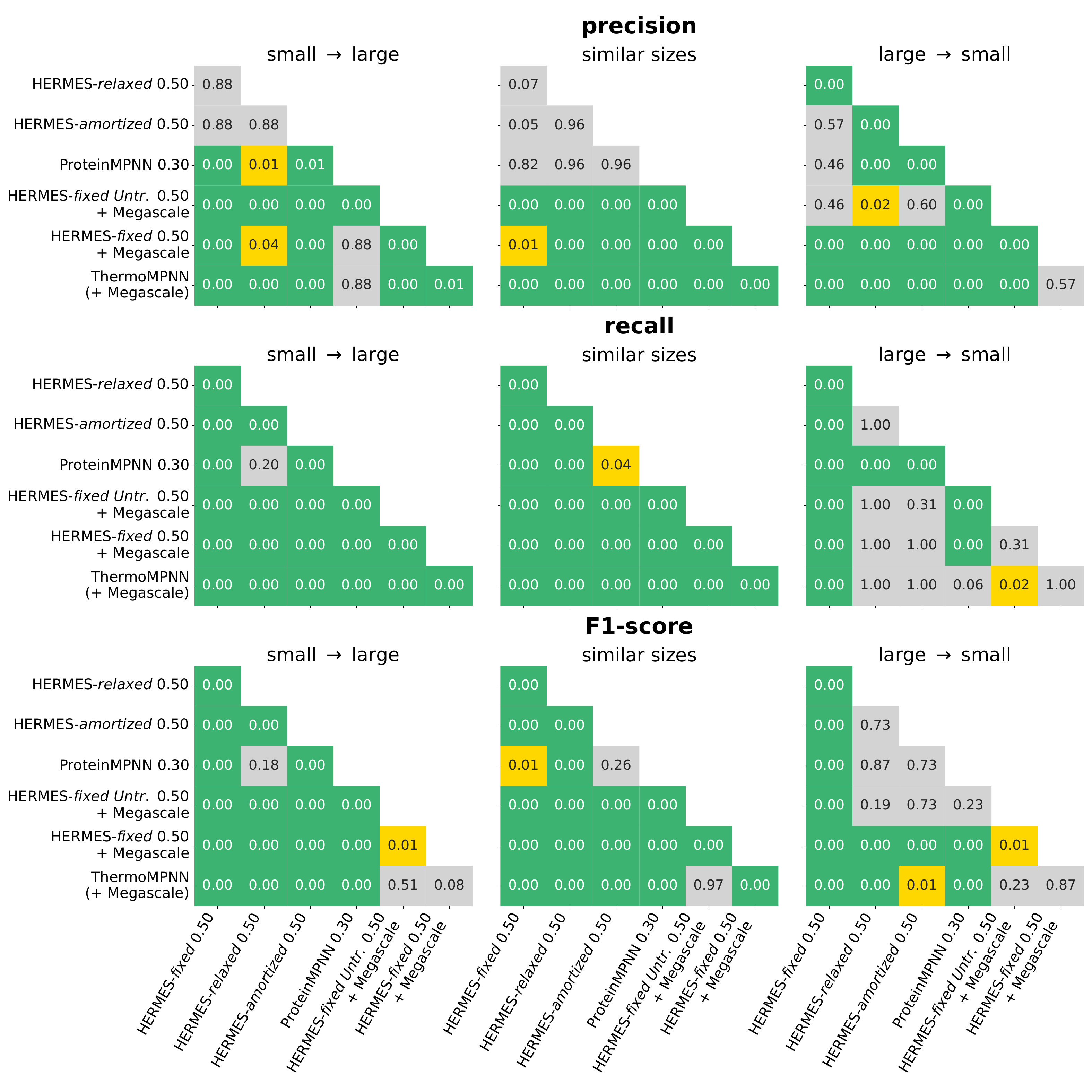}
    \caption{
    {\bf Statistical significance of model performance differences on stabilizing mutation identification.} Shown are two-tailed p-values for differences in model performance on identifying stabilizing mutations across Megascale test set subsets. P-values correspond to the performance comparisons shown in Fig.~\ref{fig:megascale_precision_recall_f1_by_size_cutoff_0p0}. Green indicates strong statistical significance ($p < 0.01$), while yellow indicates weaker significance ($0.01 \leq p < 0.05$). P-values were computed using a permutation test and corrected for multiple comparisons using the Holm–Bonferroni procedure within each performance metric (see Methods for details).}
\label{fig:pairwise_pvalues_permutation_bucketed_by_sizereduced_precision_recall_f1}
\end{figure}

\begin{figure}
    \centering
    \includegraphics[width=0.4\linewidth]{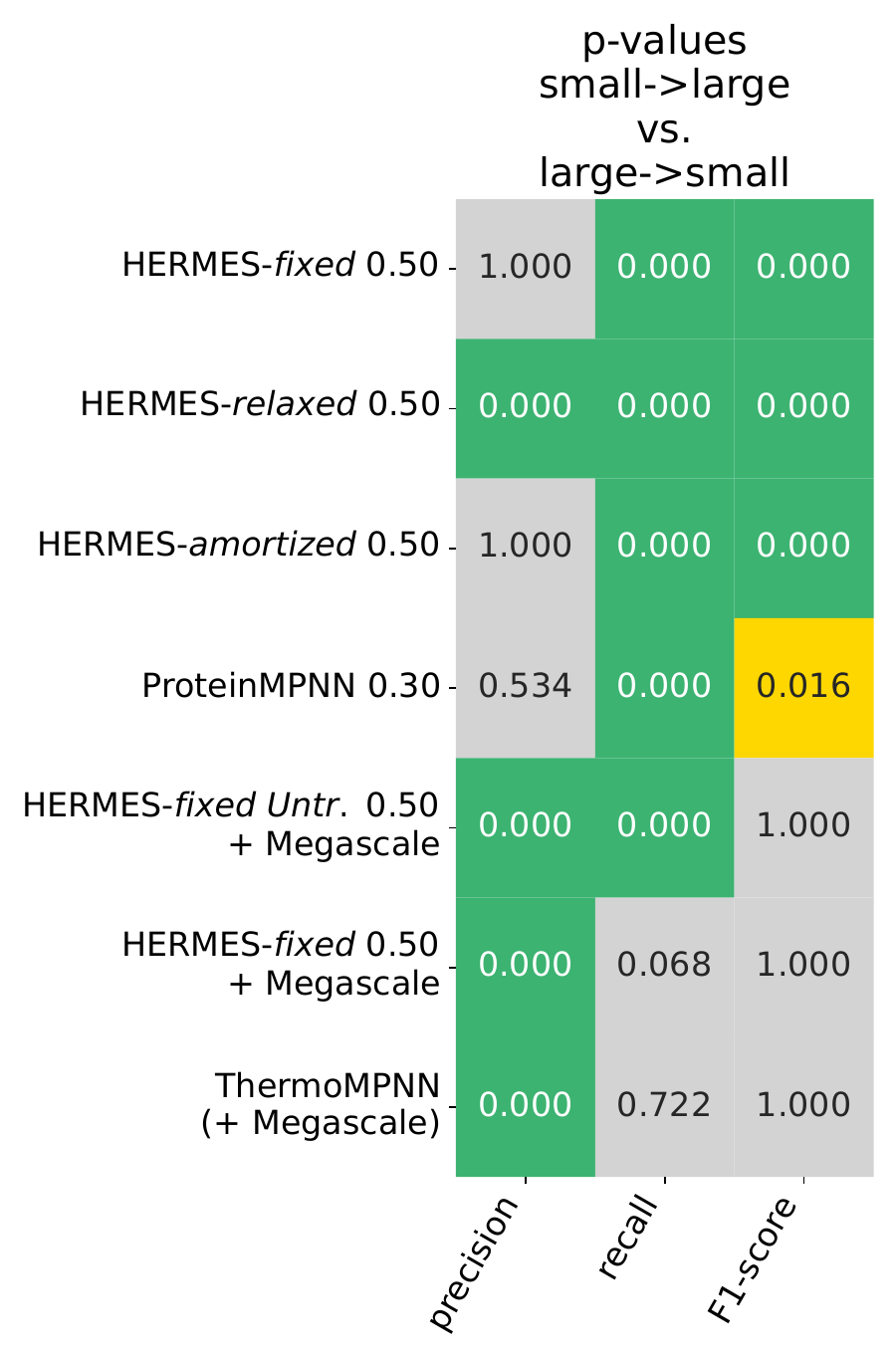}
    \caption{{\bf Statistical significance of within-model performance differences in identifying stabilizing mutations across mutational size classes.}  Shown are two-tailed P-values for within-model differences in performance when identifying stabilizing mutations from the small$\to$large vs. large$\to$small mutational subsets of the Megascale test set. P-values correspond to the performance comparisons shown in Fig.~\ref{fig:megascale_precision_recall_f1_by_size_cutoff_0p0}. Green indicates strong statistical significance ($p < 0.01$), while yellow indicates weaker significance ($0.01\leq p < 0.05$). P-values were computed using a bootstrap test and corrected for multiple comparisons using the Holm–Bonferroni procedure within each performance metric (see Methods for details).}    \label{fig:pvalues__bucketed_by_size__between_small_large_buckets__vertical}
\end{figure}

\begin{figure}
    \centering
    \includegraphics[width=1.0\linewidth]{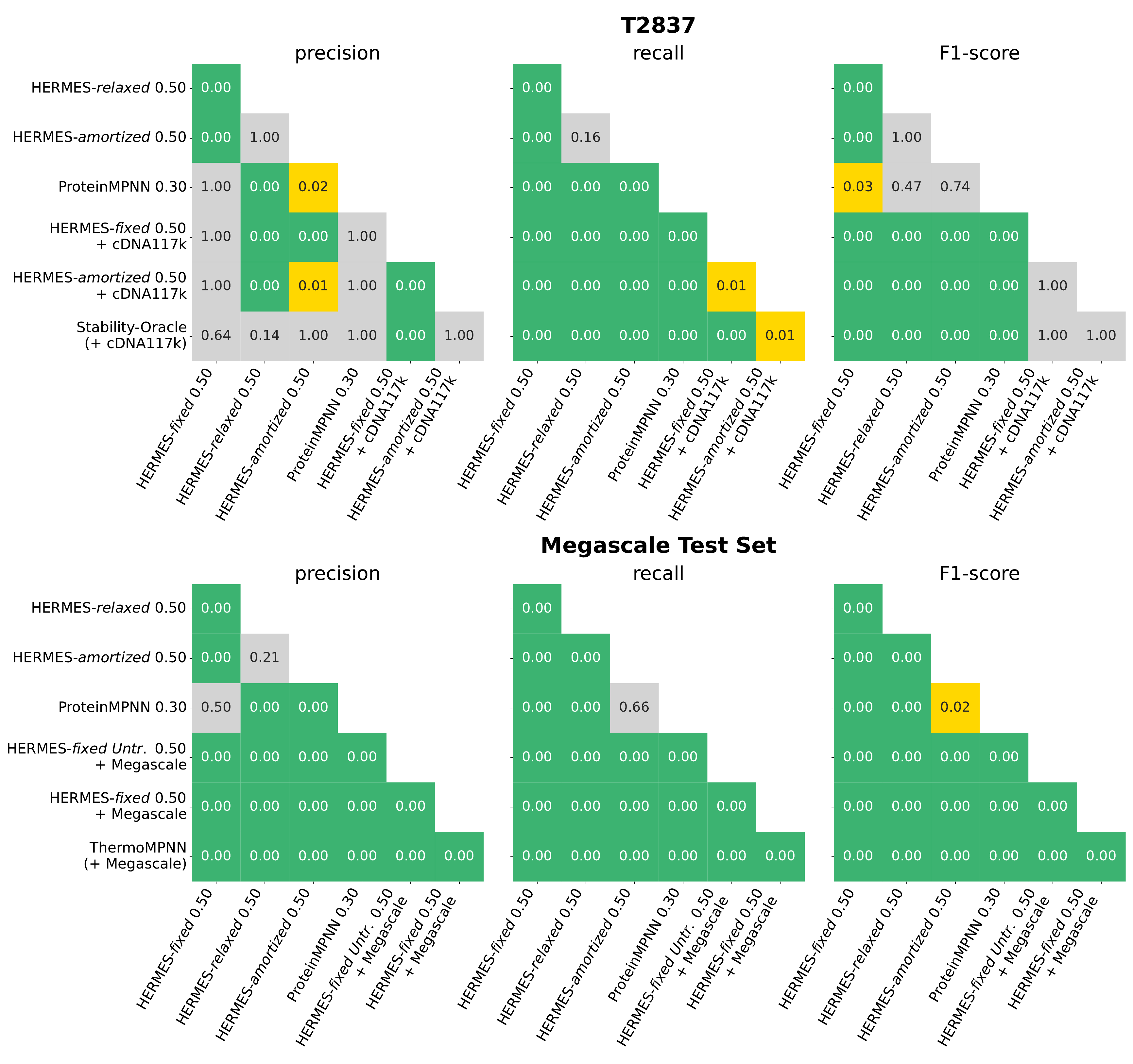}
    \caption{
    {\bf Statistical significance of model performance differences in identifying stabilizing mutations on the full test set.} 
     Shown are two-tailed p-values for differences in model performance when identifying stabilizing mutations on the full test set. P-values correspond to the performance comparisons shown in Fig.~\ref{fig:radial_plots__noise}. Green indicates strong statistical significance ($p < 0.01$), while yellow indicates weaker significance ($0.01\leq p < 0.05$).  P-values were computed using a permutation test and corrected for multiple comparisons using the Holm–Bonferroni procedure within each performance metric (see Methods for details).}
    \label{fig:t2837_and_megascale_pairwise_pvalues_permutation}
\end{figure}

\begin{figure}
    \centering
    \includegraphics[width=0.7\linewidth]{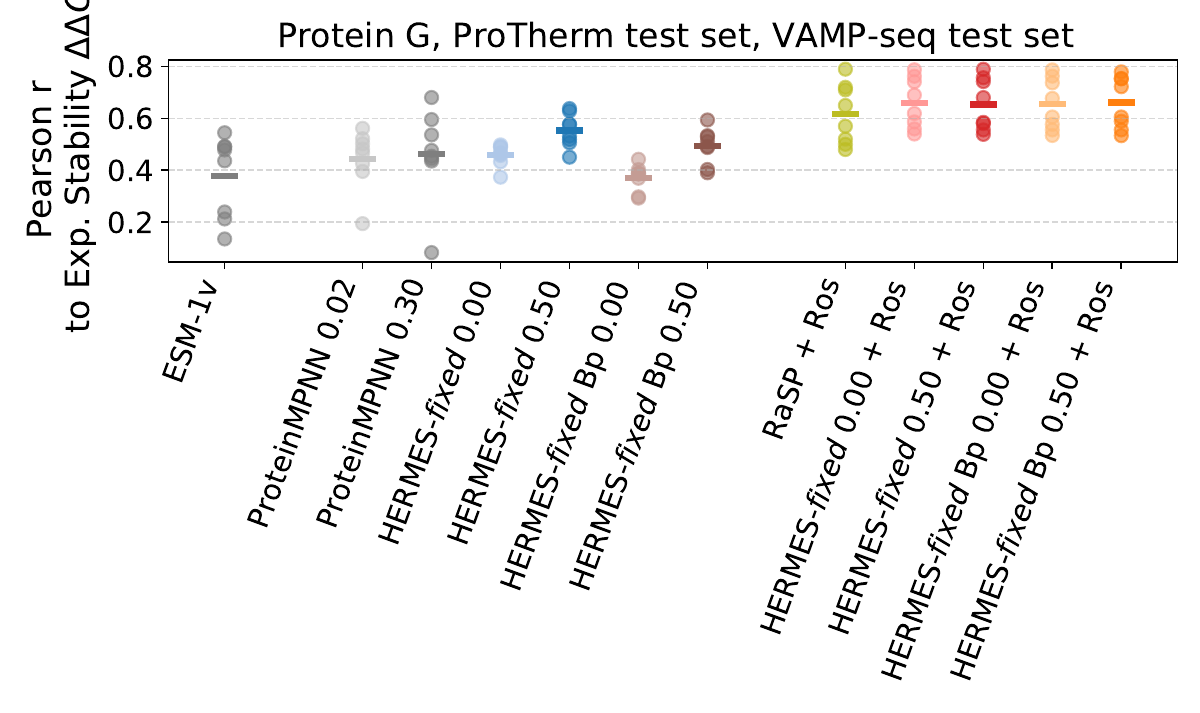}
    \caption{\textbf{Pearson correlation between model predictions and experimental stability effects on the RaSP test set (8 proteins)~\citep{blaabjerg_rapid_2023}.} 
    Each dot represents one protein, and the horizontal bar indicates the mean correlation across proteins. Model labels specify the architecture, the coordinate-noise amplitude, and, when applicable, the fine-tuning dataset (denoted after ``$+$"). ``Bp" indicates the use of our open-source Biopython-based protein Pre-processing. See Methods for details on the RaSP dataset.}
    \label{fig:rasp_exp}
\end{figure}

\begin{figure}
    \centering
    \includegraphics[width=0.9\linewidth]{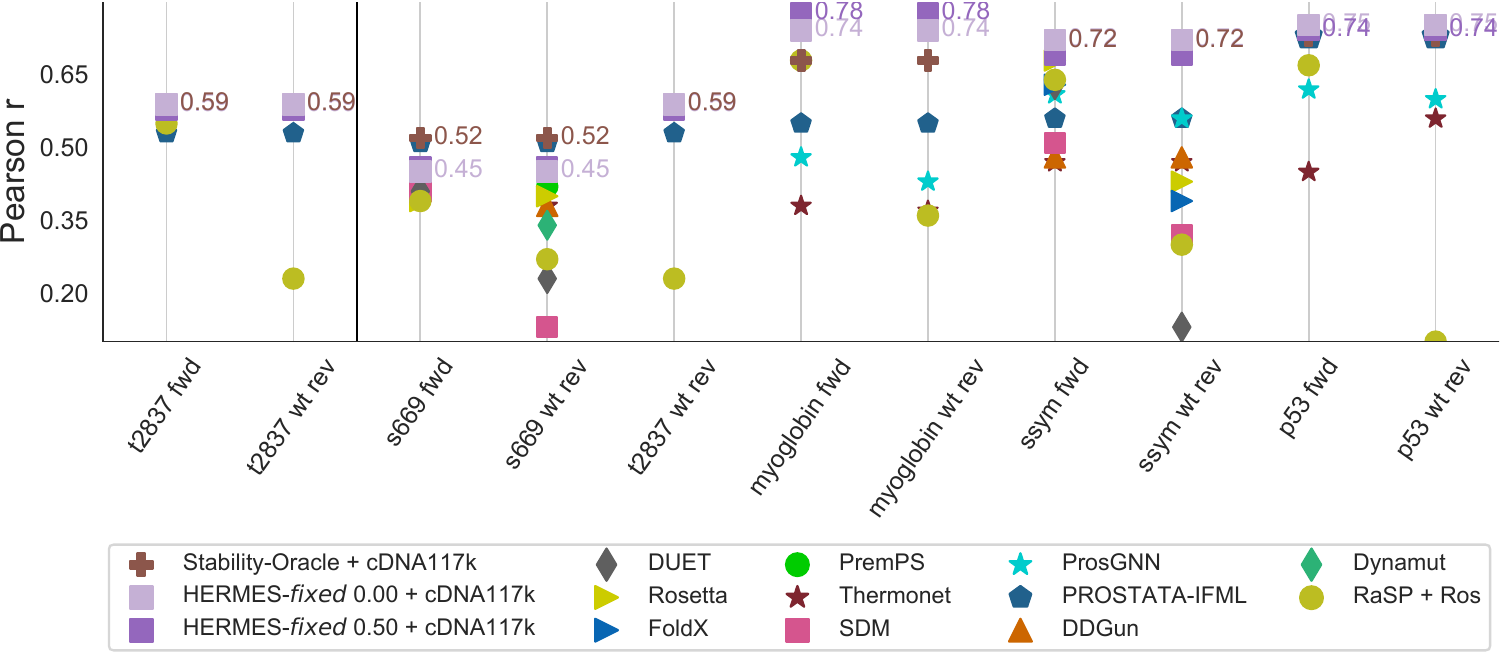}
    \caption{{\bf Pearson correlation between model predictions and experimental stability effects on the T2837 dataset and its subsets.}
Pearson correlation values for all models other than HERMES are taken from~\citep{diaz_stability_2024}. This figure closely replicates a figure from ref.~\citep{diaz_stability_2024}, with the key difference that predictions for ``reverse" mutations are computed here by conditioning on wild-type structures (denoted as ``wt rev"). This distinction is made to avoid confusion with ``reverse" mutation predictions computed on mutant structures in the Ssym dataset (Fig.~\ref{fig:ssym_figure}). For each dataset (x-axis), we denote in text the performance of the HERMES models as well as that of the best-performing model overall.}
    \label{fig:t2837_broken_down_and_comparison}
\end{figure}

\begin{figure}
    \centering
    \includegraphics[width=0.5\linewidth]{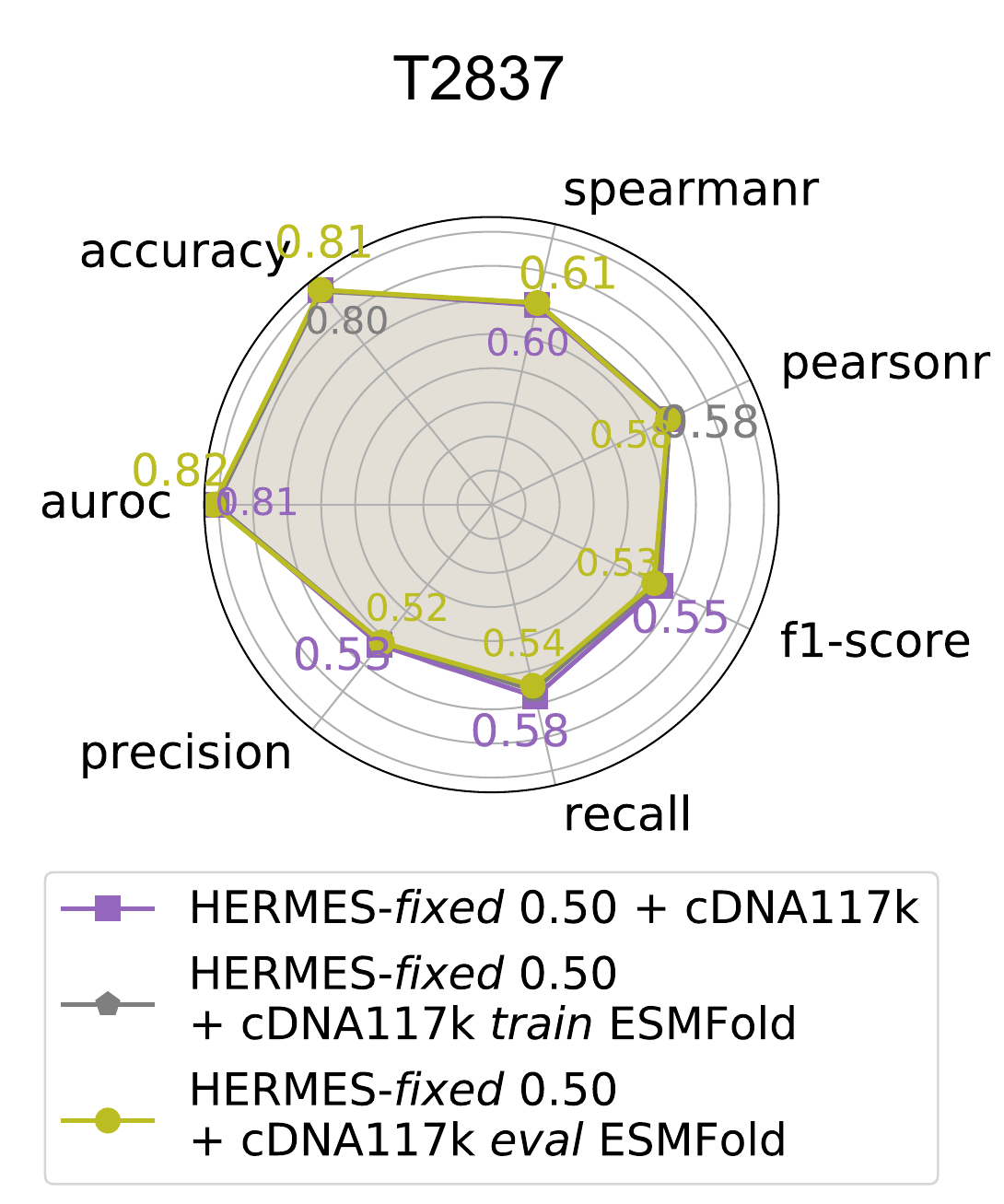}
    \caption{\textbf{Predicting mutational effects on thermodynamic stability using ESMfold predicted structures for fine-tuning or testing.} Stabilizing-versus-destabilizing classification metrics are computed using $\Delta\Delta G < 0$ (experimental) and $\Delta \log p > 0$ (predicted) as  cutoffs for  stabilizing mutations. We report results on the T2837 dataset, after fine-tuning models on cDNA117k. We consider models fine-tuned and evaluated on crystal structures (purple), models fine-tuned on ESMfold predicted structures and evaluated on crystal structures (grey, ``\textit{train} ESMfold" in the model name), and models fine-tuned on crystal structures and evaluated on ESMfold predicted structures (olive, ``\textit{eval} ESMfold" in the model name).
    }
    \label{fig:radial_plots__esmfold}
\end{figure}

\begin{figure*}
    \centering
    \includegraphics[width=0.95\textwidth]{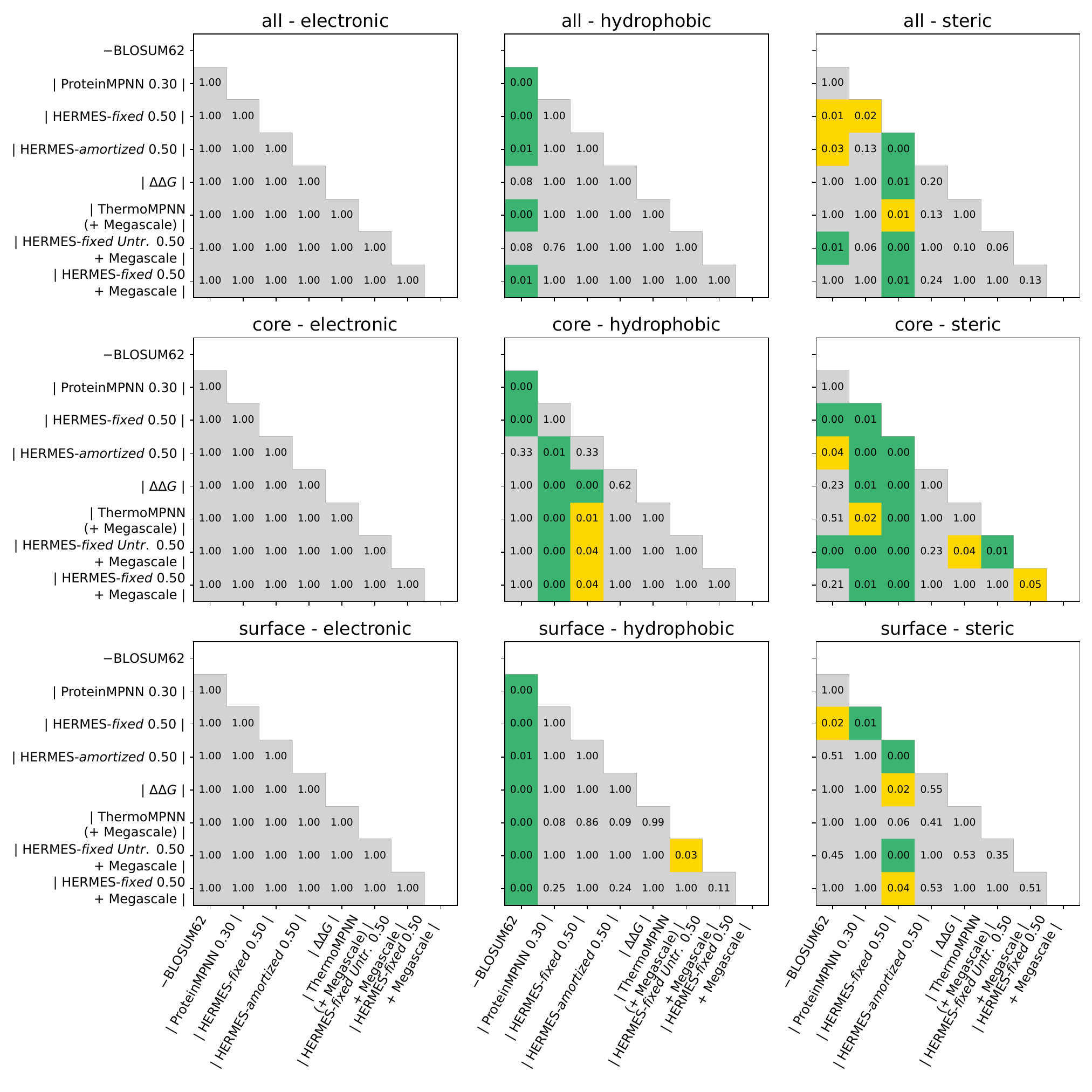}
    \caption{\textbf{Significance testing results for Fig.~\ref{fig:correlations_to_aa_properties}B.}
    Shown are p-values of two-tailed t-tests comparing the distributions of spearman correlations between each model's substitution matrix, and amino-acid properties of a particular class (electronic, hydrophobic, steric). The Holm-Bonferroni method was used to correct p-values for multiple testing error. Entries corresponding to pairs of distributions with p-value $\leq 0.01$ are colored in green, p-value $\leq 0.05$ are colored in yellow, and p-value $> 0.05$ are colored in gray.}
    \label{fig:correlation_to_aa_properties__significance}
\end{figure*}

\begin{figure*}
    \centering
    \includegraphics[width=0.6\linewidth]{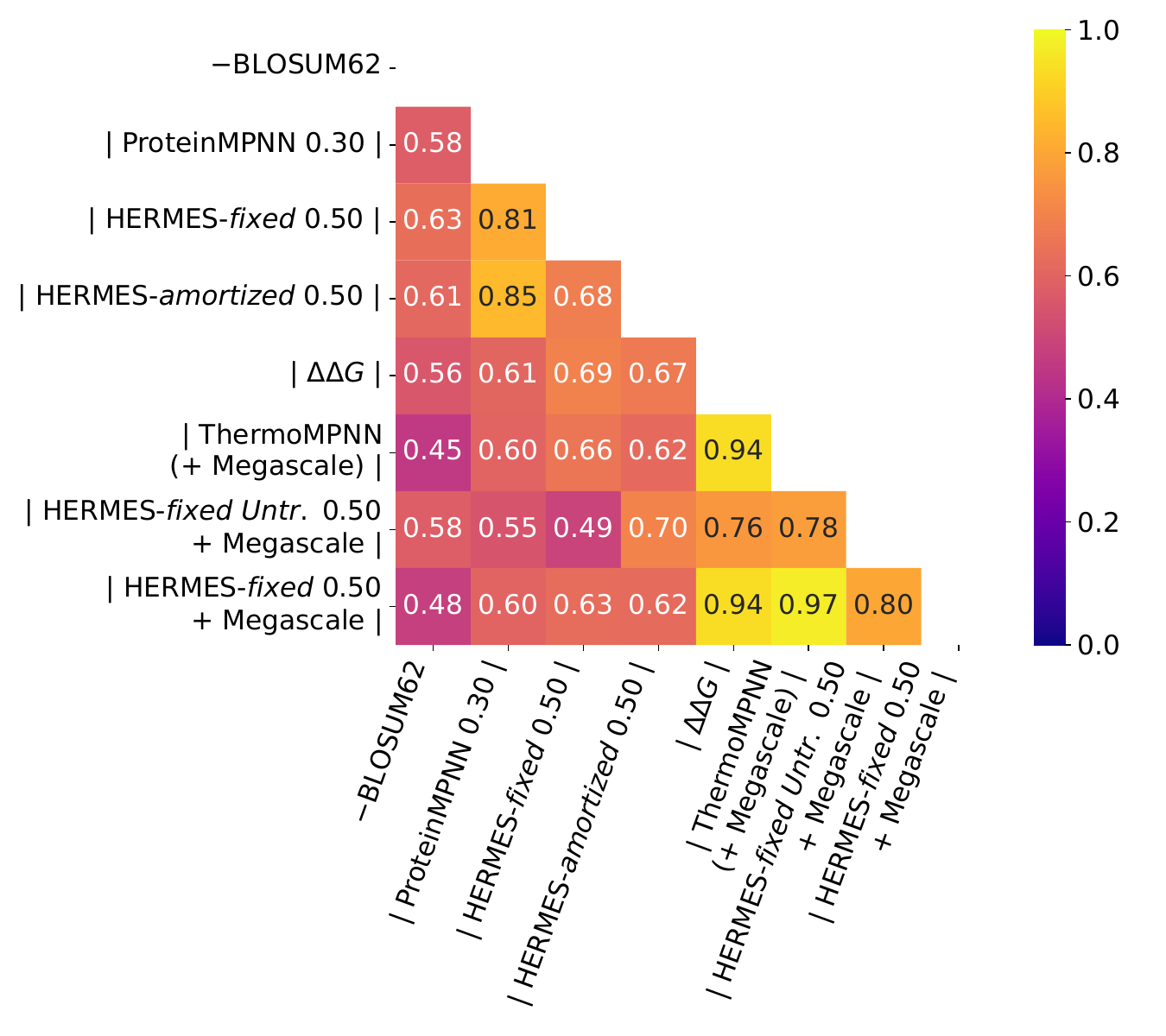}
    \caption{\textbf{Spearman correlations between pairs of model-average substitution matrices $M^{\model}$.} The heatmaps for the underlying model-predicted substitution matrices are shown in Fig.~\ref{fig:correlations_to_aa_properties}.}
    \label{fig:spearmanr_between_model_average_substitution_matrices}
\end{figure*}

\begin{figure*}
    \centering
    \includegraphics[width=0.8\textwidth]{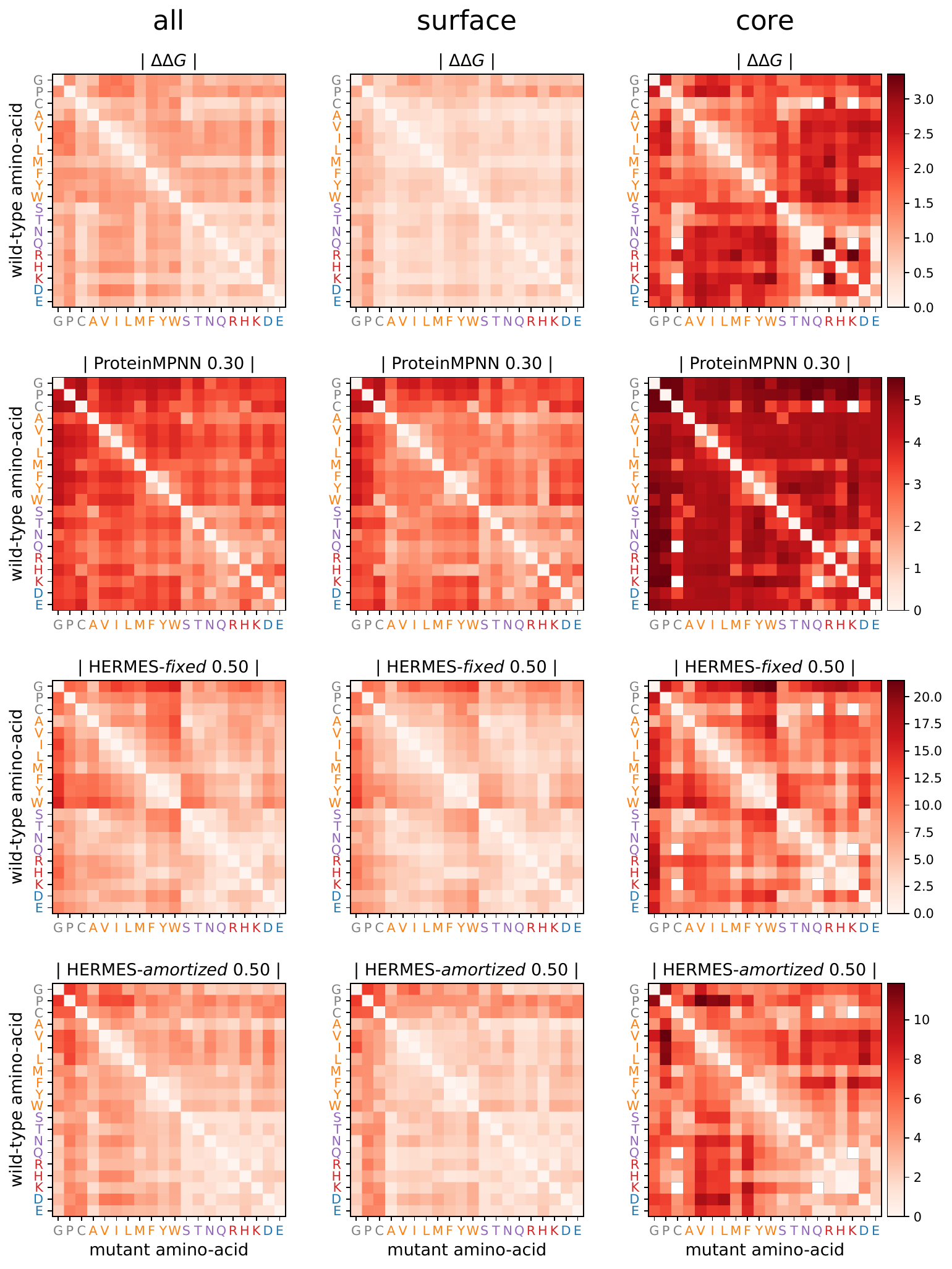}
    \caption{\textbf{Model-averaged substitution matrices stratified by protein core and surface residues for zero-shot models.}
Shown are model-averaged substitution matrices $M^{\model}$ for different zero-shot models (rows 2-4), computed from subsets of sites in the Megascale test set. The first row shows the  experimental matrices for mean $|\Delta \Delta G|$ values across mutation subsets. Columns correspond to all residues (left), core residues with solvent-accessible surface area SASA $< 1 \AA^2$ (center), and surface residues with SASA $> 3 \AA^2$ (right).}
    \label{fig:average_prediction_matrices__abs_symm__part_1}
\end{figure*}

\begin{figure*}
    \centering
    \includegraphics[width=0.8\textwidth]{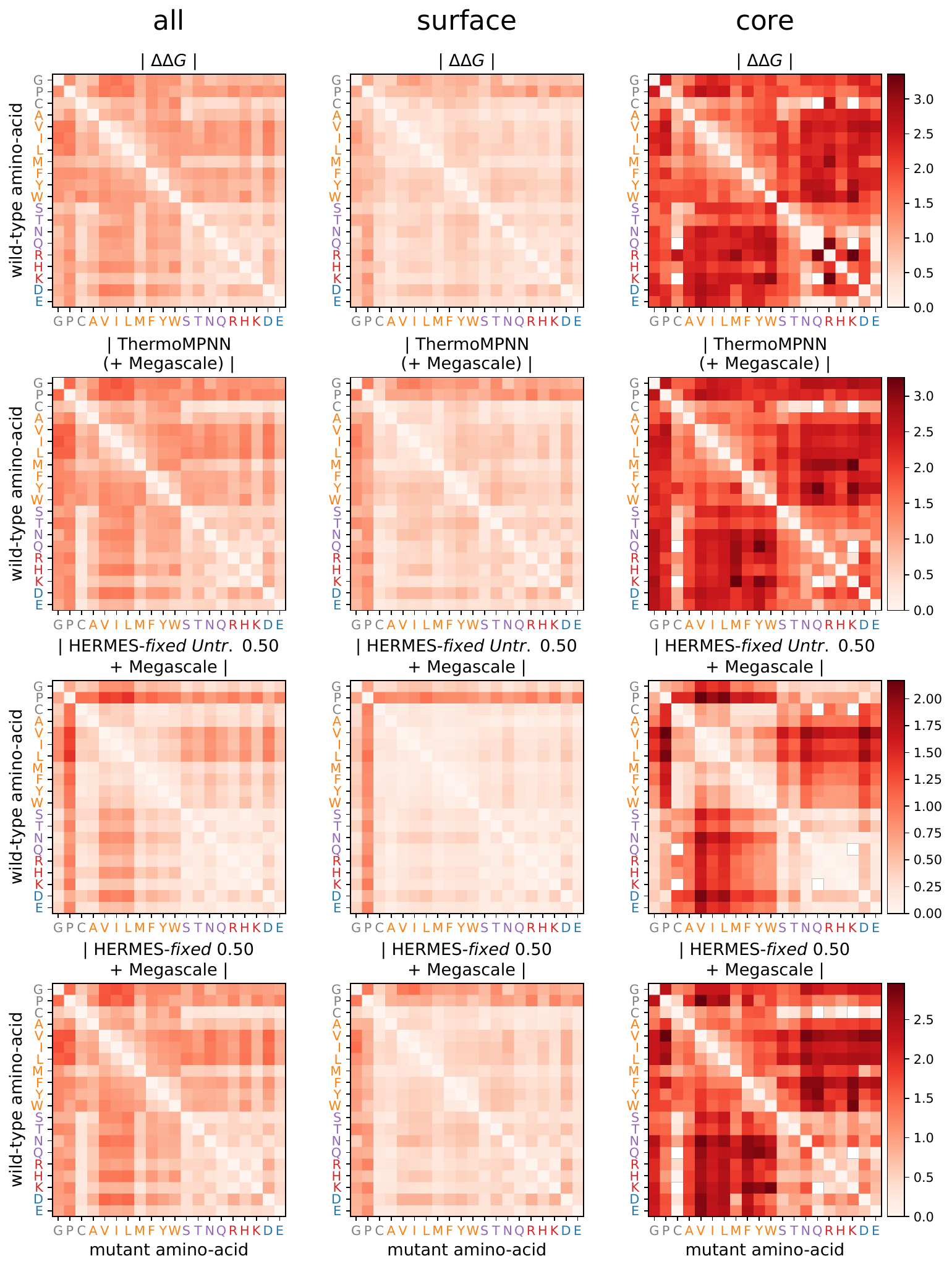}
    \caption{\textbf{Model-averaged substitution matrices stratified by protein core and surface residues for stability fine-tuned models.}
     Similar to Fig.~\ref{fig:average_prediction_matrices__abs_symm__part_1} but for models fine-tuned on the Megascale dataset.}
    \label{fig:average_prediction_matrices__abs_symm__part_2}
\end{figure*}

\begin{figure*}
    \centering
    \includegraphics[width=0.9\linewidth]{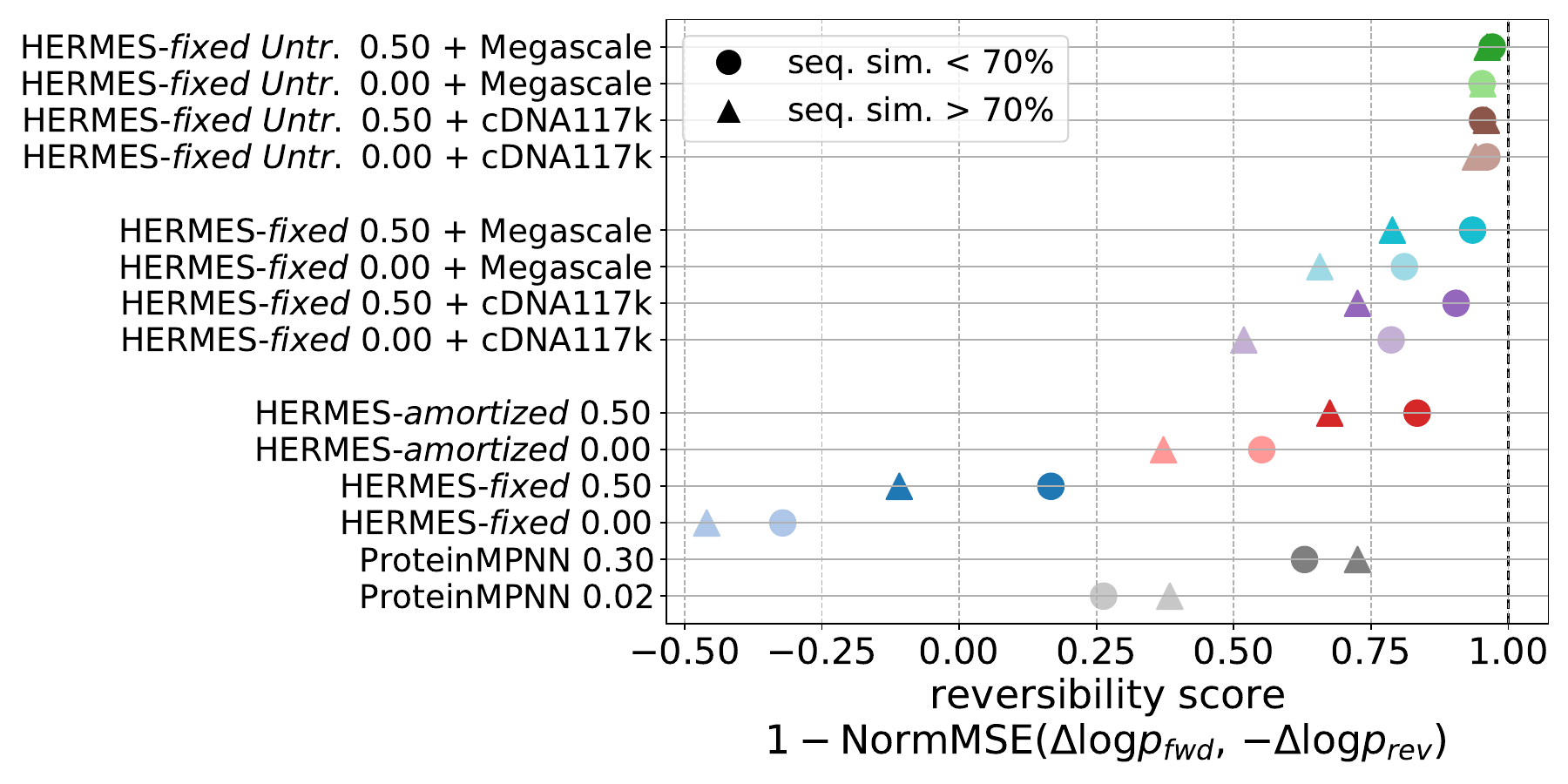}
    \caption{\textbf{Model reversibility scores on the Ssym dataset.}
    Reversibility is quantified as one minus the mean squared error between the forward mutational effect $\Delta\log p_{fwd}$ and the negated reverse effect $-\Delta\log p_{rev}$ for different models (rows), where model predictions are conditioned on the protein structure containing the outgoing amino acid; The resulting score is normalized to lie  between -1 and 1, with higher values indicating a greater degree of reversibility: $1 - mean((\Delta\log p_{fwd}+\Delta\log p_{rev})^2) / (mean({\Delta\log p_{fwd}}^2)+mean({\Delta\log p_{rev}}^2))$. 
    }
    \label{fig:ssym_antisymmetry_reversibility_score_results}
\end{figure*}

\begin{figure}
    \centering
\includegraphics[width=0.95\linewidth]{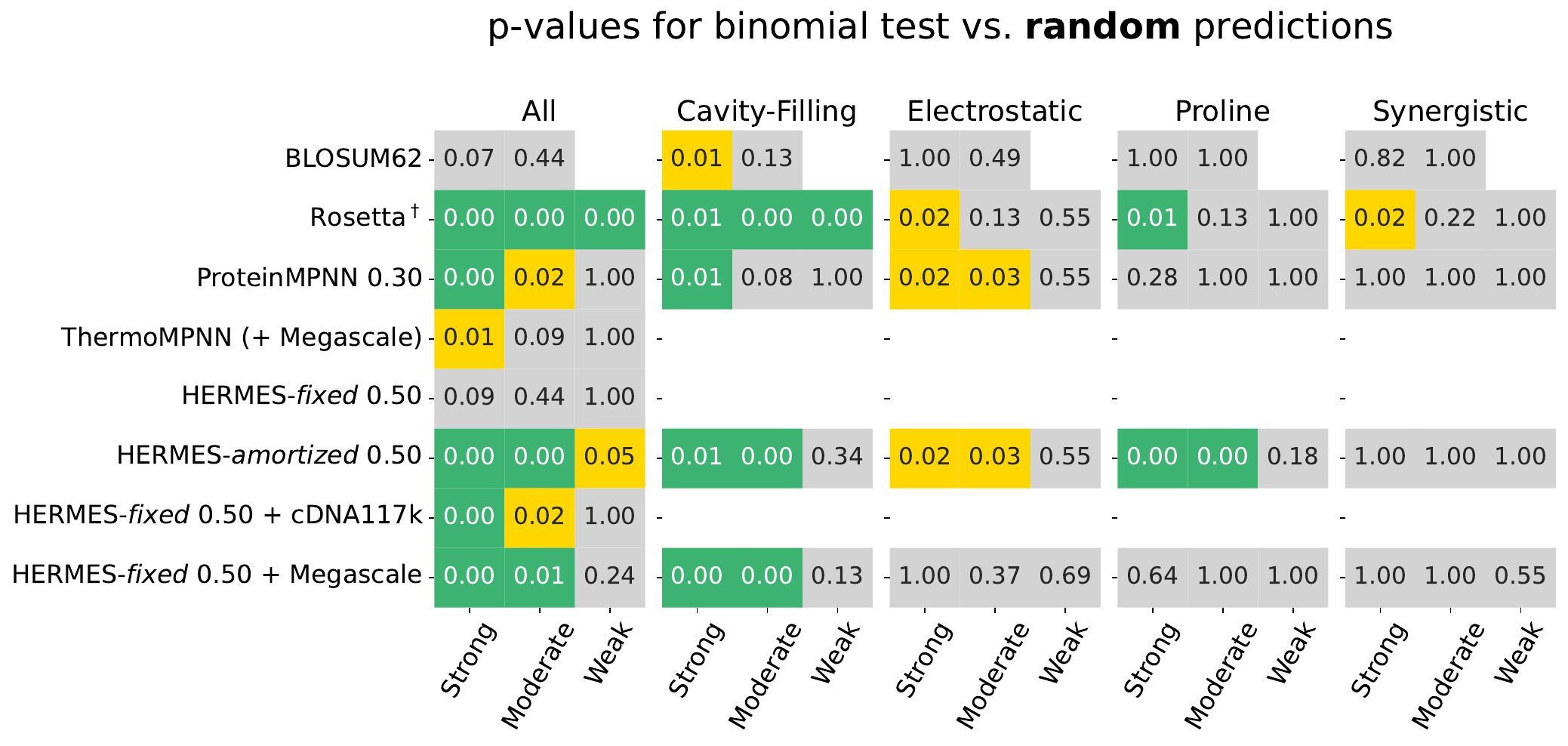}\vspace{0.5cm}
\includegraphics[width=0.95\linewidth]{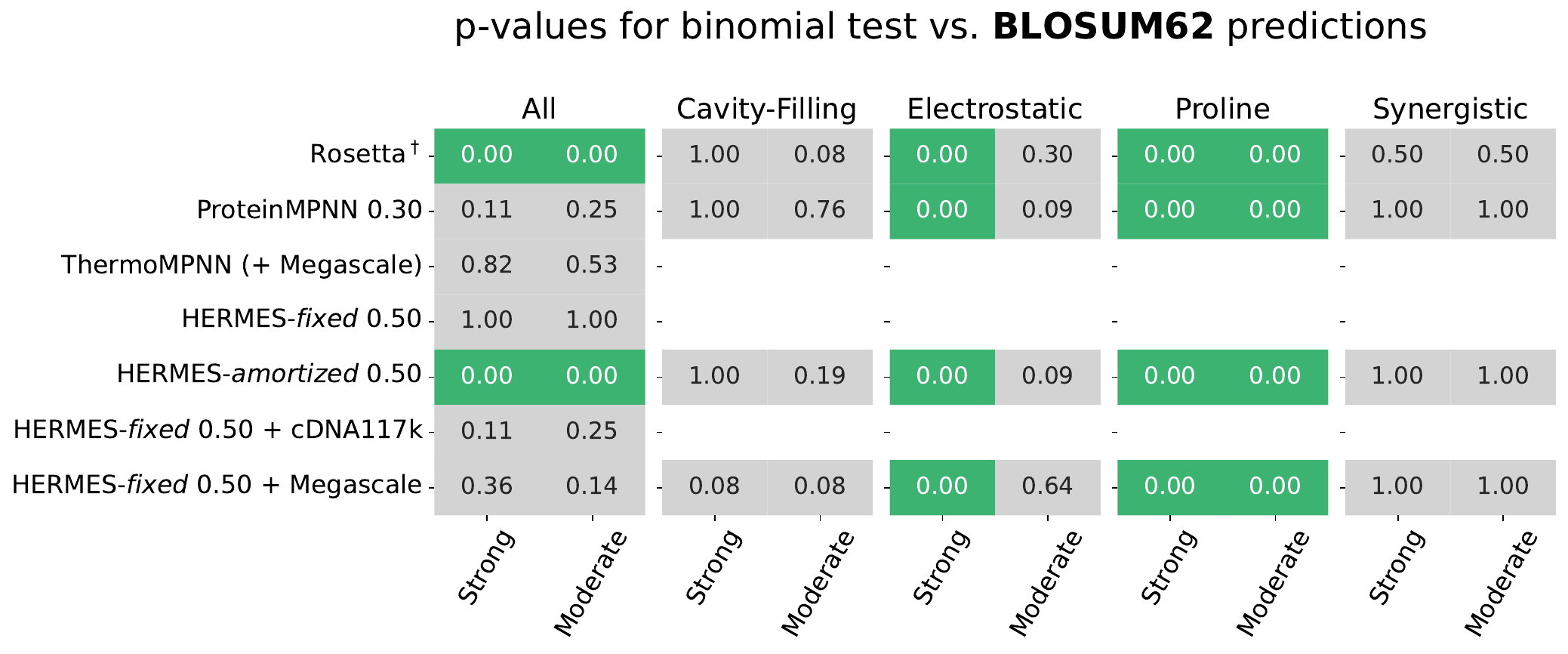}
    \caption{
    {\bf Statistical significance for the number of retrieved antigen-stabilizing mutations.}
 Shown are p-values from binomial tests comparing the number of antigen-stabilizing mutations retrieved by each model (rows) against random expectation (top) and the BLOSUM62 predictions (bottom). These significance tests correspond to the results reported in Figs.~\ref{fig:antigen_figure_main_results},~\ref{fig:antigen_figure_of_different_types_of_mutations} and Table~\ref{table:antigen_results}; see Methods for details of p-value computation.}
    \label{fig:antigen_pvalues_vs_random}
    \label{fig:antigen_pvalues_vs_blosum62}
\end{figure}

\begin{figure}
    \centering
    \includegraphics[width=0.8\linewidth]{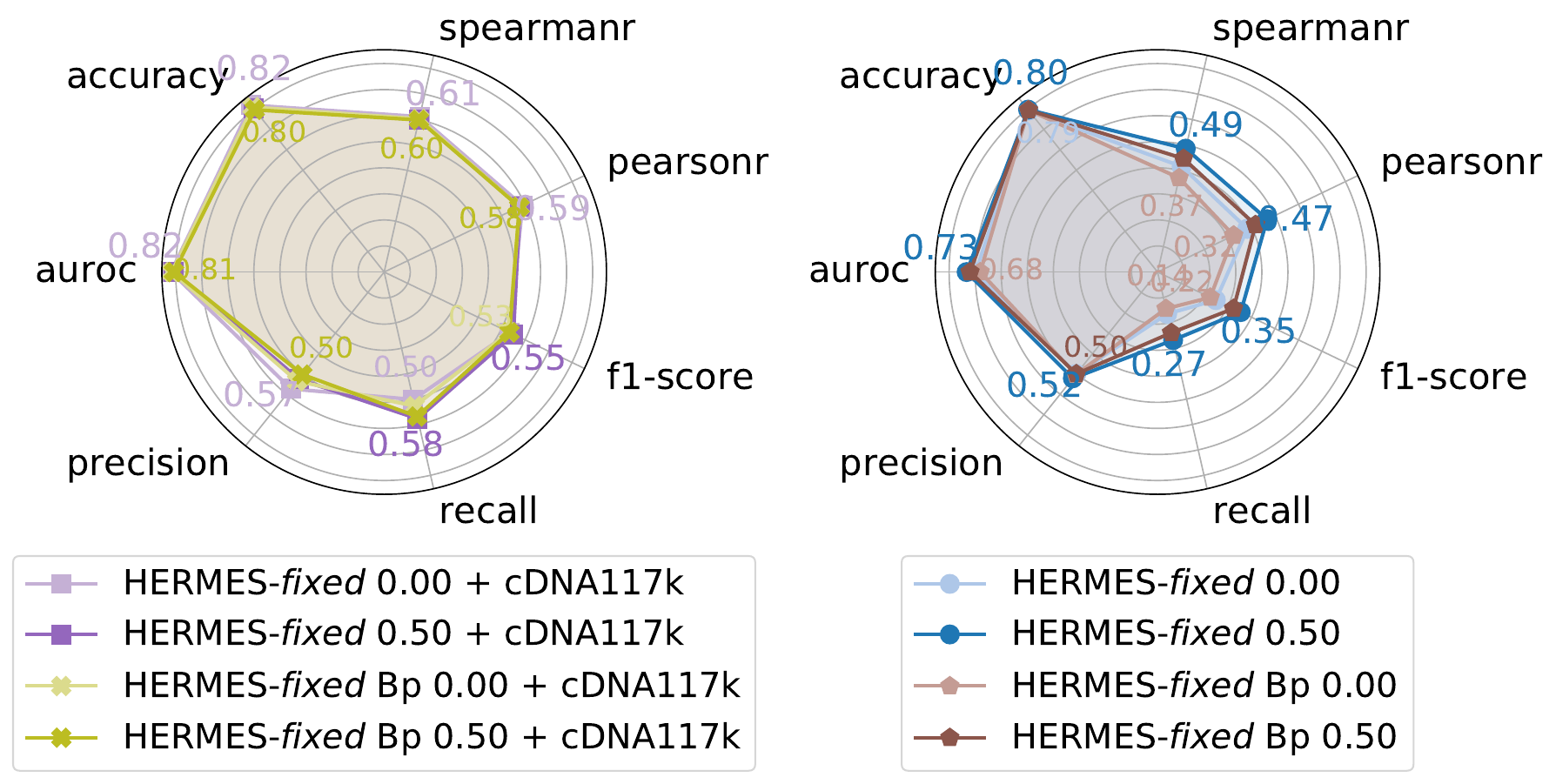}
    \caption{\textbf{Comparison of PyRosetta and Biopython Pre-processing pipelines for predicting mutation stability effects on T2837.} Classification accuracy metrics, analogous to those shown in Fig.~\ref{fig:radial_plots__noise}, are reported for fine-tuned models (left) and zero-shot models (right). In each case, models trained using PyRosetta-based Pre-processing are compared with those using Biopython-based Pre-processing (denoted by “Bp” in the model name).
    Model labels specify the architecture, the coordinate-noise amplitude, and, when applicable, the fine-tuning dataset (listed after ``+"). Consistent with results on the RaSP dataset (Fig.~\ref{fig:rasp_exp}), Biopython-Pre-processed models show slightly reduced performance relative to PyRosetta-Pre-processed models; however, this difference becomes statistically insignificant after fine-tuning.}
    \label{fig:t2837_py_vs_bp}
\end{figure}

\begin{figure*}
    \centering
    \includegraphics[width=1.0\linewidth]{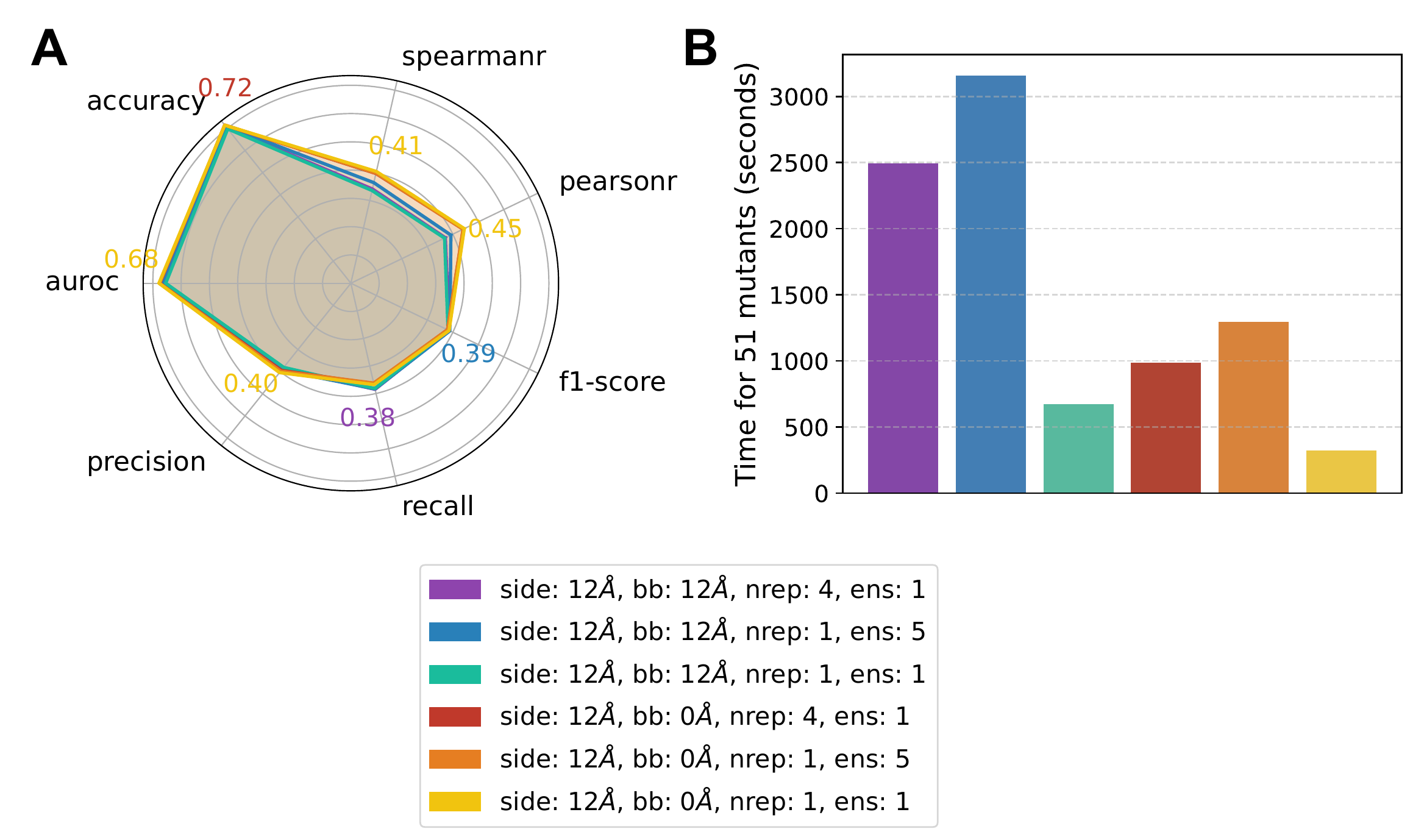}
    \caption{\textbf{Ablation of PyRosetta fastrelax parameters for HERMES-\textit{relaxed} 0.50 on the cDNA117k dataset.}
    \textbf{(A)} Classification accuracy metrics, analogous to those shown in Fig.~\ref{fig:radial_plots__noise}, are reported for HERMES-{\em relaxed} models evaluated on relaxed mutant structures using different PyRosetta fastrelax parameters (indicated by color).
    HERMES-{\em relaxed} scores a mutation as the log-probability difference between the mutant and wild-type amino acids. The wild-type log-probability is evaluated on the wild-type structure, while the mutant log-probability is evaluated on the wild-type structure after introducing the mutation and performing local relaxation. We use the PyRosetta fastrelax protocol and vary the following parameters, noting that the procedure is stochastic: (1) {\bf side}, the distance cutoff for side-chain relaxation; (2) {\bf bb}, the distance cutoff for backbone relaxation; (3) {\bf nreps}, the number of protocol repetitions, with the lowest-energy conformation retained; (4) {\bf ens}, the ensemble size, where predictions are averaged over relaxations obtained with different random seeds.
    {\textbf{(B)} Inference speed for predicting mutational effects on 51 mutants across the same PyRosetta fastrelax parameters as in  (A) (colors). A single NVIDIA A40 GPU and a single CPU with 64G of memory were used for all parameters.}
    }
    \label{fig:relaxations_ablation}
\end{figure*}

\end{document}